%% file: higher-order.tex
\documentclass[a4paper]{article}
\usepackage{a4wide}

\usepackage{cancel}
\usepackage{pict2e}
\usepackage{graphicx}
\usepackage{xcolor}
\usepackage{newtxtext}
\usepackage{newtxmath}
\usepackage{natbib}
\usepackage{hyperref}
\hypersetup{
    colorlinks = true,
    urlcolor   = blue,
    citecolor  = black,
}

\newcommand{\eps}{\varepsilon}
\newcommand{\be}{\begin{equation}}
\newcommand{\ee}[1]{\label{#1}\end{equation}}
\newcommand{\de}[2]{\frac{\partial #1}{\partial #2}}
\newcommand{\im}{\mathrm{i}}
\newcommand{\cg}[1]{{\color{gray} #1}}

\renewcommand{\d}{\mathrm{d}}
\renewcommand{\Phi}{\varPhi}
\renewcommand{\Psi}{\varPsi}
\newcommand\revised[1]{#1}
\newcommand{\dc}[1]{#1}

\newcommand\upi{\pi}
\newcommand\bnabla{\boldsymbol{\nabla}}
\newcommand{\backsection}[1]{\noindent\textbf{#1.}~}

\title{Higher-order homogenised riblet boundary conditions}

\author{Paolo Luchini$^+$ and Daniel Chung$^*$\\
\normalsize
$^+$Dipartimento di Ingegneria Industriale, Universit\`a di Salerno, 84084 Fisciano, Italy\\
\normalsize
$^*$Department of Mechanical Engineering, University of Melbourne, Victoria 3010, Australia}

\numberwithin{equation}{section}

\begin{document}
\maketitle

\begin{abstract}
The description of riblets and other drag-reducing devices has long used the concept of longitudinal and transverse protrusion heights, both as a means to predict the drag reduction itself and as equivalent boundary conditions to simplify numerical simulations by transferring the effect of riblets onto a flat virtual boundary. The limitation of this idea is that it stems from a first-order approximation in the riblet-size parameter $s^+$, and as a consequence it cannot predict other than a linear dependence of drag reduction upon $s^+$; in other words, the initial slope of the drag-reduction curve. Here the concept is extended to a full asymptotic expansion using matched asymptotics, which consistently provides higher-order \revised{protrusion coefficients} and higher-order equivalent boundary conditions on a virtual flat surface. \revised{While the majority of our results, though nonlinear in $s^+$, remain linear in velocity, and therefore we shall not directly address the shape of the drag-reduction curve, this procedure will} also allow us to explore the way nonlinearities of the Navier-Stokes equations first enter the $s^+$-expansion, with somewhat surprising \revised{negative} results.
\end{abstract}



\section{Introduction}\label{intro}

\input{intro.tex}

\section{2D Laplace equation}\label{sec:Laplace2D}
The Laplace equation may serve as an example before handling more complex flow problems.

\begin{figure}
\centerline{{\includegraphics[clip]
{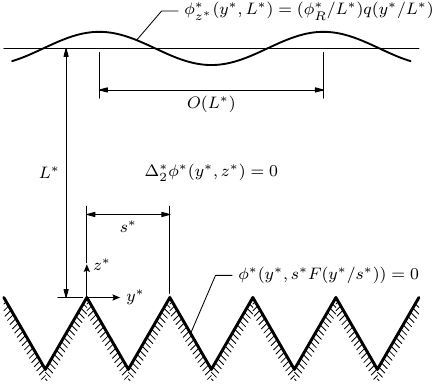}}}
\caption{Setup: $s^*$ is the riblet period/spacing, $L^* \gg s^*$ is the upper boundary condition plane (measured from the tips at $z^*=0$) where \revised{a Neumann boundary condition} on length scales O$(L^*)$ is imposed.}
\label{fig:setup}
\end{figure}
Let us consider the solution of the 2D Laplace equation
\be
\Delta_2^* \phi^* = \de{^2\phi^*}{y^{*2}} + \de{^2\phi^*}{z^{*2}} = 0
\ee{Laplace2D}
in a 2D strip characterized by a periodic arrangement of protuberances or cavities (riblets) of period $s^*$ on one wall (figure \ref{fig:setup}) and an opposing infinite flat wall at far-away distance $L^*\gg s^*$. Coordinates are named $y^*$ along the wall and $z^*$ normal to the wall with an eye to forthcoming three-dimensional applications. A $^*$ superscript is used to mark these quantities as dimensional, as opposed to plain letters used further down to denote dimensionless variables. \revised{Just as well, a notation such as $\phi^*_{z^*}$ denotes a dimensional derivative of $\phi^*$ with respect to $z^*$.} Also, later this will become a flow problem, but for the time being let the unknown $\phi^*$ be any scalar quantity that obeys the Laplace equation, say a temperature field. For definiteness let us also impose a Dirichlet boundary condition ($\phi^*=0$) on the $s^*$-periodic riblet wall, located near $z^*=0$, and a constant Neumann boundary condition ($\phi^*_{z^*}={}$const.) on the flat wall located at $z^*=L^*$. \revised{A Neumann boundary condition at the wall would require nontrivial modifications to what follows, however we shall only need a Dirichlet condition in the subsequent application to fluid flow.}

\subsection{Wavenumber domain (constant outer conditions)}\label{LaplaceWN}
The solution to the Laplace equation in this geometry is itself periodic and can be represented as a Fourier series
\begin{equation}
\phi^*(y^*,z^*) = \sum_{n=-\infty}^{\infty} \hat{\phi}^*_n(z^*) \exp(\im n \beta^*_1 y^*),
\label{Fourierphi}
\end{equation}
where $\beta^*_1 = 2\upi/s^*$\revised{, $\im$ is the imaginary unit, and $n$ numbers the harmonics}. Far from the periodic wall the mean ($n=0$) component is given in (4) of \citet{Luchini91resistance},
\be
\hat{\phi}^*_0 = a^*_0 + b^*_0 z^* = \hat{\phi}^*_{0,z}(L^*)(z^*+h^*),
\ee{wdh}
whereas all the $n\ne 0$ components decay exponentially as $\exp(-\beta^*_1 z^*)$ or faster. It follows that in most of the domain (precisely for $\beta^*_1 z^*\gg 1$, or $z^*\gg s^*$) the solution has a linear profile as if the wall were flat, and appears to cross zero at a virtual origin $z^*=-h^*$. This virtual origin, or its distance from an arbitrary reference location which often is taken to be the riblet tip, by definition is the \emph{protrusion height}, so named by \cite{Bechert89viscous} who first introduced and calculated it for the Laplace equation applied to longitudinal flow past riblets. Since the governing equation is linear, and the whole solution $\phi^*$ is proportional to the strength of the imposed flux $\hat{\phi}^*_{0,z^*}(L^*)$, the protrusion height is independent of $\hat{\phi}^*_{0,z^*}(L^*)$ and can be obtained once and for all for a given riblet geometry from a single numerical computation, several examples of which were given in \citet{Luchini91resistance}. In addition, given the scale invariance of the Laplace equation, $h^*$ is proportional to $s^*$ and can be written as $h^*=h_1 s^*$, where $h_1$ is a dimensionless number calculated for period equal to unity.

It is also apparent, since the $n\ne 0$ Fourier components decay exponentially with distance and thus also with the ratio $L^*/s^*$, that the approximation provided by substituting the linear profile \eqref{wdh} for the actual solution \eqref{Fourierphi} is exponentially accurate (its error decays faster that any finite power of $s^*/L^*$), and a power expansion in the small parameter $\eps=s^*/L^*$ at fixed $z^*/L^*$, such as will be considered in the next section, contains no higher powers of $\eps$ than the first-order term with coefficient $h_1$ already included in \eqref{wdh}.

\subsection{Physical domain (slowly varying outer conditions)}\label{LaplacePD}
Let us now consider a situation where the outer imposed flux is not exactly constant with $y^*$ but slowly varying on a length scale $L^*\gg s^*$. Under these conditions the solution of the Laplace equation will no longer be exactly periodic, and its expansion in powers of $s^*/L^*$ will contain non-trivial higher-order corrections. (Equivalently the period $s^*$ itself could be assumed to be slowly varying, but we shall not consider this extension at present.)
When the main field is characterized by a length scale $L^*$ and the wall texture by a period $s^*$, the problem lends itself to matched asymptotic expansions in the parameter $\eps = s^*/L^*$.
 Notice that, even if the Laplace equation is linear with respect to forcing, it is not linear with respect to geometry, and therefore the expansion is allowed to contain higher powers of $\eps$.
 
In order to introduce matched asymptotic expansions, in the style in which they are usually introduced in boundary layers \citep{vanDyke}, we need inner coordinates $Y, Z$, made dimensionless with the riblet period $s^*$, and outer coordinates $y=y_0+\eps Y, z=\eps Z$ made dimensionless with the outer scale $L^*$ in a neighbourhood of a generic boundary point $(y_0,0)$. \revised{As a mnemonic, big $Z$ is a numerically big dimensionless ordinate, $z^*$ divided by the inner reference length, and small $z$ is a small dimensionless ordinate, $z^*$ divided by the outer reference length, so that $z=\eps Z$.} The, now dimensionless, dependent variable $\phi$ can be expanded in an inner series
\be
\phi=\Phi_0(Y,Z)+\eps\Phi_1(Y,Z)+\cdots = \sum_{i=0}^\infty \eps^i\Phi_i(Y,Z) ,
\ee{phi_inner}
and an outer series
\be
\phi=\phi_0(y,z)+\eps\phi_1(y,z)+\cdots = \sum_{i=0}^\infty \eps^i\phi_i(y,z) .
\ee{phi_outer}
In consequence of the Laplace equation's being linear and scale-invariant, each term of both the inner and outer expansions will independently obey its own Laplace equation:
\[
\Delta_2 \Phi_i(Y,Z) = 0, \quad \Delta_2 \phi_i(y,z) = 0 \quad \forall i .
\]
In addition, $\Phi_i$ will obey a Dirichlet boundary condition on a curve $Z=F(Y)$, $\Phi_i[Y,F(Y)]=0$, with the riblet shape $F(Y)$ periodic of period $1$ (because $Y$ has been made dimensionless with the period), and $\phi_i$ will obey a variable Neumann boundary condition at the outer wall $z=1$:
\be
\phi_{0,z}(y,1)=q(y);\quad \
\phi_{i,z}(y,1)=0\;\text{for }i\ne 0 .
\ee{outer_phi_bc}

In order to match the inner to the outer solution in a neighbourhood of a generic \revised{boundary position $(y_0,0)$},
we can set $y=y_0 + \eps Y$, $z =\eps Z$, and expand each term of the outer series \eqref{phi_outer} in a 2D Taylor series in $\eps Y$ and $\eps Z$:
\be
\begin{split}
\phi={}&\phi_0(y_0,0)+\eps Y\de{\phi_0}{y}(y_0,0)+\eps Z\de{\phi_0}{z}(y_0,0)+\eps\phi_1(y_0,0)+{}\\
&\eps^2\frac{Y^2}{2}\de{^2\phi_0}{y^2}(y_0,0)+\eps^2YZ\de{^2\phi_0}{y \partial z}(y_0,0)+\eps^2\frac{Z^2}{2}\de{^2\phi_0}{z^2}(y_0,0)+{}\\
&\eps^2 Y\de{\phi_1}{y}(y_0,0)+\eps^2 Z\de{\phi_1}{z}(y_0,0)+\eps^2\phi_2(y_0,0)+{}\cdots .
\end{split}
\ee{phi_outer_inner}
According to the principle of matched asymptotic expansions, as applied to the \emph{small-roughness} limit, the inner and outer expansions will not be uniformly valid but only match in an intermediate distinguished limit, in which $Y$ and $Z$ tend to infinity at the same time as $z\rightarrow 0$ and $y\rightarrow y_0$ with $\eps\rightarrow 0$.
In this limit we shall term by term equate coefficients of equal powers of $\eps$ between \eqref{phi_inner} and \eqref{phi_outer_inner}. So doing yields the following hierarchy:
\subsubsection{Order zero}
At inner order 0, matching yields that $\Phi_{0,Z}(Y,\infty) = 0$, because the $\eps^0$ term of \eqref{phi_outer_inner} is just a constant and has zero derivative; being a solution of the Laplace equation with all zero boundary conditions, $\Phi_0$ is identically null (and the inner expansion could just as well have started from $i=1$).

At outer order 0 we need to solve the Laplace equation for $\phi_0(y,z)$ with the boundary condition \eqref{outer_phi_bc} at the outer wall and $\phi_0(y,0)=0$ at the flat inner wall $z=0$, to which the periodic wall reduces for $\eps=0$. From the result we can extract $\phi_{0,z}(y,0)$ (which also includes information about all of its $y$-derivatives) for later use.

\subsubsection{First order}
At inner order $1$ from \eqref{phi_outer_inner} we obtain
\be
\Phi_1(y_0;Y,Z) \sim Z\phi_{0,z}(y_0,0)+\phi_1(y_0,0)
\ee{matching_1}
where $y_0$ is to be treated as a parameter as far as the Laplace equation for $\Phi_1$ is concerned, and symbol ${}\sim{}$ \revised{(``behaves as'')} is here precisely defined to mean that the difference between its \textit{l.h.s.} and \textit{r.h.s.} must tend to zero when $Z\rightarrow\infty$. This definition may imply more than one actual equation, since the \textit{l.h.s.} and \textit{r.h.s.} are asymptotically polynomials in $Z$, and each coefficient of these polynomials must individually match for the difference to tend to zero. Notice also that the second term of \eqref{phi_outer_inner}, $Y\phi_{0,y}(y_0,0)$, has been omitted from \eqref{matching_1} because $\phi_0(y,0) = 0$ for all $y$, and its derivatives are thus also zero. \revised{With ${}\sim{}$ given the above meaning,} asymptotic condition \eqref{matching_1} \revised{is equivalent to} the two equations
\be
\Phi_{1,Z}(y_0;Y,\infty) = \phi_{0,z}(y_0,0)
\ee{inner_1_bc}
and
\be
\phi_1(y_0,0) = \lim_{Z\rightarrow\infty} \left[\Phi_1(y_0;Y,Z) - Z\phi_{0,z}(y_0,0)\right]
\ee{outer_1_bc}
Equation \eqref{inner_1_bc} (the $r.h.s.$ of which will be a non-trivial function of $y_0$ unless, as in the previous section, $\phi_{0,z}(y,1)$ is constant) closes the Laplace problem for the order-1 inner solution $\Phi_1$. Equation \eqref{outer_1_bc} in turn provides the wall boundary condition for the first-order outer solution $\phi_1$. Since $\Phi_1\propto \phi_{0,z}(y_0,0)$ (because of \eqref{inner_1_bc} and the linearity of the Laplace problem), \eqref{outer_1_bc} can also be written as
\be
\phi_1(y_0,0) = h_1 \phi_{0,z}(y_0,0)
\ee{outer_1_bc_h}
where constant $h_1$, independent of $y_0$, is the same protrusion height as already introduced in \S\ref{LaplaceWN}, and is the only quantity in \eqref{outer_1_bc_h} which depends on the actual wall texture $Z=F(Y)$. In fact, introducing the normalized $\overline{\Phi_{11}}=\Phi_1(y_0;Y,Z)/\phi_{0,z}(y_0,0)$, one realizes that $\overline{\Phi_{11}}$ is independent of $y_0$, and obeys the Laplace equation $\Delta_2 \overline{\Phi_{11}} = 0$ with the normalized boundary conditions
\be
\overline{\Phi_{11}}[Y,F(Y)]=0,\quad \overline{\Phi_{11}}_{,Z}(Y,\infty) = 1 ,
\ee{Phi1}
Substituting $\overline{\Phi_{11}}(Y,Z)$ in \eqref{outer_1_bc} and then comparing with \eqref{outer_1_bc_h}, one can obtain 
\be
h_1=\lim_{Z\rightarrow\infty} \left[\overline{\Phi_{11}}(Y,Z) - Z\right] .
\ee{def:h1}
Whereas for theoretical purposes the inner-outer expansion is the preferred approach, in numerical computation to have to deal with the expansion is inconvenient. We can, within the same order of approximation, sum the zeroth- and first-order boundary conditions to generate
\[
\phi(y,0)=\phi_0(y,0)+\eps\phi_1(y,0) = \eps h_1 \phi_{0,z}(y,0) +\text{O}(\eps^2),
\]
and, within a similar error O$(\eps^2)$, replace $\eps \phi_{0,z}(y,0)$ by $\eps \phi_{z}(y,0)$ \citep[analogously, in a sense, to what in Aerodynamics is called ``interactive boundary layer theory'', \textit{e.g.}][]{cebeci1999engineering}. There results a first-order equivalent boundary condition for the total $\phi$ without any \revised{remaining} reference to its series expansion:  
\be
\phi(y,0) = \eps h_1 \phi_{z}(y,0) +\text{O}(\eps^2),
\ee{sliplength}
or in dimensional form:
\be
\phi^*(y^*,0) \simeq s^* h_1 \phi^*_{z^*}(y^*,0),
\ee{sliplength_dim}

This is the \emph{slip-length} boundary condition often encountered in the literature on the computational solution of the outer equation, with the slip length $h^*=s^*h_1$ being just another name for the protrusion height.
But in addition, as will now be seen, the present matched asymptotic expansion allows the analysis to be continued to higher and higher orders.

\subsubsection{Second order}
Matching the second-order terms of \eqref{phi_inner} and \eqref{phi_outer_inner} gives:
\be
\Phi_2(y_0;Y,Z) \sim YZ\phi_{0,yz}(y_0,0)+Y\phi_{1,y}(y_0,0)+Z\phi_{1,z}(y_0,0)+\phi_2(y_0,0)
\ee{matching_2}
where $\phi_{0,yy}$ has been omitted because $\phi_0(y,0)$ is identically zero, and $\phi_{0,zz}$ has also been omitted for the same reason, because from the Laplace equation it equals $-\phi_{0,yy}$. We notice also that, from the $y$-derivative of \eqref{outer_1_bc_h}, $\phi_{1,y}(y_0,0)$ can be replaced by $h_1\phi_{0,yz}(y_0,0)$.

A solution of the Laplace equation for $\Phi_2$ such that it grows linearly in $Y$ at large $Y$, like \eqref{matching_2} requires, can be written as
\be
\Phi_2 = \Phi_{21}(y_0;Y,Z)+Y\,\Phi_{211}(y_0;Y,Z),
\ee{Phi2}
where $\Phi_{21}$ and $\Phi_{211}$ are periodic functions of $Y$ (of period $1$) obeying
\be
\Delta_2 \Phi_{211}(y_0;Y,Z) = 0; \quad \Delta_2 \Phi_{21}(y_0;Y,Z) = -2\Phi_{211,Y}(y_0;Y,Z)
\ee{Phi20}
and
\begin{gather}
\Phi_{211,Z}(y_0;Y,\infty) = \phi_{0,yz}(y_0,0) ,
\label{inner_21_bc} \\
\Phi_{21,Z}(y_0;Y,\infty) = \phi_{1,z}(y_0,0) .
\label{inner_20_bc}
\end{gather}
This is the second-order inner solution. Owing to linearity of the Laplace equation with respect to its inhomogeneous boundary condition, $\Phi_{211}$ can be written in normalized form as $\Phi_{211}(y_0;Y,Z)=\overline{\Phi_{211}}(Y,Z)\,\phi_{0,yz}(y_0,0)$. But $\overline{\Phi_{211}}$ turns out to obey the same equation and boundary conditions as $\overline{\Phi_{11}}$, and thus coincides with it. In addition, using \eqref{outer_1_bc_h} to replace $\phi_{1,y}$, and $\overline{\Phi_{211}}=\overline{\Phi_{11}}\sim Z+h_1$ from \eqref{def:h1}, it can be shown that $\Phi_{211}$ cancels out on both sides of \eqref{matching_2} as follows:
\[
\begin{split}
&\lim_{Z\rightarrow\infty}\left[Y\Phi_{211}(y_0;Y,Z) - YZ\phi_{0,yz}(y_0,0)-Y\phi_{1,y}(y_0,0)\right] =\\
&\lim_{Z\rightarrow\infty}\left[ Y(Z+h_1) \phi_{0,yz}(y_0,0) - YZ\phi_{0,yz}(y_0,0)- Yh_1\phi_{0,yz}(y_0,0)\right] = 0 .
\end{split}
\]
  The completion of \eqref{matching_2} then gives
\be
\phi_2(y_0,0) = \lim_{Z\rightarrow\infty} \left[\Phi_{21}(y_0;Y,Z) - Z\phi_{1,z}(y_0,0)\right]
\ee{outer_2_bc}
as the wall boundary condition for the second-order outer solution.

 $\Phi_{21}$, on the other hand, is comprised of two normalized contributions, one from the \textit{r.h.s.} of \eqref{Phi20} and one from the boundary condition \eqref{inner_20_bc}, $\Phi_{21}=\overline{\Phi_{21}}\phi_{0,yz}(y_0,0)+\overline{\Phi_{212}}\phi_{1,z}(y_0,0)$. It turns out that $\overline{\Phi_{212}}$ again coincides with $\overline{\Phi_{11}}$, whereas $\overline{\Phi_{21}}$ is the solution of
\be
\Delta_2\overline{\Phi_{21}} = - 2\overline{\Phi_{11}}_{,Y}
\ee{eq_200}
with homogeneous boundary conditions $\overline{\Phi_{21}}[Y,F(Y)]=0$, $\overline{\Phi_{21}}_{,Z}(Y,\infty) = 0$. Upon defining a new constant
\be
h_2 = \lim_{Z\rightarrow\infty} \overline{\Phi_{21}}(Y,Z) ,
\ee{def:h2}
the second-order outer boundary condition \eqref{outer_2_bc} can be rewritten as
\be
\phi_2(y,0)=h_1 \phi_{1,z}(y,0) + h_2 \phi_{0,yz}(y,0),
\ee{outer_2_bc_h}
and the compound equivalent boundary condition as
\be
\phi(y,0)= \eps h_1 \phi_{z}(y,0) + \eps^2 h_2 \phi_{yz}(y,0),
\ee{outer_2_bc_h_eq}
or in dimensional form as
\be
\phi^*(y^*,0)= s^* h_1 \phi^*_{z^*}(y^*,0) + s^{*2} h_2 \phi^*_{y^*z^*}(y^*,0).
\ee{outer_2_bc_h_dim}
The new dimensionless coefficient $h_2$ \revised{can be called ``second-order protrusion coefficient'' (or ``second-order protrusion height'', but we refrain from this denomination because $s^{*2} h_2$ has physical dimensions of the square of a height, and subsequent $n^\text{th}$-order coefficients will be the $n^\text{th}$ power of a height)}. Also, it can be noted that if $\overline{\Phi_{21}}$ happens to be an odd function of $Y$, a limit such as \eqref{def:h2}, which must then equal its opposite under the transformation $Y\leftarrow -Y$, is necessarily zero. More about this symmetry at the end of \S\ref{compoundLaplace}. 

Now a pattern begins to appear.
\subsubsection{Third order}\label{Laplace3rd}
The third-order matching condition is
\[
\Phi_3(y_0;Y,Z) \sim \left(\frac{Y^2Z}{2}-\frac{Z^3}{6}\right)\phi_{0,yyz} + \frac{Y^2-Z^2}{2}\phi_{1,yy} + YZ\phi_{1,yz}+Y\phi_{2,y}+Z\phi_{2,z}+\phi_3 ,
\]
where second and higher $z$-derivatives of $\phi_0$ and $\phi_1$ have been substituted from the Laplace equation, and subsequently $y$-derivatives of $\phi_0$ have been cancelled. $\phi_{1,yy}$ and $\phi_{2,y}$ can be replaced from \eqref{outer_1_bc} and \eqref{outer_2_bc} as
\[
\phi_{1,yy}(y_0,0) = h_1 \phi_{0,yyz}(y_0,0)
\]
\[
\phi_{2,y}(y_0,0)=h_1 \phi_{1,yz}(y_0,0) + h_2 \phi_{0,yyz}(y_0,0) ,
\]
to yield
\be
\Phi_3(y_0;Y,Z) \sim \left[\frac{Y^2}{2}(Z+h_1)+Yh_2-\frac{Z^3}{6} - \frac{Z^2}{2}h_1\right]\phi_{0,yyz} + Y\left(Z+h_1\right)\phi_{1,yz}+Z\phi_{2,z}+\phi_3 ,
\ee{matching_3}

By analogy with \eqref{Phi2} one can make the following ansatz:
\be
\begin{split}
\Phi_3 = \left[\overline{\Phi_{31}}(Y,Z)+Y\,\overline{\Phi_{21}}(Y,Z)+(Y^2/2)\,\overline{\Phi_{11}}(Y,Z)\right]\phi_{0,yyz}+{}\\
\left[\,\overline{\Phi_{21}}(Y,Z)+Y\overline{\Phi_{11}}(Y,Z)\right]\phi_{1,yz}+{}\\
\overline{\Phi_{11}}(Y,Z)\,\phi_{2,z}.
\end{split}
\ee{Phi3}
The second and third line here are nothing else than the previous-order normalized solutions of the Laplace equation multiplied by $\phi_{1,yz}$ instead of $\phi_{0,yz}$, and $\phi_{2,z}$ instead of $\phi_{0,z}$. The first line, once inserted into the Laplace equation for $\Phi_{3}$, gives
\be
\Delta_2 \overline{\Phi_{31}} + 2\overline{\Phi_{21}}_{,Y} + \overline{\Phi_{11}} = 0
\ee{eq_300}
as the determining equation for the new periodic function $\overline{\Phi_{31}}$, with boundary conditions
\[
\overline{\Phi_{31}}[Y,F(Y)]=0,\quad \lim_{Z\rightarrow\infty}\left[\overline{\Phi_{31}}_{,Z} + Z^2/2 + h_1 Z\right] = 0 .
\]

From \eqref{eq_300} and the already known asymptotic behaviour of $\overline{\Phi_{21}}_{,Y}$ and $\overline{\Phi_{11}}$ one can get the asymptotic behaviour for $Z\rightarrow\infty$ of $\overline{\Phi_{31}}$ as
\[
 \overline{\Phi_{31}} \sim -Z^3/6 - h_1 Z^2/2 + h_3.
\]
which, in addition to making \eqref{matching_3} satisfied, of course implies the definition
\be
h_3 = \lim_{Z\rightarrow\infty} \left[\overline{\Phi_{31}}  +Z^3/6 + h_1 Z^2/2\right] .
\ee{def:h3}
Extracting $\phi_3(y_0,0)$ from \eqref{matching_3} then gives the sought-after third-order outer boundary condition:
\be
\phi_3(y_0,0)=h_1 \phi_{2,z}(y_0,0) + h_2 \phi_{1,yz}(y_0,0) + h_3 \phi_{0,yyz}(y_0,0) .
\ee{outer_3_bc_h}

\subsubsection{Compound equivalent boundary condition}\label{compoundLaplace}
The above process can be continued to higher and higher orders, and for each $n$ will provide an outer boundary condition of the form
\be
\phi_n(y,0)= \sum_{i=1}^n h_i \de{^{i}\phi_{n-i}}{y^{i-1}\partial z}(y,0) .
\ee{outer_n_bc_h}
The boundary conditions of different orders can be multiplied by $\eps^n$ and summed together, if desired, to get the total outer $\phi$ at the virtual wall:
\be
\phi(y,0)=\sum_{n=0}^N \eps^n\phi_n(y,0)+\text{O}(\eps^{N+1}) =\sum_{n=1}^N \sum_{i=1}^n \eps^n h_i \de{^{i}\phi_{n-i}}{y^{i-1}\partial z}(y,0) +\text{O}(\eps^{N+1}) .
\ee{pps}
This equation looks like the product of two power series. In fact, by reordering summations it can be rewritten as
\[
\phi(y,0)=\sum_{i=1}^N h_i \de{^{i}}{y^{i-1}\partial z} \sum_{n=i}^N \eps^{n} \phi_{n-i}(y,0) +\text{O}(\eps^{N+1}) .
\]
But $\sum_{n=i}^N \eps^{n} \phi_{n-i} =\eps^i\sum_{n=0}^{N-i} \eps^{n} \phi_{n} = \eps^i\phi+\text{O}(\eps^{N+1})$, and therefore, within an error $\text{O}(\eps^{N+1})$,
\be
\phi(y,0)=\sum_{i=1}^N \eps^i h_i \de{^{i}\phi}{y^{i-1}\partial z} +\text{O}(\eps^{N+1}) ,
\ee{outer_N_slip-len}
or, in dimensional form,
\be
\phi^*(y^*,0)=\sum_{i=1}^N s^{*i} h_i \de{^{i}\phi^*}{y^{*i-1}\partial z^*} +\text{O}(\eps^{N+1}) .
\ee{outer_N_slip-len_dim}

Just as \eqref{sliplength} was at first order, this single implicit $N^\text{th}$-order boundary condition is free of any reference to the outer series expansion, and may be preferred to \eqref{outer_n_bc_h} for computational purposes. One can also observe that if the wall shape $Z=F(Y)$ happens to be left-right symmetric, \textit{i.e.} invariant under the \revised{mirror} transformation $Y\leftarrow -Y$, only even $y$-derivatives have the same symmetry and are allowed in \eqref{outer_n_bc_h}--\eqref{outer_N_slip-len_dim}. Odd $y$-derivatives are forbidden by symmetry, and the corresponding higher-order \revised{protrusion coefficients} must automatically come out as zero. Notice that this does not imply that the corresponding inner solutions are zero, only that they shall have the appropriate symmetry for these coefficients to vanish. For instance, when riblets are \revised{mirror} symmetric, $\overline{\Phi_{21}}$ of \eqref{eq_200} is an odd function of $Y$, because so is $\overline{\Phi_{11}}_{,Y}$, and \revised{thus} its mean value, which when $Z\rightarrow\infty$ provides $h_2$, is zero.

\section{Longitudinal Stokes flow past riblets}\label{longitudinal}
In the analysis of the viscous sublayer near riblets \citep{Luchini91resistance} we want to solve Stokes' equations:
\begin{subequations}
\begin{align}
\bnabla^*\boldsymbol{\cdot}\boldsymbol{u}^* &= 0, \\
-\bnabla^*p^* + \mu^* \nabla^{*2} \boldsymbol{u}^* &= \boldsymbol{0},
\end{align}
\label{Stokes_3D}
\end{subequations}
with boundary conditions
$\boldsymbol{u}^* = \boldsymbol{0}$ at the solid riblet wall with riblet period $s^*$, and an imposed velocity gradient for $z^*\rightarrow\infty$.
The riblet longitudinal direction is $x^*$, the spanwise direction is $y^*$ and the wall-normal direction is $z^*$.

For classical straight riblets described by $z^*/s^* = F(y^*/s^*)$ (independent of $x^*$) the system of equations \eqref{Stokes_3D} can be split into longitudinal and transverse components \citep[see, \textit{e.g.},][]{Luchini91resistance}. The longitudinal component is governed by a Poisson equation
\be
\Delta^*_2 u^* = p^*_{x^*}/\mu^* ,
\ee{long_eq}
where the, typically negative, pressure gradient $p^*_{x^*}$ is of an order of magnitude dictated by the outer scale \revised{(O$(\eps^2)$ in inner variables, as will appear below)}, and for this reason was neglected by \cite{Bechert89viscous} and by \cite{Luchini91resistance}. When the pressure gradient is neglected, the longitudinal velocity $u^*$ behaves exactly like the quantity $\phi^*$ of \S\ref{sec:Laplace2D}.

At higher than first order in $\eps$, the pressure gradient can no longer be neglected and, as will be seen, it provides a correction even for constant outer conditions. \revised{Exactly where in the expansion this correction appears will depend on the particular distinguished limit adopted. For the purpose of a flow where the pressure gradient retains the same order of magnitude it had on a flat wall (\textit{i.e.} where the ratio of the pressure gradient to outer scales is kept constant while $\eps\rightarrow 0$),} we shall assume the pressure gradient $p^*_{x^*}/\mu^*$ to be proportional to $u^*/L^{*2}$, so that \eqref{long_eq} is balanced when both sides are expressed in outer units. In other words, the outer dimensionless  pressure gradient $p_x=L^{*2}p^*_{x^*}/\mu^*u_R^*$ (where $u_R^*$ is a reference velocity that will be specified later but for the time being can be arbitrary) is \revised{kept constant and} of order unity. Then the dimensionless form of \eqref{long_eq} on the scale of the period $s^*$ (\textit{i.e.} in inner units) is
\be
\Delta_2 U = \eps^2 p_x,
\ee{long_eq_s}
where $U=u^*/u_R^*$ and coordinates are $Y=y^*/s^*$ and $Z=z^*/s^*$.

\subsection{Wavenumber domain (constant outer conditions)\label{pressurewd}}
For constant $p_x$ and constant velocity gradient at $Z=L=1/\eps$, the solution of \eqref{long_eq_s} can be expanded in a Fourier series, just like the solution of the Laplace equation \eqref{Laplace2D}, \textit{i.e.}
\begin{equation}
U(Y,Z) = \sum_{n=-\infty}^{\infty} \hat{U}_n(Z) \exp(\im n 2\upi Y).
\end{equation}
The only difference is that now $\hat{U}_{0,ZZ}$ equals $\eps^2 p_x$, and instead of \eqref{wdh} the mean ($n=0$) Fourier component becomes
\[
\hat{U}_0 = a_0 + b_0 Z +\eps^2p_x Z^2/2.
\]

The coefficients $a_0$ and $b_0$ in this expression could be determined, like in \S\ref{LaplaceWN} from the value of the velocity gradient at large distance. However, it must be noted that while in \eqref{wdh} $\hat\phi^*_{0,z^*}$ is constant with $z^*$, and $\hat\phi^*_{0,z^*}(L^*)$ directly equals $b_0$, in the presence of a pressure gradient the mean velocity profile becomes a quadratic, and to determine $b_0=\hat{U}_{0,Z}(0)$ is more complicated. If one wants to write an equivalent far-from-the-wall boundary condition transported in $Z=0$:
\be
a_0 = b_0 h_1 + \eps^2 p_x h^{(p_x)}_2 ,
\ee{wdh20}
by linearity, the two constants $h_1$ and $h^{(p_x)}_2$ can be determined through once applying a suitable non-zero boundary condition on $\hat{U}_{0,Z}$ at $Z=L$ with $p_x=0$ (and this shows that $h_1$ is precisely the same protrusion height as defined in \S\ref{sec:Laplace2D}), and then the boundary condition $\hat{U}_{0,Z}=-\eps^2p_x L$ at $Z=L$, to make $b_0=0$ with $p_x\ne 0$. Since
$
b_0 = 0 ,
$
and $
\hat{U}_0(L)= \cancel{b_0(h_1+L)} +\eps^2p_x ( h^{(p_x)}_2 + L^2/2) ,
$
one can obtain
\[
h^{(p_x)}_2 = \frac{\hat{U}_0(L)}{\eps^2p_x}- \frac{L^2}{2}
\]
and this value is actually independent of both $p_x$ and $L$.

In outer coordinates, \eqref{wdh20} provides an equivalent boundary condition
\be
u(y,0)=\eps h_1 u_z(y,0)+ \eps^2 p_x h_2^{(p_x)}
\ee{wdh2}
which includes the second-order effect of the pressure gradient.
Again no higher powers of $\eps$ are involved, and \eqref{wdh2} is exponentially accurate, when outer conditions are constant and the \revised{inner} solution is perfectly periodic. 

\subsection{Physical domain (slowly varying outer conditions)}
In the presence of a slowly varying outer shear $u_{0,z}(y,1) = q(y)$ and dimensionless pressure gradient $p_x(y,z)$, the outer problem becomes a boundary-value problem with in addition an equivalent wall boundary condition at $z=0$, which in full generality may be written by assigning $u(y,0)$ as a functional of $u_z(y,0)$ and $p_x(y,0)$. In order to obtain the equivalent boundary condition without a priori solving the outer problem, one can observe that if $u(y,0)$, $u_z(y,0)$ and $p_x$ were all simultaneously known, the outer problem could be in principle solved as an initial-value problem by Cauchy series. As long as the solutions of the boundary-value problem and the initial-value problem are the same, it is therefore sufficient that the equivalent boundary condition be consistent with this Cauchy series.

To remind what a Cauchy series is, it is in fact a Taylor series in a neighbourhood of a generic wall point $(y_0,0)$, where $z$-derivatives higher that the first have been replaced using the differential equation, differentiated zero or more times in $z$:
\be
u(y,z)=u(y_0,z)+\de{u}{y}(y-y_0)+\de{u}{z}z + \de{^2 u}{y^2}\frac{(y-y_0)^2}{2}+\de{^2 u}{y\partial z}(y-y_0)z+\left(p_x - \de{^2 u}{y^2}\right)\frac{z^2}{2} + \cdots
\ee{Cauchy_long}
We, in fact, already did a similar replacement when, in \eqref{matching_2}, we obtained the second derivative $\phi_{0,zz}$ from the Laplace equation as $\phi_{0,zz}=-\phi_{0,yy}$ and thus determined it was zero.

This device was originally invented by Cauchy in order to express the general solution of an initial-value problem in time, but here it will be used to reformulate our boundary-value problem as an initial-value problem in $z$. Whereas for an elliptic equation the initial-value problem is notoriously ill-posed, the trick here will be to formally apply the Cauchy series to initial values obtained from the two-sided boundary-value problem itself. This will allow us to reformulate the inner problem as an equivalent boundary condition, without (or before) actually solving the outer problem.

For the purposes of matched asymptotic expansions, the Cauchy series will replace the outer solution that the inner solution must match. If in \eqref{Cauchy_long} $y-y_0$ is replaced by $\eps Y$ and $z$ by $\eps Z$, \eqref{Cauchy_long} becomes
\be
u(y,z) = u(y_0,0)+ \eps\left(\de{u}{y} Y+\de{u}{z} Z\right) +\eps^2\left[\de{^2 u}{y^2}\frac{Y^2}{2} + \de{^2 u}{yz}YZ +\left(p_x-\de{^2 u}{y^2}\right)\frac{Z^2}{2}\right]+\cdots 
\ee{Cauchy_long_inner}
and can be reordered as a power series in $\eps$. This is the asymptotic behaviour that the inner solution must match, term by term in $\eps$ and uniformly in $Y$ (or at least for large $Y$, but then automatically for any $Y$), when $Z\rightarrow\infty$.

Since the problem depends linearly on $p_x$, and the equivalent boundary condition can be written as a superposition of a $u(y_0,0)$ function of $u_z(y_0,0)$ and a $u(y_0,0)$ function of $p_x(y_0,0)$, it is evident that the $u_z(y,0)$-dependent part will be identical to the one calculated in \S\ref{sec:Laplace2D} and given by \eqref{outer_2_bc_h_eq}. $p_x(y_0,0)$, unsurprisingly, first appears in the expansion at order $\eps^2$ in \eqref{Cauchy_long_inner} and multiplying $Z^2/2$. Therefore the corresponding inner problem is
\be
\Delta_2 \overline{U_{22}} = 1
\ee{Upx}
with boundary conditions $\overline{U_{22}}[Y,F(Y)]=0$ and $\lim_{Z\rightarrow\infty}[\overline{U_{22}}_{,Z}(Y,Z)- Z] =0$. From the solution of this inner problem one extracts
\be
h_2^{(p_x)} = \lim_{Z\rightarrow\infty} [\overline{U_{22}}(Y,Z)-Z^2/2]
\ee{def:h2px}
and \eqref{outer_2_bc_h} acquires an additional term $h_2^{(p_x)} p_{0,x}(y,0)$, consistently with \eqref{wdh2}. \eqref{outer_2_bc_h_eq} then becomes
\be
u(y,0) = \eps h_1 u_{z} + \eps^2 h_2 u_{yz} + \eps^2 h_2^{(p_x)} p_{x} .
\ee{outer_2_bc_U}

Consideration of the next order will be delayed until \S\ref{sec:NS3} since it involves coupling the longitudinal flow with the crossflow.

\section{Transverse 2D Stokes problem}
\begin{figure}
\centerline{{\includegraphics[clip]
{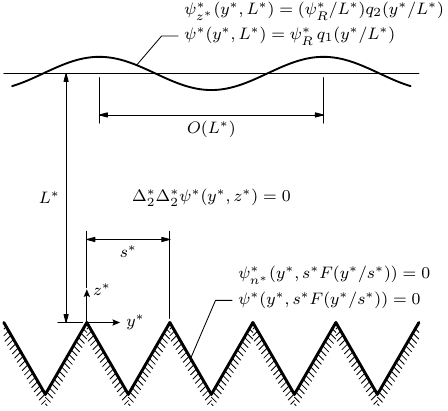}}}
\caption{\dc{Setup: $s^*$ is the riblet period/spacing, $L^* \gg s^*$ is the upper boundary condition plane (measured from the tips at $z^*=0$) where flow on length scales O$(L^*)$ is imposed.}}
\label{fig:setup_psi}
\end{figure}
In the ($y^*$-$z^*$) plane orthogonal to the riblet direction $x^*$, the Stokes equations reduce to
\begin{subequations}
\begin{align}
\frac{\partial v^*}{\partial y^*} + \frac{\partial w^*}{\partial z^*} &= 0, \\
-\frac{\partial p^*}{\partial y^*} + \mu^*\left(\frac{\partial^2 v^*}{\partial y^{*2}} + \frac{\partial^2 v^*}{\partial z^{*2}}\right) &= 0, \\
-\frac{\partial p^*}{\partial z^*} + \mu^*\left(\frac{\partial^2 w^*}{\partial y^{*2}} + \frac{\partial^2 w^*}{\partial z^{*2}}\right) &= 0,
\end{align}
\label{2DStokes}
\end{subequations}
with $v^* = w^* = 0$ at the riblet wall.
By use of the streamfunction $\psi^*$ such that $v^* = \partial\psi^*/\partial z^*$, $w^* = -\partial \psi^*/\partial y^*$, continuity is automatically satisfied, and system \eqref{2DStokes} reduces to the biharmonic equation
\be
\left(\frac{\partial^2}{\partial y^{*2}} + \frac{\partial^2}{\partial z^{*2}}\right)^2 \psi^* = 0.
\ee{biharmonic}
This is a $4^\text{th}$-order elliptic equation requiring 4 boundary conditions, 2 at each side of a strip-like domain in $z^*$. At the solid riblet wall we have
 $\partial \psi^*/\partial z^* = \partial \psi^*/\partial y^*=0$, which is the same as $\partial \psi^*/\partial n^* = 0$ (\revised{derivative normal to the riblet wall provides the tangential velocity, \textit{i.e.} no slip}) and $\partial \psi^*/\partial s^* = 0$ (\revised{derivative parallel to the riblet wall provides the normal velocity, \textit{i.e.} impermeability}). \revised{$\partial \psi^*/\partial s^* = 0$} can be integrated to $\psi^* = \psi^*_\text{w}$, an arbitrary constant which on one of the two walls can without loss of generality be taken as $\psi^*_\text{w}=0$.

Aim is to get 2 equivalent/homogenised boundary conditions at a reference plane $z^*=0$, say at the riblet crest, the most general linear form of which (compare \eqref{outer_1_bc_h}) can be written in matrix form as
\begin{equation}\label{eq:bc}
\left[\begin{matrix}
\psi^*_{z^*}(y^*,0) \\
\psi^*(y^*,0)
\end{matrix}\right]
=\left[\begin{matrix}
a^* & b^* \\
c^* & d^* \\
\end{matrix}\right]
\left[\begin{matrix}
\psi^*_{z^*z^*}(y^*,0) \\
\psi^*_{z^*z^*z^*}(y^*,0)
\end{matrix}\right] .
\end{equation}
These are written so as to provide two generic linear constraints among $\psi^*$ and its first three normal derivatives at the wall; making these constraints explicit in $\psi^*_{z^*}(y^*,0)$ and $\psi^*(y^*,0)$ has the additional advantage that $a^*=b^*=c^*=d^*=0$ corresponds to the standard conditions on a flat solid wall.

A crucial observation, which marks an important difference with the scalar case, is that the $a^*,b^*,c^*,d^*$ coefficients have different dimensions, and each behave differently upon rescaling. If dimensionless versions of these coefficients are defined in a setting where lengths are scaled with the riblet period $s^*$,
\be
\left[\begin{matrix}
s^*\psi^*_{z^*}(y,0) \\
\psi^*(y^*,0)
\end{matrix}\right]
=\underbrace{\left[\begin{matrix}
a_1 & b_2 \\
c_2 & d_3 \\
\end{matrix}\right]}_{\text{dimensionless}}
\left[\begin{matrix}
s^{*2}\psi^*_{z^*z^*}(y^*,0) \\
s^{*3}\psi^*_{z^*z^*z^*}(y^*,0)
\end{matrix}\right] ,
\ee{abcd}
and then powers of $s^*$ are moved inside the center block, \eqref{eq:bc} becomes
\be
\left[\begin{matrix}
\psi^*_{z^*}(y^*,0) \\
\psi^*(y^*,0)
\end{matrix}\right]
=\left[\begin{matrix}
a_1s^* & b_2s^{*2} \\
c_2s^{*2} & d_3s^{*3} \\
\end{matrix}\right]
\left[\begin{matrix}
\psi^*_{z^*z^*}(y^*,0) \\
\psi^*_{z^*z^*z^*}(y^*,0)
\end{matrix}\right],
\ee{eq:dimlbc}
where the subscripts $1,2,3$ of the dimensionless coefficients $a_1,b_2,c_2,d_3$ have been assigned according to their scaling.

Because of their different scaling, which is clearly tied to the simultaneous presence of derivatives of different order in the boundary conditions, the $a_1,b_2,c_2,d_3$ coefficients will turn up at different orders in the $\eps$-expansion, and only the $a_1$ coefficient appears at first order. Since the condition $\psi^*_{z^*} = a_1 s^*\psi^*_{z^*z^*}$ is the same as $v^* = a_1 s^* v^*_{z^*}$, $a_1$ can easily be recognized to coincide with the scalar, first-order, transverse protrusion height originally introduced in \citet{Luchini91resistance}. A complete expansion accounting for higher-order terms will be provided in the following \S\ref{StokesPD}.

\subsection{Wavenumber domain (constant outer conditions)}\label{StokesWN}
In Fourier space, the solution to the biharmonic equation is
\begin{equation}
\psi^*(y^*,z^*) = \sum_{n=-\infty}^{\infty}
\hat\psi^*_n(z^*) \exp(\mathrm{i} n \beta^*_1 y^*),
\end{equation}
where the far-field (mean, $n=0$) component is a third-degree polynomial given in (5) of \citet{Luchini91resistance}, 
\be
\begin{split}
&\hat{\psi}^*_0 = A^*_0 + B^*_0 z^* + C^*_0 z^{*2} + D^*_0 z^{*3} = {}\\
&         \hat{\psi}^*_0(L^*)
          +\hat{\psi}^*_{0,z^*}(L^*) (z^*-L^*)
          +\hat{\psi}^*_{0,z^*z^*}(L^*) \frac{1}{2}(z^*-L^*)^2
          +\hat{\psi}^*_{0,z^*z^*z^*}(L^*) \frac{1}{6}(z^*-L^*)^3,
\end{split}
\ee{farfield}
and all the other components decay at least as fast as $z^*\exp(-\beta^*_1 z^*)=z^*\exp(-2\upi z^*/s^*)$ with $z^*\rightarrow\infty$.

When the problem is linear and constant-coefficient, the $a^*,b^*,c^*,d^*$ coefficients can be exactly recovered from a set of two numerical calculations over a domain $0\le z^* \le L^*$ (with $L^*\gg s^*$):
\begin{enumerate}
    \item Set $\psi^*_{z^*z^*}(y^*,L^*) = 1$ (arbitrary constant), $\psi^*_{z^*z^*z^*}(y^*,L^*) = 0$,
    $\psi^*_{n^*}|_\text{w} = 0$ (no slip), $\psi^*|_\text{w} = 0$ (impermeable), and measure
   $\psi^*(y^*,L^*)$, $\psi_z(y^*,L^*)$ (constants with $y^*$), so we know $\boldsymbol{p}_1^T|_L \equiv (\hat{\psi}^*_0, \hat{\psi}^*_{0,z^*},\hat{\psi}^*_{0,z^*z^*},\hat{\psi}^*_{0,z^*z^*z^*})_1(L^*)$. 
    \item Set $\psi^*_{z^*z^*}(y^*,L^*) = 0$, $\psi^*_{z^*z^*z^*}(y^*,L^*) = 1$ (arbitrary constant),
 $\psi^*_{n^*}|_\text{w} = 0$ (no slip), $\psi^*|_\text{w} = 0$ (impermeable), and measure
   $\psi^*(y^*,L^*)$, $\psi_z(y^*,L^*)$ (constants with $y^*$), so we know $\boldsymbol{p}_2^T|_L \equiv  (\hat{\psi}^*_0, \hat{\psi}^*_{0,z^*},\hat{\psi}^*_{0,z^*z^*},\hat{\psi}^*_{0,z^*z^*z^*})_2(L^*)$. 
\end{enumerate}
We can get the far-field solution \eqref{farfield} and its derivatives propagated from $z^*=L^*$ to $z^*=0$ by
\begin{equation}\label{transfer}
\underbrace{\left[\begin{matrix}
\hat{\psi}^*_0 \\
\hat{\psi}^*_{0,z^*} \\
\hat{\psi}^*_{0,z^*z^*} \\
\hat{\psi}^*_{0,z^*z^*z^*}
\end{matrix}\right](0)}_{\boldsymbol{p}|_0}
= \left[\begin{matrix}
1 & -L^* & (-L^*)^2/2 & (-L^*)^3/6 \\
0 & 1 & -L^* & (-L^*)^2/2 \\
0 & 0 & 1 & -L^* \\
0 & 0 & 0 & 1 \\
\end{matrix}\right]
\underbrace{\left[\begin{matrix}
\hat{\psi}^*_0 \\
\hat{\psi}^*_{0,z^*} \\
\hat{\psi}^*_{0,z^*z^*} \\
\hat{\psi}^*_{0,z^*z^*z^*}
\end{matrix}\right](L^*)}_{\boldsymbol{p}|_L},
\end{equation}
so we now know how to calculate $\boldsymbol{p}_1|_0$ and $\boldsymbol{p}_2|_0$, which are linearly independent.
From evaluating the two boundary-condition equations in (\ref{eq:bc}) for each one of $\boldsymbol{p}_1|_0$ and $\boldsymbol{p}_2|_0$, we can write the four equations
\begin{align*}
\left[\begin{matrix}
(\hat{\psi}^*_{0,z^*z^*})_1|_0 & (\hat{\psi}^*_{0,z^*z^*z^*})_1|_0 \\
(\hat{\psi}^*_{0,z^*z^*})_2|_0 & (\hat{\psi}^*_{0,z^*z^*z^*})_2|_0
\end{matrix}\right]
\left[\begin{matrix}
a^* \\
b^*
\end{matrix}\right]
&=\left[\begin{matrix}
(\hat{\psi}^*_{0,z^*})_1|_0 \\
(\hat{\psi}^*_{0,z^*})_2|_0
\end{matrix}\right], \\
\left[\begin{matrix}
(\hat{\psi}^*_{0,z^*z^*})_1|_0 & (\hat{\psi}^*_{0,z^*z^*z^*})_1|_0 \\
(\hat{\psi}^*_{0,z^*z^*})_2|_0 & (\hat{\psi}^*_{0,z^*z^*z^*})_2|_0
\end{matrix}\right]
\left[\begin{matrix}
c^* \\
d^*
\end{matrix}\right]
&=\left[\begin{matrix}
(\hat{\psi}^*_0)_1|_0\\
(\hat{\psi}^*_0)_2|_0
\end{matrix}\right],
\end{align*}
and then solve for the four unknowns $a^*, b^*, c^*, d^*$. The resulting equivalent boundary condition \eqref{eq:bc}, containing terms up to third order, will be exponentially accurate, and never contain powers of $\eps$ higher than the third, for constant outer boundary conditions.

\subsection{Physical domain (slowly varying outer conditions)}\label{StokesPD}
Just like in \S\ref{LaplacePD}, the dimensionless dependent variable $\psi$ can be expanded in inner coordinates $Y=(y^*-y_0^*)/s^*$ and $Z=z^*/s^*$, $s^*$ being the wall-texture period:
\be
\psi=\Psi_0(Y,Z)+\eps\Psi_1(Y,Z)+\cdots = \sum_{i=0}^\infty \eps^i\Psi_i(Y,Z) ,
\ee{psi_inner}
and outer coordinates $y=y^*/L^*$ and $z=z^*/L^*$, $L^*$ being an outer reference length (typically, the distance to the opposing wall):
\be
\psi=\psi_0(y,z)+\eps\psi_1(y,z)+\cdots = \sum_{i=0}^\infty \eps^i\psi_i(y,z) .
\ee{psi_outer}
The expansion parameter $\eps$ is the length ratio $s^*/L^*$.

Because the biharmonic equation is linear and scale-invariant, each term of both the inner and outer expansions will independently obey its own biharmonic equation:
\[
\Delta_2 \Delta_2 \Psi_i(Y,Z) = 0, \quad \Delta_2 \Delta_2 \psi_i(y,z) = 0 \quad \forall i .
\]
In addition, $\Psi_i$ will obey solid-wall boundary conditions on a curve $Z=F(Y)$,
\[
\Psi_i[Y,F(Y)]=\de{\Psi_i}{n}[Y,F(Y)]=0 ,
\]
with $F(Y)$ periodic of period $1$ in inner coordinates, and $\psi_i$ will obey two generic boundary conditions at an outer wall $z=1$.

The problem will be completed by matching conditions imposed in an intermediate distinguished limit where $Z\rightarrow\infty$ and $z\rightarrow 0$ as $\eps \rightarrow 0$.

\subsubsection{Order zero}
At outer order 0 we need to solve the biharmonic equation for $\psi_0(y,z)$ with whatever boundary conditions are imposed at the outer wall and $\psi_0(y,0)=\psi_{0,z}(y,0)=0$ at the flat inner wall $z=0$, to which the periodic wall reduces for $\eps=0$. From the (probably numerical, when the outer boundary conditions are varying in a generic way) solution of this equation we obtain, among other things, the values of $\psi_{0,zz}(y,0)$ and $\psi_{0,zzz}(y,0)$.

With $\psi(y,0)$, $\psi_{z}(y,0)$, $\psi_{zz}(y,0)$, and $\psi_{zzz}(y,0)$ \revised{all} known, we can formally reconstruct the outer solution by Cauchy series (actually by Taylor series until, at fourth order, we begin to see the effect of the biharmonic differential equation in eliminating fourth and higher $z$-derivatives):
\be
\begin{split}
&\psi(y,z) = \psi(y_0,0)+ (y-y_0)\psi_y(y_0,0)+z\psi_z(y_0,0) +{}\\
&\frac{(y-y_0)^2}{2}\psi_{yy}(y_0,0) +(y-y_0)z\psi_{yz}(y_0,0)+\frac{z^2}{2}\psi_{zz}(y_0,0)+{}\\
&\frac{(y-y_0)^3}{6}\psi_{yyy}(y_0,0) +\frac{(y-y_0)^2z}{2}\psi_{yyz}(y_0,0) +\frac{(y-y_0)z^2}{2}\psi_{yzz}(y_0,0)+{}\\
&\frac{z^3}{6}\psi_{zzz}(y_0,0)+
\frac{z^4}{24}\left[-\psi_{yyyy}(y_0,0)-2\psi_{yyzz}(y_0,0)\right]+\cdots
\end{split}
\ee{psi_Cauchy}
For the purpose of matching, we insert the outer expansion \eqref{psi_outer} into \eqref{psi_Cauchy}, transform it in inner variables $Y=(y-y_0)/\eps$, $Z=z/\eps$, and reorder in powers of $\eps$:
\be
\begin{split}
&\psi(y,z) = \cancel{\psi_0(y_0,0)}+ \eps\left(Y\cancel{\psi_{0,y}}+Z\cancel{\psi_{0,z}}+\psi_1(y_0,0)\right) +{}\\
&\eps^2\left(\frac{Y^2}{2}\cancel{\psi_{0,yy}} +YZ\cancel{\psi_{0,yz}}+\frac{Z^2}{2}\psi_{0,zz}+Y\psi_{1,y}+Z\psi_{1,z}+\psi_2(y_0,0)\right)+{}\\
&\eps^3\left(\frac{Y^3}{6}\cancel{\psi_{0,yyy}} +\frac{Y^2Z}{2}\cancel{\psi_{0,yyz}} +\frac{YZ^2}{2}\psi_{0,yzz}+\frac{Z^3}{6}\psi_{0,zzz}+{}\right.\\
&\left.\frac{Y^2}{2}\psi_{1,yy} +YZ\psi_{1,yz}+\frac{Z^2}{2}\psi_{1,zz}+Y\psi_{2,y}+Z\psi_{2,z}+\psi_3(y_0,0)\right)+\cdots
\end{split}
\ee{psi_matching}
The cancellations are due to $\psi_0$ and $\psi_{0,z}$ being zero, together with their $y$-derivatives, on the flat wall that corresponds to $\eps=0$.
This is the sequence that must be matched, term by term according to powers of $\eps$, by the inner expansion \eqref{psi_inner}.

The general matching principle is that the two higher $Z$-derivatives, $\partial^2 \Psi_{i}/\partial Z^2$ and $\partial^3 \Psi_{i}/\partial Z^3$, associated with the dominant terms in the asymptotic expansion at large $Z$, are set by the outer solution and drive the inner solution as boundary conditions in the limit of $Z\rightarrow\infty$, just as $\partial\Phi_i/\partial Z$ did in \S\ref{LaplacePD}, whereas the two lower $z$-derivatives, $\psi_i$ and $\partial \psi_{i}/\partial z$, are set by the inner solution and drive the outer solution at $z=0$, just as $\phi_i$ did in \S\ref{LaplacePD}.

At inner order 0, matching yields all zero boundary conditions on $\Psi_0(Y,Z)$; as a result $\Psi_0$ is identically null (and the inner expansion could just as well have started from $i=1$).
\subsubsection{First order}
At inner order $1$ we obtain from \eqref{psi_matching}
\be
\Psi_1(y_0;Y,Z) \sim \psi_1(y_0,0).
\ee{psi_matching_1}
where, just as in \S\ref{LaplacePD}, $y_0$ is to be treated as a parameter insofar as the biharmonic equation for $\Psi_1$ is concerned, and symbol ${}\sim{}$ is precisely defined to mean that the difference between its \textit{l.h.s.} and \textit{r.h.s.} must tend to zero when $Z\rightarrow\infty$. Second and third $Z$-derivatives of $\Psi_1$ are absent (\textit{i.e.} zero) in the \textit{r.h.s.} of \eqref{psi_matching_1}, and therefore $\Psi_1$, for the same reasons as $\Psi_0$, is identically null.

At outer order 1, \eqref{psi_matching_1} consequently gives $\psi_1(y_0,0)=0$. The biharmonic equation, however, requires two boundary conditions at each end, and in order to find the second boundary condition for $\psi_1$ we need to simultaneously examine the $\eps^2$ term of \eqref{psi_matching}:
\be
\Psi_2(y_0;Y,Z) \sim\frac{Z^2}{2}\psi_{0,zz}+Z\psi_{1,z}+\psi_2(y_0,0)
\ee{psi_matching_2}
where $Y\psi_{1,y}(y_0,0)$ has been omitted because $\psi_1(y,0) = 0$ for all $y$ and its derivative is thus also zero.

According to \eqref{psi_matching_2} the second $Z$-derivative of $\Psi_2(y_0;Y,Z)$ is non-zero at infinity, and therefore $\Psi_2$ is the first non-trivial term of the inner expansion. It must be found as a solution of the biharmonic equation with the boundary conditions 
\[
\Psi_2[Y,F(Y)]=\Psi_{2,N}[Y,F(Y)]=0,\quad \Psi_{2,ZZ}[Y,\infty]= \psi_{0,zz}(y_0,0),\quad \Psi_{2,ZZZ}[Y,\infty] =0
\]
or equivalently, because of linearity, by means of a normalized $\overline{\Psi_{21}}$ obeying the biharmonic equation with boundary conditions
\be
\overline{\Psi_{21}}[Y,F(Y)]=\overline{\Psi_{21}}_{,N}[Y,F(Y)]=0,\quad \overline{\Psi_{21}}_{,ZZ}[Y,\infty]= 1,\quad \overline{\Psi_{21}}_{,ZZZ}[Y,\infty]=0 , 
\ee{Psi21}
as
\be
\Psi_2(y_0;Y,Z)=\psi_{0,zz}(y_0,0)\,\overline{\Psi_{21}}(Y,Z) .
\ee{def:Psi21}
Once $\overline{\Psi_{21}}$ has been determined, the first derivative of \eqref{psi_matching_2} provides the second wall boundary condition for the first-order outer solution $\psi_1$,
\be
\psi_{1,z}(y_0,0)=\lim_{Z\rightarrow\infty}\left[\Psi_{2,Z} - Z\psi_{0,zz}(y_0,0) \right] = a_1 \psi_{0,zz}(y_0,0),
\ee{bc_psi_1}
where
\be
a_1=\lim_{Z\rightarrow\infty}\left(\overline{\Psi_{21}}_{,Z} - Z\right).
\ee{def:a1}

Constant $a_1$, which coincides with $a_1$ of \S\ref{StokesWN} and is calculated in an equivalent manner, is the transverse protrusion height of \citet{Luchini91resistance}.
The present matching method embeds it into an expansion that can be continued to higher and higher orders.

\subsubsection{Second order}\label{StokesPD2}
Matching condition \eqref{psi_matching_2} did not yet exhaust its function. Enforcing \eqref{psi_matching_2} directly, in addition to its derivative, yields one of the boundary conditions for the second-order outer solution $\psi_2$:
\be
\psi_2(y_0,0)=\lim_{Z\rightarrow\infty}\left[\Psi_2(y_0;Y,Z) -\frac{Z^2}{2}\psi_{0,zz}-a_1 Z\psi_{0,zz}\right] = c_2\psi_{0,zz}(y_0,0),
\ee{bc_psi_2}
where
\be
c_2=\lim_{Z\rightarrow\infty}\left[\overline{\Psi_{21}}(Y,Z) -\frac{Z^2}{2}-a_1 Z\right].
\ee{def:c2}

The other boundary condition for $\psi_2$, as should by now be expected, is obtained from the third-order matching condition
\be
\Psi_3 \sim  \frac{YZ^2}{2}\psi_{0,yzz}+\frac{Z^3}{6}\psi_{0,zzz}+
 YZ\psi_{1,yz}+\frac{Z^2}{2}\psi_{1,zz}+Y\psi_{2,y}+Z\psi_{2,z}+\psi_3(y_0,0).
\ee{psi_matching_3}
Preliminarily the third-order inner solution $\Psi_3$ must be calculated, and it can be seen to be the superposition of several contributions. The first term proportional to $\psi_{0,yzz}$ can be combined with the third using \eqref{bc_psi_1} and the fifth using \eqref{bc_psi_2}, to give $Y(Z^2/2+a_1Z+c_2)\psi_{0,yzz}$. This asymptotic behaviour can be satisfied, in analogy to \eqref{Phi2}, by a compound solution of the form
\[
\Psi_{31} = \left[\overline{\Psi_{31}}(Y,Z)+Y\overline{\Psi_{21}}(Y,Z)\right]\psi_{0,yzz},
\]
where $\overline{\Psi_{21}}(Y,Z)$ \revised{is the same as} was defined in \eqref{Psi21}, and $\overline{\Psi_{31}}(Y,Z)$ is a \revised{new} periodic function of $Y$ obeying the equation
\be
\Delta_2 \Delta_2\overline{\Psi_{31}} + 4 \Delta_2\overline{\Psi_{21}}_{,Y} = 0
\ee{Psi31}
with boundary conditions
\[
\overline{\Psi_{31}}[Y,F(Y)]=\overline{\Psi_{31}}_{,N}[Y,F(Y)]=0,\quad\overline{\Psi_{31}}_{,ZZ}[Y,\infty]= \overline{\Psi_{31}}_{,ZZZ}[Y,\infty] =0 .
\]
The second term of \eqref{psi_matching_3} proportional to $Z^3/6$ can be matched by a normalized contribution\footnote{\revised{Here and in what follows we shall adopt a convention where contributions are numbered by a first digit denoting order and a second sequential digit.}} $\overline{\Psi_{32}}$ which obeys the biharmonic equation with boundary conditions
\be
\overline{\Psi_{32}}[Y,F(Y)]=\overline{\Psi_{32}}_{,N}[Y,F(Y)]=0,\quad \overline{\Psi_{32}}_{,ZZ}[Y,\infty]\sim Z,\quad \overline{\Psi_{32}}_{,ZZZ}[Y,\infty] =1 .
\ee{Psi32}
The fourth term of \eqref{psi_matching_3} is analogous to the term multiplying $\psi_{0,zz}$ in \eqref{psi_matching_2}, and when used as a boundary condition produces the same normalized solution $\overline{\Psi_{21}}$. The sixth and seventh terms do not contribute to the second or third $Z$-derivatives, and thus have no role in determining $\Psi_3$.

Combining all three contributions together provides
\be
\Psi_3 =\left[\overline{\Psi_{31}}(Y,Z)+Y\overline{\Psi_{21}}(Y,Z)\right]\psi_{0,yzz}+ \overline{\Psi_{32}}(Y,Z)\psi_{0,zzz}+\overline{\Psi_{21}}(Y,Z)\psi_{1,zz}
\ee{def:Psi3}

Injecting $\Psi_3$ back into the $Z$-derivative of \eqref{psi_matching_3} now gives
\be
\psi_{2,z}(y_0,0)=a_2\psi_{0,yzz}+b_2\psi_{0,zzz}+a_1\psi_{1,zz},
\ee{psi_outer_2}
where
\be
a_2=\lim_{Z\rightarrow\infty}\overline{\Psi_{31}}_{,Z}(Y,Z),\quad \
b_2=\lim_{Z\rightarrow\infty}\left( \overline{\Psi_{32}}_{,Z} -\frac{Z^2}{2}\right).
\ee{def:a2b2}

Equation \eqref{psi_outer_2} closes the problem for the second-order outer solution $\psi_2$. Constants $b_2$ in \eqref{def:a2b2} and $c_2$ in \eqref{def:c2} have been so defined they coincide with $b_2$ and $c_2$ of \S\ref{StokesWN}. $a_2$ on the other hand, the analog here of $h_2$ in \S\ref{LaplacePD}, did not exist in \S\ref{StokesWN}, and can only be discovered by matched asymptotic expansions when $\psi_{0,zz}$ is a non-constant function of $y$. Owing to variational properties of the biharmonic equation, discussed in appendix \ref{symmetrysec}, it also turns out that $b_2=-c_2$.

\subsubsection{Third order}\label{Stokes3rd}
Just as happened for the second order, enforcing \eqref{psi_matching_3} as is yields one of the boundary conditions for the third-order outer solution $\psi_3$:
\be
\begin{split}
\psi_3(y_0,0)=&\lim_{Z\rightarrow\infty}\left[\Psi_3 - \frac{YZ^2}{2}\psi_{0,yzz}-\frac{Z^3}{6}\psi_{0,zzz}-
 \frac{Z^2}{2}\psi_{1,zz}-Y\psi_{2,y}-Z\psi_{2,z}\right] =\\
 &c_3\psi_{0,yzz} +c_2\psi_{1,zz} + d_3\psi_{0,zzz},
\end{split}
\ee{bc_psi_3}
where
\be
c_3=\lim_{Z\rightarrow\infty}\left[\overline{\Psi_{31}}(Y,Z) -a_2 Z\right],
\ee{def:c3}
\be
d_3=\lim_{Z\rightarrow\infty}\left[\overline{\Psi_{32}}(Y,Z) -\frac{Z^3}{6}-b_2 Z\right].
\ee{def:d3}
Constant $d_3$, just as well as its previous sister constants, coincides with $d_3$ of \S\ref{StokesWN}. 

The other boundary condition, specifying $\psi_{3,z}(y_0,0)$, must be obtained from the fourth-order matching condition
\[
\begin{split}
&\Psi_4(y_0;Y,Z) \sim  \left(\frac{Y^2Z^2}{4}-\frac{Z^4}{12}\right)\psi_{0,yyzz}+\frac{YZ^3}{6}\psi_{0,yzzz}+\frac{Y^2Z}{2}\psi_{1,yyz}+\frac{YZ^2}{2}\psi_{1,yzz}+{}\\
&\frac{Z^3}{6}\psi_{1,zzz}+
 \frac{Y^2}{2}\psi_{2,yy}+YZ\psi_{2,yz}+\frac{Z^2}{2}\psi_{2,zz}+Y\psi_{3,y}+Z\psi_{3,z}+\psi_4(y_0,0).
 \end{split}
\]

Two terms of this expression can be simplified using the previous boundary conditions, $\psi_{1,z}=a_1\psi_{0,zz}$, $\psi_2=c_2\psi_{0,zz}$, to give
\be
\begin{split}
&\Psi_4(y_0;Y,Z) \sim  \left[\frac{Y^2}{2}\left(\frac{Z^2}{2}+a_1Z+c_2\right)-\frac{Z^4}{12}\right]\psi_{0,yyzz}+\frac{YZ^3}{6}\psi_{0,yzzz}+{}\\
&\frac{YZ^2}{2}\psi_{1,yzz}+\frac{Z^3}{6}\psi_{1,zzz}+
 YZ\psi_{2,yz}+\frac{Z^2}{2}\psi_{2,zz}+Y\psi_{3,y}+Z\psi_{3,z}+\psi_4(y_0,0).
 \end{split}
\ee{psi_matching_4}
In addition, the second line of \eqref{psi_matching_4} can be recognized to have the same structure as $\Psi_3$ of \eqref{psi_matching_3}, with only a shift from $\psi_0$ to $\psi_1$, from $\psi_1$ to $\psi_2$ and from $\psi_3$ to $\psi_4$, and the corresponding terms of $\Psi_4$ can be written out accordingly.

The inner solution $\Psi_{41}$ corresponding to the first term of \eqref{psi_matching_4} has the structure
\[
\Psi_{41} = \left[\overline{\Psi_{41}}(Y,Z)+Y\overline{\Psi_{31}}(Y,Z)+\frac{Y^2}{2}\overline{\Psi_{21}}(Y,Z)\right]\psi_{0,yyzz},
\]
where $\overline{\Psi_{21}}$ and $\overline{\Psi_{31}}$ were defined earlier, and $\overline{\Psi_{41}}(Y,Z)$ is a \revised{new} periodic function of $Y$ obeying the equation
\[
\Delta_2 \Delta_2\overline{\Psi_{41}}+ 4\Delta_2\overline{\Psi_{31}}_{,Y} + 6\overline{\Psi_{21}}_{,YY}+2\overline{\Psi_{21}}_{,ZZ} = 0
\]
with boundary conditions
\[
\overline{\Psi_{41}}[Y,F(Y)]=\overline{\Psi_{41}}_{,N}[Y,F(Y)]=0,\quad\overline{\Psi_{41}}_{,ZZ}[Y,\infty]\sim -Z^2,\quad \overline{\Psi_{41}}_{,ZZZ}[Y,\infty] \sim -2Z ,
\]
the latter two coming from the $-Z^4/12$ behaviour at $Z\rightarrow\infty$ expected of $\overline{\Psi_{41}}$ in \eqref{psi_matching_4}.

The second term of \eqref{psi_matching_4} asymptotically proportional to $YZ^3/6$ in turn can be written as
\[
\Psi_{42} = \left[\overline{\Psi_{42}}(Y,Z)+Y\overline{\Psi_{32}}(Y,Z)\right]\psi_{0,yzzz},
\]
where $\overline{\Psi_{32}}$ was defined earlier, and $\overline{\Psi_{42}}(Y,Z)$ is \revised{yet another} periodic function of $Y$ obeying the equation
\[
\Delta_2 \Delta_2\overline{\Psi_{42}}+ 4\Delta_2\overline{\Psi_{32}}_{,Y} = 0
\]
with boundary conditions
\[
\overline{\Psi_{41}}[Y,F(Y)]=\overline{\Psi_{41}}_{,N}[Y,F(Y)]=0,\quad\overline{\Psi_{42}}_{,ZZ}[Y,\infty]= \overline{\Psi_{42}}_{,ZZZ}[Y,\infty] =0 ,
\]
the latter two coming from the lack of any $Y$-independent asymptotic part in the second term of \eqref{psi_matching_4}.

Combining all contributions together provides
\be
\begin{split}
\Psi_4 =& \left[\overline{\Psi_{41}}(Y,Z)+Y\overline{\Psi_{31}}(Y,Z)+\frac{Y^2}{2}\overline{\Psi_{21}}(Y,Z)\right]\psi_{0,yyzz}+{}\\
&\left[\overline{\Psi_{42}}(Y,Z)+Y\overline{\Psi_{32}}(Y,Z)\right]\psi_{0,yzzz}+{}\\
&\left[\overline{\Psi_{31}}(Y,Z)+Y\overline{\Psi_{21}}(Y,Z)\right]\psi_{1,yzz}+ \overline{\Psi_{32}}(Y,Z)\psi_{1,zzz}+\overline{\Psi_{21}}(Y,Z)\psi_{2,zz}
\end{split}
\ee{Psi4combined}

Injecting $\Psi_4$ back into the $Z$-derivative of \eqref{psi_matching_4} now gives
\be
\psi_{3,z}(y_0,0)=a_3\psi_{0,yyzz}+b_3\psi_{0,yzzz}+a_2\psi_{1,yzz}+b_2\psi_{1,zzz}+a_1\psi_{2,zz},
\ee{psi_outer_3}
where
\be
a_3=\lim_{Z\rightarrow\infty}\left( \overline{\Psi_{41}}_{,Z}(Y,Z)+2Z^3/3\right),\quad b_3=\lim_{Z\rightarrow\infty}\overline{\Psi_{42}}_{,Z}(Y,Z) .
\ee{def:a3b3}
Equation \eqref{psi_outer_3}, together with \eqref{bc_psi_3}, closes the problem for the third-order outer solution $\psi_3$. Owing to special symmetries of the biharmonic equation, it also turns out that $b_3=c_3$.

\subsubsection{Higher orders and equivalent boundary condition}
No apparent obstacle, aside from becoming overwhelmed with calculations, exists to continuing the procedure to higher and higher orders. Each inner contribution $\Psi_n$ must in turn be asymptotically matched to an expression derived from the Cauchy series of the outer solution, in the precise sense that the difference between the two must tend to zero for $Z\rightarrow\infty$. This matching implies a growing number of numerical equalities, because both sides are asymptotically polynomials of growing degree and every coefficient of these polynomials must individually match for the two polynomials to coincide.

Each $n^\text{th}$ matching condition will contain two new contributions, one proportional to $\partial^n\psi_0/\partial y^{n-2}\partial z^2$ and one proportional to $\partial^n\psi_0/\partial y^{n-3}\partial z^3$, in addition to a sequence of other terms which mimic the $(n-1)^\text{th}$ matching condition with $\psi_i$ replaced by  $\psi_{i+1}$. The two new contributions will produce two independent normalized inner solutions $\overline{\Psi_{n1}}(Y,Z)$ and $\overline{\Psi_{n2}}(Y,Z)$, and eventually produce constant numerical coefficients $a_{n-1}$, $b_{n-1}$, $c_{n}$, and $d_{n}$ for the $n^\text{th}$ outer boundary condition
\begin{equation}\label{eq:bcn}
\left[\begin{matrix}
\displaystyle
\de{\psi_{n}}{z}(y,0) \\
\psi_n(y,0)
\end{matrix}\right]
=\left[\begin{matrix}
\displaystyle
\sum_{i=1}^n a_i\de{^{i+1}\psi_{n-i}}{y^{i-1}\partial z^2}(y,0) + \sum_{i=2}^n b_i \de{^{i+1}\psi_{n-i}}{y^{i-2}\partial z^3}(y,0)\\
\displaystyle
\sum_{i=2}^n c_i\de{^{i}\psi_{n-i}}{y^{i-2}\partial z^2}(y,0) + \sum_{i=3}^n d_i \de{^{i}\psi_{n-i}}{y^{i-3}\partial z^3}(y,0) \\
\end{matrix}\right] .
\end{equation}

In a similar manner to \eqref{pps}, \eqref{eq:bcn} can be multiplied by $\eps^n$ and summed over $n=[0,N]$, so as to yield $\psi = \sum_{n=0}^{N} \eps^n\psi_n + \mathrm{O}(\eps^{N+1})$, and then recast as the formal product of two power series which only contains the total $\psi$ on the \textit{r.h.s} as well as the \textit{l.h.s}:
\begin{equation}\label{eq:bcN}
\left[\begin{matrix}
\displaystyle
\de{\psi}{z}(y,0) \\
\psi(y,0)
\end{matrix}\right]
=\left[\begin{matrix}
\displaystyle
\sum_{i=1}^N \eps^i a_i\de{^{i+1}\psi}{y^{i-1}\partial z^2}(y,0) + \sum_{i=2}^N \eps^i b_i \de{^{i+1}\psi}{y^{i-2}\partial z^3}(y,0)\\
\displaystyle
\sum_{i=2}^N \eps^i c_i\de{^{i}\psi}{y^{i-2}\partial z^2}(y,0) + \sum_{i=3}^N \eps^i d_i \de{^{i}\psi}{y^{i-3}\partial z^3}(y,0) \\
\end{matrix}\right] + \mathrm{O}(\eps^{N+1}) .
\end{equation}

The implicit boundary condition \eqref{eq:bcN}, analogous to \eqref{outer_N_slip-len} for the Laplace problem, may be preferred in numerical computations. In dimensional form it can immediately be written using the wall-texture period $s^*$ in place of $\eps$, as
\begin{equation}\label{eq:bcNdim}
\left[\begin{matrix}
\displaystyle
\de{\psi^*}{z^*}(y^*,0) \\
\psi^*(y^*,0)
\end{matrix}\right]
=\left[\begin{matrix}
\displaystyle
\sum_{i=1}^N s^{*i} a_i\de{^{i+1}\psi^*}{y^{*i-1}\partial z^{*2}}(y^*,0) + \sum_{i=2}^N s^{*i} b_i \de{^{i+1}\psi^*}{y^{*i-2}\partial z^{*3}}(y^*,0)\\
\displaystyle
\sum_{i=2}^N s^{*i} c_i\de{^{i}\psi^*}{y^{*i-2}\partial z^{*2}}(y^*,0) + \sum_{i=3}^N s^{*i} d_i \de{^{i}\psi^*}{y^{*i-3}\partial z^{*3}}(y^*,0) \\
\end{matrix}\right] .
\end{equation}
This is the generalization to slowly varying outer conditions of \eqref{eq:dimlbc} which was valid for constant outer conditions.

Also, as was true in \S\ref{compoundLaplace}, when the geometry is left-right symmetric \eqref{eq:bcn}--\eqref{eq:bcNdim} must enforce the same symmetry, which means that only even $y$-derivatives of $\psi$ can appear. The $a_i,b_i,c_i,d_i$ coefficients corresponding to odd derivatives will automatically turn out to be zero in this case.

\section{Homogenisation of the 2D Navier--Stokes equations}\label{2DNavierStokes}

Up to this point we have only considered linear problems, but the matched asymptotic expansions also apply to nonlinear ones with relatively little modification, as will soon be seen. The wavenumber domain that was adopted in the previous sections when outer conditions were constant, on the other hand, loses applicability. In fact, in the original application of the protrusion height, the aim was to consider a high-Reynolds number and indeed a turbulent flow, with riblets embedded in the near-wall region known as the viscous sublayer. There should be no doubt that the reduction of the viscous sublayer from a Navier-Stokes to a Stokes problem is not a separate step but itself an application of matched asymptotic expansions, and must be consistently performed in one and the same process.

Flow past a riblet surface can be considered as having three characteristic lengths: the riblet period $s^*$, the distance to the opposing wall $L^*$ and a dynamic length which will be taken to be the viscous length $\ell^* = \nu^*/u^*_\tau$. For definiteness here $u^*_\tau$ will be the characteristic perturbation velocity that in a turbulent flow is usually defined from the mean wall-shear stress, but for the purpose of what follows any other reference velocity $u^*_R$ would do as well, as long as it correctly represents the order of magnitude of \revised{near-wall} velocities. \revised{Another defining feature of the distinguished limit that we adopt here is that all three components of velocity scale proportionally to each other, \textit{i.e.} share the same reference velocity $u^*_R=u^*_\tau$. This is ordinarily the case in a near-wall turbulent flow but not, for instance, in a laminar boundary layer.} The ratio $L^*/\ell^*=L^*u^*_\tau/\nu^*$ is one of the possible definitions of Reynolds number, and decides the type of flow. When $L^* \ll \ell^*$ the Reynolds number is low, $\ell^*$ becomes irrelevant, and we are in the Stokes regime dealt with in the previous sections. When $\ell^* \ll L^*$, the Reynolds number is high and we are in the Navier-Stokes, possibly turbulent, regime. Near the wall $L^*$ now becomes irrelevant, and the asymptotic-expansion parameter $\eps$ must be chosen to be the ratio $\eps =s^*/\ell^*$, the same parameter that is frequently denoted as $s^+$ (or riblet period in \emph{wall units}). \revised{This discussion only shows, one could argue, that $\eps$ must be \emph{proportional} to $s^+$, but in the argument of a power expansion a fixed multiplying coefficient (say, $\eps=s^*/\ell^*$ versus $\eps=s^*/(20\ell^*)$) makes no difference.}

In a schematization where turbulence is presumed to be dominated by quasi-streamwise vortices ($\partial/\partial x^* = 0$), we can start from an assumption where the flow is governed by the unsteady 2D Navier-Stokes equations in primitive variables $v^*(y^*,z^*,t^*)$, $w^*(y^*,z^*,t^*)$, $p^*(y^*,z^*,t^*)$:
\be
\begin{split}
v^*_{t^*} + v^*v^*_{y^*} + w^*v^*_{z^*} +p^*_{y^*}/\rho^* &= \nu^* (v^*_{y^*y^*}+v^*_{z^*z^*}), \\
w^*_{t^*} + v^*w^*_{y^*} + w^*w^*_{z^*} +p^*_{z^*}/\rho^* &= \nu^* (w^*_{y^*y^*}+w^*_{z^*z^*}), \\
v^*_{y^*} + w^*_{z^*} &= 0.
\end{split}
\ee{prim}
It is in fact reasonable that, especially near riblets where the largest variations occur in the $y^*$ and $z^*$ directions, $x^*$-derivatives should be of a higher order. Considerations of how high this order is, and the consequent effects of three-dimensionality, will be deferred to the next section.

In a nondimensionalization with reference length $\ell^*$, reference velocity $u^*_\tau$, reference time $\ell^*/u^*_\tau$, and reference pressure $\mu^* u_\tau^{*}/\ell^* = \rho^* u_\tau^{*2}$, all constant dimensional coefficients in \eqref{prim} disappear and these equations become
\be
v_t + vv_y + wv_z + p_y = v_{yy}+v_{zz},
\quad w_t + vw_y + ww_z + p_z = w_{yy}+w_{zz},
\quad v_y + w_z = 0.
\ee{dlessprim}
This system can be further reduced to a single equation, by introducing the streamfunction $\psi(y,z,t)$ such that $v=\psi_z$, $w=-\psi_y$. Continuity is then automatically satisfied, and the only non-zero component of the curl of the momentum equations gives
\be
\Delta_2 \Delta_2 \psi = \Delta_2 \psi_t + \psi_z \Delta_2\psi_y - \psi_y  \Delta_2\psi_z
\ee{nlpsieq}

The highest derivatives in this equation are the same as in the transverse Stokes equation \eqref{biharmonic}, and the required boundary conditions are also the same. In particular, two boundary conditions are required at each of the two walls of a strip-like domain, one of which is textured, and a solid wall is characterized by $\psi =\psi_\text{w} ={}$constant (where the constant may, on one of the two walls, be assumed to be zero) and $\partial\psi/\partial n = 0$.

\subsection{Outer expansion}
Just as in \S\ref{StokesPD}, the outer solution will see equivalent boundary conditions assigned at a virtual flat wall which is chosen as the origin $z=0$.
The outer expansion is written as
\be
\psi(y,z,t) = \sum_{i=0}^\infty \eps^i \psi_i(y,z,t)
\ee{nlouterexp}
where the small parameter is, as already discussed, $\eps=s^*/\ell^*=s^+$. Upon inserting \eqref{nlouterexp} into \eqref{nlpsieq} and collecting equal powers of $\eps$, $\psi_0$ still obeys the same equation \eqref{nlpsieq}, but with no-slip and impermeability conditions on the virtual flat wall, $\psi_0(y,0,t) =\psi_{0,z}(y,0,t) =0$ \revised{(as explained near \eqref{biharmonic})}. From the (numerical, say) solution of the $\psi_0$-equation we can obtain, among other things, the values of $\psi_{0,zz}(y,0,t)$ and $\psi_{0,zzz}(y,0,t)$. With this information available, we can formally reconstruct the solution by Cauchy series. The beginning of this series will be identical to \eqref{psi_Cauchy} up to third order, and only start to differ in the fourth-order term which is derived from the governing differential equation:
\be
\begin{split}
&\psi(y,z,t) = \psi(y_0,0)+ (y-y_0)\psi_y(y_0,0)+z\psi_z(y_0,0) +{}\\
&\frac{(y-y_0)^2}{2}\psi_{yy}(y_0,0) +(y-y_0)z\psi_{yz}(y_0,0)+\frac{z^2}{2}\psi_{zz}(y_0,0)+{}\\
&\frac{(y-y_0)^3}{6}\psi_{yyy}(y_0,0) +\frac{(y-y_0)^2z}{2}\psi_{yyz}(y_0,0) +\frac{(y-y_0)z^2}{2}\psi_{yzz}(y_0,0)+\frac{z^3}{6}\psi_{zzz}(y_0,0)+{}\\
&\frac{z^4}{24}\left[-\psi_{yyyy}(y_0,0)-2\psi_{yyzz}(y_0,0)+ \psi_z \Delta_2\psi_y - \psi_y  \Delta_2\psi_z + \Delta_2 \psi_t\right]+\cdots
\end{split}
\ee{NS2_Cauchy}

\subsection{Inner expansion}
The inner expansion is $\psi(y_0+\eps Y,\eps Z,t)=\Psi(y_0,t;Y,Z)$, where
\be
\Psi(y_0,t;Y,Z) = \sum_{i=0}^\infty \eps^i \Psi_i(y_0,t; Y,Z) .
\ee{nlinnerexp}
There is no rescaling in the time coordinate insofar as the riblet shape does not depend on time; in the inner solution time plays the role of a parameter just like $y_0$ does.

Since \eqref{nlpsieq} is homogeneous with respect to space coordinates, when this equation is transformed in inner variables it remains visibly unchanged save the time derivative:
\be
\Delta_2 \Delta_2 \Psi = \eps^2\Delta_2 \Psi_t + \Psi_Z \Delta_2\Psi_Y - \Psi_Y  \Delta_2\Psi_Z.
\ee{innernlpsieq}
\eqref{innernlpsieq} may then give the impression that $\Psi_0$ shall obey a nonlinear equation. However, it must be remembered that in \S\ref{StokesPD} $\Psi_0$ and $\Psi_1$ turned out to be identically zero, and the expansion effectively started with $\Psi_2$. There is a simple qualitative explanation for this behaviour: owing to the boundary conditions $\psi=\psi_z=0$, the streamfunction $\psi$ is O$(z^2)$ near the wall $z=0$, and thus proportional to $\eps^2 Z^2$ in inner coordinates, and of the second order in $\eps$. This estimate is unchanged in the Navier-Stokes problem. With the expansion \eqref{nlinnerexp} starting to be non-zero from $i=2$, \eqref{innernlpsieq} produces the following hierarchy:
\begin{subequations}
\begin{align}
\Delta_2 \Delta_2 \Psi_2 &= 0\\
\Delta_2 \Delta_2 \Psi_3 &= 0\\
\Delta_2 \Delta_2 \Psi_4 &= \Delta_2 \Psi_{2,t} + \Psi_{2,Z} \Delta_2\Psi_{2,Y} - \Psi_{2,Y}  \Delta_2\Psi_{2,Z}
\label{innerhierarchy4} \\
&\cdots\cdots\cdots \nonumber
\end{align}
\label{innerhierarchy}
\end{subequations}
It can be observed that each equation in \eqref{innerhierarchy} has the same, linear, \textit{l.h.s.}, nonlinearity being confined to the \textit{r.h.s.} which at every stage is a known function of previous terms.

\subsection{Orders 0, 1, 2}
It begins to appear that, \eqref{NS2_Cauchy} and \eqref{innerhierarchy} being identical to the corresponding equations in \S\ref{StokesPD} up to third order, nothing changes in the formulation up to inner order 3 and outer order 2 (we recall that outer order 3 partially depends upon inner order 4). This has a number of consequences:
\begin{enumerate}
\item (\emph{expected}) All the previous literature that used the reduced, steady Stokes equations to describe the viscous sublayer of an otherwise general and unsteady flow was rational in doing so: this reduction can be justified by a formal asymptotic expansion, and indeed correctly represents the Navier--Stokes equations at leading order in feature size $\eps = s^+$.
\item (\emph{maybe less expected}) The approximation is not only good at leading order (inner order 2, outer order 1), but also at inner-order 3, outer order 2. Therefore, second-order outer boundary conditions \eqref{bc_psi_2} and \eqref{psi_outer_2} are linear and Stokes-like as well. 
\item (\emph{however}) The above second-order boundary condition involves several different contributions on equal footing, and may be more complex than previously thought (and even more complex it will become in \S\ref{sec:NS3}). 
\end{enumerate}

The derivation of the inner solutions $\Psi_2=\overline{\Psi_{21}}\,\psi_{0,zz}$ and $\Psi_3=(\overline{\Psi_{31}}+Y\overline{\Psi_{21}})\,\psi_{0,yzz}+\overline{\Psi_{32}}\,\psi_{0,zzz}$ is identical to \eqref{Psi21}, \eqref{Psi31}, \eqref{Psi32} and won't be repeated. The corresponding equivalent boundary conditions for the outer problem are
\begin{subequations}
\begin{align}
\text{First order: }&\left\{ \begin{array}{l}
\psi_z(y,0,t)=\eps a_1 \psi_{zz} \\
\psi(y,0,t)=0
\end{array}\right. \\
\text{Second order: }&\left\{ \begin{array}{l}
\psi_z(y,0,t)=\eps a_1 \psi_{zz}+\eps^2 a_2 \psi_{yzz}+\eps^2 b_2 \psi_{zzz} \\
\psi(y,0,t)=\eps^2 c_2 \psi_{zz}
\end{array}\right.
\end{align}
\label{firstsecond}
\end{subequations}
where numerical constants $a_1$, $a_2$, $b_2$ and $c_2=-b_2$ are defined in \eqref{def:a1}, \eqref{def:a2b2}, \eqref{def:c2} respectively, and $a_2$ is expected to be zero for left-right-symmetric riblets (and in fact for any riblets, as will later be shown to be the case).

It can be practically useful to reformulate \eqref{firstsecond} in primitive variables. Using $v=\psi_z$ and $w=-\psi_y$, a straightforward transformation of \eqref{firstsecond} gives
\begin{align*}
\text{First order: }&\left\{ \begin{array}{l}
v(y,0,t)=\eps a_1 v_{z} \\
w(y,0,t)=0
\end{array}\right. \\
\text{Second order: }&\left\{ \begin{array}{l}
v(y,0,t)=\eps a_1 v_{z}-\eps^2 a_2 w_{zz}+\eps^2 b_2 v_{zz} \\
w(y,0,t)=\eps^2 c_2 w_{zz}
\end{array}\right.
\end{align*}
This is not the most useful form, however. Whereas in \eqref{firstsecond} up to the third $z$-derivative of $\psi$ is allowed to appear, in primitive variables $v_{zz}$ and $w_z$ can be substituted from the momentum and continuity equations \eqref{dlessprim}, and actually \emph{must} be substituted in order to maintain compatibility with the Cauchy expansion in the developments to be presented later in the 3D extensions of this argument. (When the continuity equation, for instance, becomes 3-dimensional as $u_x+v_y+w_z=0$, $v_y$ and $-w_z$ become no longer interchangeable; $u_x,v_y$, without $w_z$, is the pair that appears in the Cauchy expansion and formulas derived from it.)

In this minimal-$z$-derivative form, the equivalent boundary conditions become
\begin{subequations}
\begin{align}
\text{First order: }&\left\{ \begin{array}{l}
v(y,0,t)=\eps a_1 v_{z} \\
w(y,0,t)=0
\end{array}\right. \\
\text{Second order: }&\left\{ \begin{array}{l}
v(y,0,t)=\eps a_1 v_{z}+\eps^2 a_2 v_{yz}+\eps^2 b_2 p_{y} \\
w(y,0,t)=-\eps^2 c_2 v_{yz}
\end{array}\right. \label{secondw}
\end{align}
\label{firstsecondvw}
\end{subequations}
where $v_{0,zz}$ has been eliminated through the momentum equation $v_{0,zz}+v_{0,yy}=p_{0,y}$ combined with $v_{0,yy}=0$, and $w_{0,zz}$ has been eliminated from the $z$-derivative of the continuity equation, $w_{0,zz}=-v_{0,yz}$. 

The first-order condition contains, as already stressed, the classical \revised{transverse} protrusion height \revised{$a_1$} for the wall-parallel velocity and zero for the wall-normal velocity. The wall-parallel second-order condition contains the wall-parallel pressure gradient in a similar manner to the longitudinal condition \eqref{outer_2_bc_U}; however, just as \eqref{outer_2_bc_U} contains an additional second-order term proportional to $u_{yz}$, both of \eqref{secondw} contain an additional term proportional to $v_{yz}$. One of the coefficients of these terms, $a_2$, vanishes for left-right-symmetric riblets (and in reality for any kind of riblets, as will be discovered later), but not $-c_2$ which always equals $b_2$ (see appendix \ref{symmetrysec}).
  
The same boundary conditions can be cast in dimensional form by simply replacing $\eps$ by the riblet period $s^*$.
  
\subsection{Third order}\label{2DNS3}
We recall from \S\ref{Stokes3rd} that one of the third-order conditions, the one for $\psi_{3}(y,0)$, was obtained from the third-order inner solution. This condition remains unchanged in the Navier-Stokes problem, and is still
\be
\psi_3(y,0,t)=c_2\psi_{1,yzz} + c_3\psi_{0,yzz} +d_3\psi_{0,zzz},
\ee{psi_bc3rep}
with constants $c_2$, $c_3$ and $d_3$ defined by \eqref{def:c2}, \eqref{def:c3} and \eqref{def:d3} respectively.

The other boundary condition, specifying $\psi_{3,z}(y,0,t)$, must be obtained from the fourth-order inner solution, which now obeys \eqref{innerhierarchy4}, \textit{i.e.}, on account of \eqref{def:Psi21}:
\[
\begin{split}
&\Delta_2 \Delta_2 \Psi_4(y_0,t;Y,Z) =\Delta_2 \overline{\Psi_{21}}(Y,Z)\,\psi_{0,zzt}(y_0,0,t)+{} \\
&\left[\overline{\Psi_{21}}_{,Z}(Y,Z) \Delta_2\overline{\Psi_{21}}_{,Y}(Y,Z)- \overline{\Psi_{21}}_{,Y}(Y,Z) \Delta_2\overline{\Psi_{21}}_{,Z}(Y,Z)\right]\psi^2_{0,zz}(y_0,0,t) .
\end{split}
\]
Since the \textit{l.h.s.} of this equation is still linear, its solution can be obtained by simply adding to the expression \eqref{Psi4combined} of $\Psi_4$ the two new contributions $\overline{\Psi_{43}}(Y,Z)\,\psi_{0,zzt}(y_0,0,t)+\overline{\Psi_{44}}(Y,Z)\,\psi^2_{0,zz}(y_0,0,t)$ \revised{(with an incremental second subscript according to the convention set forth before)}, where $\overline{\Psi_{43}}$ obeys the equation
\be
\Delta_2 \Delta_2 \overline{\Psi_{43}} = \Delta_2 \overline{\Psi_{21}}
\ee{Psi43}
with boundary conditions
\[
\overline{\Psi_{43}}[Y,F(Y)]=\overline{\Psi_{43}}_{,N}[Y,F(Y)]=0,\quad\overline{\Psi_{43}}_{,ZZ}[Y,\infty]\sim Z^2/2,\quad \overline{\Psi_{43}}_{,ZZZ}[Y,\infty] \sim Z ,
\]
and $\overline{\Psi_{44}}$ obeys the equation
\be
\Delta_2 \Delta_2 \overline{\Psi_{44}} = \overline{\Psi_{21}}_{,Z}(Y,Z) \Delta_2\overline{\Psi_{21}}_{,Y}(Y,Z)- \overline{\Psi_{21}}_{,Y}(Y,Z) \Delta_2\overline{\Psi_{21}}_{,Z}(Y,Z)
\ee{Psi44}
with boundary conditions
\[
\overline{\Psi_{44}}[Y,F(Y)]=\overline{\Psi_{44}}_{,N}[Y,F(Y)]=0,\quad\overline{\Psi_{44}}_{,ZZ}[Y,\infty]= \overline{\Psi_{44}}_{,ZZZ}[Y,\infty] =0 .
\]

Injecting $\Psi_4$ back into the $Z$-derivative of the matching condition \eqref{psi_matching_4} now gives, instead of \eqref{psi_outer_3},
\[
\psi_{3,z}(y_0,0)=a_3\psi_{0,yyzz}+b_3\psi_{0,yzzz}+a_2\psi_{1,yzz}+b_2\psi_{1,zzz}+a_1\psi_{2,zz}+e_3^{(v_{zt})}\psi_{0,zzt}+e_3^{(nl)}\psi_{0,zz}^2 ,
\]
where the two new numerical constants are
\be
e_3^{(v_{zt})}=\lim_{Z\rightarrow\infty} \left[\overline{\Psi_{43}}_{,Z}(Y,Z)-Z^3/6\right],\quad e_3^{(nl)}=\lim_{Z\rightarrow\infty}\overline{\Psi_{44}}_{,Z}(Y,Z) .
\ee{def:e3}

In primitive-variable implicit form, the nonlinear third-order boundary condition is now
\be
\left\{ \begin{array}{l}
v(y,0,t)=\eps a_1 v_{z}+\eps^2 ( a_2 v_{yz}+b_2 p_y)+\eps^3 (a_3 v_{yyz}+b_3p_{yy}+e_3^{(v_{zt})}v_{zt}+e_3^{(nl)}v_{z}^2) \\
w(y,0,t)=-\eps^2 c_2 v_{yz}-\eps^3 (c_3 v_{yyz}+ d_3 p_{yy})
\end{array}\right.
\ee{nlbc3}

Note: $e_3^{(nl)}$, together with $a_2$, $b_3$ and $c_3$, is one of those coefficients which are forbidden by symmetry, and thus zero, for left-right-symmetric riblets, because when $\overline{\Psi_{21}}$ is even, the \textit{r.h.s.} of \eqref{Psi44} is odd-symmetric in $Y$ and so is $\overline{\Psi_{44}}$. It follows that for left-right-symmetric riblets nonlinearity only intervenes at an even higher than third order in $\eps$.

\section{3D Navier--Stokes equations}\label{sec:NS3}

An actual turbulent flow will be three-dimensional, even when the riblets are aligned with the streamwise direction and effectively 2D. Since the matched asymptotic expansions for the 2D Navier--Stokes problem were eventually governed by linear inhomogeneous equations, we may hope for the same to happen in the more general 3D problem,
which, in dimensionless form with reference velocity $u^*_\tau$, reference length $\ell^*=\nu^*/u^*_{\tau}$, reference time $\ell^*/u^*_\tau$, and reference pressure $\rho^* u_\tau^{*2}$, is
\begin{subequations}
\begin{align}
u_t + uu_x + vu_y + wu_z +p_x &= u_{xx}+u_{yy}+u_{zz} \\
v_t + uv_x + vv_y + wv_z +p_y &= v_{xx}+v_{yy}+v_{zz} \\
w_t + uw_x + vw_y + ww_z +p_z &= w_{xx}+w_{yy}+w_{zz} \\
u_x + v_y + w_z &= 0 .
\end{align}
\end{subequations}

In inner coordinates $Y,Z$ such that $y = y_0 + \eps Y$, $z=\eps Z$, with $x$ and $t$ unchanged, these same equations become
\begin{subequations}
\begin{align}
u_{YY}+u_{ZZ} &= \eps(vu_Y + wu_Z) + \eps^2(-u_{xx}+u_t + uu_x + p_x) \\
v_{YY}+v_{ZZ} &= \eps(vv_Y + wv_Z + p_Y) + \eps^2(-v_{xx}+v_t + uv_x) \\
w_{YY}+w_{ZZ} &= \eps(vw_Y + ww_Z + p_Z) + \eps^2(-w_{xx}+w_t + uw_x) \\
v_Y + w_Z &= - \eps u_x .
\end{align}
\label{NS3eps}
\end{subequations}

Let us now expand each of $u,v,w,p$ in a series of powers of $\eps$, using capital letters for the terms of the inner expansion as in $u=\sum \eps^i U_i$.
As suggested by the previous sections, we may expect the inner expansions of $u,v,w$ (first derivatives of $\psi$) to start at order 1 and the inner expansion of pressure (akin to a \revised{velocity gradient, or to a} second derivative of $\psi$) to start at order 0. We thus obtain the following hierarchy.

\noindent First order:
\begin{subequations}
\begin{align}
U_{1,YY}+U_{1,ZZ} &= 0,\\
V_{1,YY}+V_{1,ZZ} - P_{0,Y} &= 0,\\
W_{1,YY}+W_{1,ZZ} - P_{0,Z} &= 0,\\
V_{1,Y} + W_{1,Z} &= 0 ;
\end{align}
\label{NS3order1}
\end{subequations}
Second order:
\begin{subequations}
\begin{align}
U_{2,YY}+U_{2,ZZ} &= P_{0,x},\label{NS3umom2}\\
V_{2,YY}+V_{2,ZZ} - P_{1,Y} &=0,\label{NS3vmom2}\\
W_{2,YY}+W_{2,ZZ} - P_{1,Z} &=0,\label{NS3wmom2}\\
V_{2,Y} + W_{2,Z} &= - U_{1,x} ;\label{NS3cont2}
\end{align}\label{NS3all2}
\end{subequations}
Third order:
\begin{subequations}
\begin{align}
U_{3,YY}+U_{3,ZZ} &= P_{1,x}-U_{1,xx}+U_{1,t} + V_1U_{1,Y} + W_1U_{1,Z},\\
V_{3,YY}+V_{3,ZZ} - P_{2,Y} &= -V_{1,xx}+V_{1,t} + V_1V_{1,Y} + W_1V_{1,Z},\\
W_{3,YY}+W_{3,ZZ} - P_{2,Z} &= -W_{1,xx}+W_{1,t} + V_1W_{1,Y} + W_1W_{1,Z},\\
V_{3,Y} + W_{3,Z} &= - U_{2,x} .\label{continuity3}
\end{align}
\label{NS3order3}
\end{subequations}
\vspace{-\baselineskip}
\[\boldsymbol{\cdots\cdots\cdots}\]

One can observe as much: i) each order is governed by the same, linear, system of \textit{l.h.s.} equations, driven in various ways by the solutions of previous orders; ii) this system of equations neatly decouples into a 2D Poisson problem for $U_n$ and a steady 2D Stokes problem for $V_n,W_n,P_{n-1}$, exactly those problems that were solved in the previous sections of this paper; iii) with respect to the previous sections there are some new forcing terms, including those in the continuity equation; iv) \revised{just as in \S\ref{2DNavierStokes}, nonlinear terms make their first appearance at third order (and not at second order as might have been inferred from their quadratic nature); v)}
since the hierarchical problem is linear we can exploit superposition, taking the already acquired \revised{solutions of the homogeneous problem} for granted and \revised{adding in the contribution of each new \textit{r.h.s.} term} in turn.

\subsection{First order}
Nothing new to add. The first-order equations are homogeneous just as they were in the previous sections, and driven by the matching conditions only. They lead to the longitudinal and transverse protrusion heights of \citet{Luchini91resistance} according to \eqref{def:h1} and \eqref{def:a1}.

\subsection{Second order}
Here the pressure gradient makes its appearance in the $U_2$-equation \eqref{NS3umom2}, just as it was taken into account in \S\ref{longitudinal}. However, rather than being constant with $Y,Z$, the pressure gradient is shaped by the first-order transverse Stokes solution (\ref{NS3order1}\textit{b--d}). More precisely
\[
P_0=\overline{P_{01}}(Y,Z)v_{0,z}(x,y_0,0,t)+p_0(x,y_0,0,t)
\]
where $\overline{P_{01}}(Y,Z)$ satisfies the Cauchy-Riemann relations
\be
\overline{P_{01}}_{,Y}=(\Delta_2\overline{\Psi_{21}})_Z,\quad \overline{P_{01}}_{,Z}=-(\Delta_2\overline{\Psi_{21}})_Y
\ee{CauchyRiemann}
with $\overline{P_{01}}(Y,\infty)=0$, so that $P_0(Y,\infty)=p_0(x,y_0,0,t)$, and $\overline{\Psi_{21}}$ being defined by \eqref{Psi21}; $v_{0,z}(x,y_0,0,t), p_0(x,y_0,0,t)$ are the corresponding components of the zeroth-order outer solution at the wall. Therefore
\be
U_2=\overline{\Phi_{21}}(Y,Z)u_{0,yz}(x,y_0,0,t)+\overline{U_{22}}(Y,Z)p_{0,x}(x,y_0,0,t)+\overline{U_{23}}(Y,Z)v_{0,xz}(x,y_0,0,t)
\ee{def:U2}
($p_{0,x}$ being the same as the constant pressure gradient formerly considered in \S\ref{longitudinal}), where $\overline{\Phi_{21}}$ was defined in \eqref{Phi2}, $\overline{U_{22}}$ was defined in \eqref{Upx}, and \revised{the new} $\overline{U_{23}}$ obeys
\be
\Delta_2 \overline{U_{23}} = \overline{P_{01}}
\ee{U22}
with boundary conditions
$
 \overline{U_{23}}[Y,F(Y)]=0,\;\overline{U_{23}}_{,Z}[Y,\infty]=0 . 
$

The corresponding second-order outer boundary condition \eqref{wdh2} now becomes
\be
u_2(x,y,0,t) = h_1 u_{1,z} + h_2 u_{0,yz} + h^{(p_x)}_2 p_{0,x} + h_2^{(v_{xz})} v_{0,xz},
\ee{wdh2rev}
where $h_1,h_2,h^{(p_x)}_2$ were defined in \eqref{def:h1}, \eqref{def:h2}, \eqref{def:h2px}, and
\be
h_2^{(v_{xz})}=\lim_{Z\rightarrow\infty}\overline{U_{23}}(Y,Z) .
\ee{def:h2vxz}

As to the \textit{r.h.s.} of \eqref{NS3cont2}, it makes a streamfunction impossible to use to derive the velocity field, and the simplest choice is to solve (\ref{NS3vmom2}--$d$) numerically in primitive form. Nonetheless, their solution can be written as the linear superposition of a solution with non-zero forcing from $v_{0,z}(x,y_0,0,t)$ and zero \textit{r.h.s.}, which \revised{can be derived from a streamfunction and} is exactly the one already dealt with in \S\ref{StokesPD2}, and a solution with $v_{0,z}(x,y_0,0,t)=0$ and non-zero \textit{r.h.s.} of \eqref{NS3cont2}, which can be represented through a new normalized triplet $\overline{V_{23}},\overline{W_{23}},\overline{P_{13}}$ that solves
\begin{subequations}
\begin{align}
\overline{V_{23}}_{,YY}+\overline{V_{23}}_{,ZZ} - \overline{P_{13}}_{,Y} &=0,\\
\overline{W_{23}}_{,YY}+\overline{W_{23}}_{,ZZ} - \overline{P_{13}}_{,Z} &=0,\\
\overline{V_{23}}_{,Y} + \overline{W_{23}}_{,Z} &= -\overline{\Phi_{11}} ; \label{P133}
\end{align}
\label{P13}
\end{subequations}
\label{V23}
with $\overline{\Phi_{11}}$ taken from \eqref{Phi1} and boundary conditions
$
\overline{V_{23}}[Y,F(Y)]=\overline{W_{23}}[Y,F(Y)]=0,\;\overline{V_{23}}_{,Z}(Y,\infty)=0,\;\overline{P_{13}}(Y,\infty)\sim -Z .
$

Then, since $U_{1,x}(x,y_0,t; Y,Z)=\overline{\Phi_{11}}(Y,Z)\,u_{0,xz}(x,y_0,0,t)$, the second-order boundary conditions \eqref{firstsecondvw} become
\be
\left\{\begin{array}{l}
v(x,y,0,t)=\eps a_1 v_{z}+\eps^2 a_2 v_{yz}+\eps^2 b_2 p_y +\eps^2 f_2 u_{xz} \\
w(x,y,0,t)=-\eps^2 c_2 v_{yz} +\eps^2 g_2 u_{xz},
\end{array}\right.
\ee{firstsecondvwfg}
where the two newly introduced numerical constants are
\be
f_2 =\lim_{Z\rightarrow\infty} \overline{V_{23}}(Y,Z),\quad \
g_2 =\lim_{Z\rightarrow\infty} [\overline{W_{23}}(Y,Z)+Z^2/2+h_1 Z].
\ee{def:f2g2}

\subsection{Third order}
The procedure should by now be self-evident. Each stage of the $\eps$-expansion entails a growing number of contributions, which superpose linearly and can be handled independently of each other. All of the terms already included in the 2D analysis are present. As compared to the 2D Navier-Stokes equations discussed in \S\ref{2DNS3}, the third order \eqref{NS3order3} of the 3D Navier-Stokes problem produces the following additions\revised{, which we shall label according to the convention already adopted in the previous sections, a first digit denoting order and a second sequential digit}.

\subsubsection{Longitudinal strain}
Each component of the momentum equation contains a second $x$-derivative, representing viscous longitudinal strain, in a position analogous to the time derivative, as highlighted below:
\begin{align*}
U_{3,YY}+U_{3,ZZ} &= P_{1,x}\,\boxed{-U_{1,xx}+U_{1,t}} + V_1U_{1,Y} + W_1U_{1,Z},\\
V_{3,YY}+V_{3,ZZ} - P_{2,Y} &= \boxed{-V_{1,xx}+V_{1,t}} + V_1V_{1,Y} + W_1V_{1,Z},\\
W_{3,YY}+W_{3,ZZ} - P_{2,Z} &= \boxed{-W_{1,xx}+W_{1,t}} + V_1W_{1,Y} + W_1W_{1,Z},\\
V_{3,Y} + W_{3,Z} &= - U_{2,x} .
\end{align*}
Indeed, the analogy is so precise that in the transverse equivalent boundary condition \eqref{nlbc3} it suffices to replace $v_{0,zt}$ by $v_{0,zt}-v_{0,xxz}$ (and $w_{0,zt}$ by $w_{0,zt}-w_{0,xxz}$ if they were present; but, by continuity and the no-slip conditions on $u_0$ and $v_0$, $w_{0,zt} = w_{0,xxz} = 0$). In the longitudinal momentum equation, where the time derivative was not included in the analysis of \S\ref{longitudinal}, this contribution to $U_3$ can be written as $\overline{U_{33}}(Y,Z)[u_{0,zt}(x,y_0,0,t)-u_{0,xxz}(x,y_0,0,t)]$, where \revised{the new function} $\overline{U_{33}}$ is the solution of
\be
\Delta_2 \overline{U_{33}} = \overline{\Phi_{11}},
\ee{def:U33}
$\overline{\Phi_{11}}$ being defined in \eqref{Phi1}, with boundary conditions
$
\overline{U_{33}}[Y,F(Y)]=0,\;\overline{U_{33}}_{,Z}(Y,\infty)\sim Z^2/2+h_1 Z .
$

There follows an additional term to the equivalent boundary condition \eqref{outer_2_bc_U} of the form $\eps^3 h_3^{(u_{zt})} (u_{zt}-u_{xxz})$, where
\be
h_3^{(u_{zt})}=\lim_{Z\rightarrow\infty}[\overline{U_{33}}(Y,Z)-Z^3/6-h_1 Z^2/2] .
\ee{def:h3uzt}

\subsubsection{Nonlinear terms}
The nonlinear contributions to the third-order inner solution,
\begin{align*}
U_{3,YY}+U_{3,ZZ} &= P_{1,x}-U_{1,xx}+U_{1,t} + \boxed{V_1U_{1,Y} + W_1U_{1,Z}}\,,\\
V_{3,YY}+V_{3,ZZ} - P_{2,Y} &= -V_{1,xx}+V_{1,t} + \boxed{V_1V_{1,Y} + W_1V_{1,Z}}\,,\\
W_{3,YY}+W_{3,ZZ} - P_{2,Z} &= -W_{1,xx}+W_{1,t} + \boxed{V_1W_{1,Y} + W_1W_{1,Z}}\,,\\
V_{3,Y} + W_{3,Z} &= - U_{2,x} ,
\end{align*}
can be subdivided into a transverse part, which is identical to the one already studied in \S\ref{2DNS3}, and a longitudinal part which represents transport of longitudinal momentum caused by the crossflow velocities. The \textit{r.h.s.} for the longitudinal part can be written as $[\overline{\Psi_{21}}_{,Z}\overline{\Phi_{11}}_{,Y}-\overline{\Psi_{21}}_{,Y}\overline{\Phi_{11}}_{,Z}]v_{0,z}u_{0,z}$. Therefore the corresponding contribution to $U_3$ is \revised{a new function} $\overline{U_{34}}(Y,Z)v_{0,z}u_{0,z}$, where
\be
\Delta_2\overline{U_{34}} = \overline{\Psi_{21}}_{,Z}\overline{\Phi_{11}}_{,Y}-\overline{\Psi_{21}}_{,Y}\overline{\Phi_{11}}_{,Z}
\ee{def:U34}
with boundary conditions
$
\overline{U_{34}}[Y,F(Y)]=0,\;\overline{U_{34}}_{,Z}(Y,\infty)=0 .
$

There ensues an additional term to the equivalent boundary condition \eqref{outer_2_bc_U} of the form $\eps^3 h^{(nl)}_3v_{0,z}u_{0,z}$, where
\be
h^{(nl)}_3=\lim_{Z\rightarrow\infty}\overline{U_{34}}(Y,Z) .
\ee{def:h3nl}

Just as its transverse counterpart, this nonlinear contribution is forbidden (the \textit{r.h.s.} of \eqref{def:U34} is an odd function of $Y$ and so is $\overline{U_{34}}$) for left-right-symmetric wall features.

\subsubsection{Cross-coupling}
Finally, \eqref{NS3order3} contain the same cross-coupling terms between longitudinal and transverse components that made their first appearance at the second order: 
\begin{align*}
U_{3,YY}+U_{3,ZZ} &= \boxed{P_{1,x}}-U_{1,xx}+U_{1,t} + V_1U_{1,Y} + W_1U_{1,Z},\\
V_{3,YY}+V_{3,ZZ} - P_{2,Y} &= -V_{1,xx}+V_{1,t} + V_1V_{1,Y} + W_1V_{1,Z},\\
W_{3,YY}+W_{3,ZZ} - P_{2,Z} &= -W_{1,xx}+W_{1,t} + V_1W_{1,Y} + W_1W_{1,Z},\\
V_{3,Y} + W_{3,Z} &= \boxed{- U_{2,x}}\,.
\end{align*}
These cross-couplings ingenerate several third-order contributions, one for every second-order term that takes the stage on the \textit{r.h.s.} in turn.

$P_{1,x}$ can be obtained from $\Psi_3$ of \eqref{def:Psi3}, with additional contributions from $\overline{P_{13}}$ of \eqref{P13} and from the order-1 outer pressure gradient at the wall. Summing the $Z$-derivative of \eqref{def:Psi3} with the term involving $\overline{V_{23}}$ of \eqref{P13} gives
\be
\begin{split}
V_2&=\left[\overline{\Psi_{31}}_{,Z}(Y,Z)+Y\overline{\Psi_{21}}_{,Z}(Y,Z)\right]v_{0,yz}(x,y_0,0,t)+ \overline{\Psi_{32}}_{,Z}(Y,Z)p_{0,y}(x,y_0,0,t)+{} \\
&\overline{V_{23}}(Y,Z)u_{0,xz}(x,y_0,0,t)+\overline{\Psi_{21}}_{,Z}(Y,Z)v_{1,z}(x,y_0,0,t)
\end{split}
\ee{3DV2}
and correspondingly
\be
\begin{split}
P_{1,x} &=\left[\overline{P_{11}}(Y,Z)+Y\overline{P_{01}}(Y,Z)\right]v_{0,xyz}(x,y_0,0,t)+ \overline{P_{12}}(Y,Z)p_{0,xy}(x,y_0,0,t)+{} \\
&\overline{P_{13}}(Y,Z)u_{0,xxz}(x,y_0,0,t)+\overline{P_{01}}(Y,Z)v_{1,xz}(x,y_0,0,t)+p_{1,x}(x,y_0,0,t) .
\end{split}
\ee{3DP1}
To convert \eqref{3DV2} into \eqref{3DP1} it mostly suffices to apply Cauchy-Riemann relationships analogous to \eqref{CauchyRiemann} term by term. This is true, in particular, for $\overline{P_{01}}$, already available from \eqref{CauchyRiemann} itself, and for $\overline{P_{12}}$, obtained from a similar set of equations applied to $\overline{\Psi_{32}}$, whereas $\overline{P_{13}}$ directly comes from \eqref{P13}. $\overline{P_{11}}$ requires special care, owing to the presence of the term proportional to $Y$ in the square brackets\revised{, as follows}.

Let us write Cauchy-Riemann relationships for the entire square brackets in \eqref{3DV2} and \eqref{3DP1}:
\begin{align*}
\left[\overline{P_{11}}+Y\overline{P_{01}}\right]_Y = \Delta_2\left[\overline{\Psi_{31}}+Y\overline{\Psi_{21}}\right]_Z,\quad
\left[\overline{P_{11}}+Y\overline{P_{01}}\right]_Z = - \Delta_2\left[\overline{\Psi_{31}}+Y\overline{\Psi_{21}}\right]_Y .
\end{align*}
Expanding the derivatives, and taking into account that $\overline{P_{01}},\overline{\Psi_{21}} $ already obey their own Cauchy-Riemann relationships, gives
\be
\overline{P_{11}}_{,Y}+\overline{P_{01}} = \Delta_2\overline{\Psi_{31}}_{,Z}+2\overline{\Psi_{21}}_{,YZ},\quad
\overline{P_{11}}_{,Z} = - \Delta_2\overline{\Psi_{31}}_{,Y}-3\overline{\Psi_{21}}_{,YY}-\overline{\Psi_{21}}_{,ZZ} .
\ee{P11CR}
$\overline{P_{11}}$ must \revised{then} be obtained from \eqref{P11CR}, with a boundary condition $\overline{P_{11}}(Y,\infty)\sim -Z$.

The contribution to $U_3$ resulting from all the terms of \eqref{3DP1} together, to be added to \eqref{Phi3}, \revised{can be written as}
\[
\begin{split}
&\left[\overline{U_{35}}(Y,Z)+Y\overline{U_{23}}(Y,Z)\right]v_{0,xyz}(x,y_0,0,t)+ \left[\overline{U_{36}}(Y,Z)+Y\overline{U_{22}}(Y,Z)\right]p_{0,xy}(x,y_0,0,t)+{} \\
&\overline{U_{37}}(Y,Z)u_{0,xxz}(x,y_0,0,t)+\overline{U_{23}}(Y,Z)v_{1,xz}(x,y_0,0,t)+\overline{U_{22}}p_{1,x}(x,y_0,0,t)
\end{split}
\]
where $\overline{U_{22}}$ and $\overline{U_{23}}$ were already defined in \eqref{Upx} and \eqref{U22} respectively, and \revised{the new functions} $\overline{U_{35}}$, $\overline{U_{36}}$, $\overline{U_{37}}$ obey
\[
\Delta_2\overline{U_{35}}+2\overline{U_{23}}_{,Y}=\overline{P_{11}},\quad\Delta_2\overline{U_{36}}+2\overline{U_{22}}_{,Y}=\overline{P_{12}},\quad\Delta_2\overline{U_{37}}=\overline{P_{13}}
\]
with boundary conditions $\overline{U_{35}}_{,Z}(Y,\infty)\sim -Z^2/2$, $\overline{U_{36}}_{,Z}(Y,\infty)=0$, $\overline{U_{37}}_{,Z}(Y,\infty)=0$.

The equivalent boundary condition for $u(x,y,0,t)$, \eqref{outer_2_bc_U}, will then acquire new contributions
\[
\eps^3 h_3^{(v_{xyz})} v_{xyz}+\eps^3 h_3^{(p_{xy})} p_{xy}+\eps^3 h_3^{(u_{xxz})}u_{xxz}
\]
where
\be
\begin{split}
&h_3^{(v_{xyz})}=\lim_{Z\rightarrow\infty}\left[\overline{U_{35}}(Y,Z)+Z^3/6\right],\quad h_3^{(p_{xy})}=\lim_{Z\rightarrow\infty}\overline{U_{36}}(Y,Z),\\
& h_3^{(u_{xxz})}=\lim_{Z\rightarrow\infty}\overline{U_{37}}(Y,Z).
\end{split}
\ee{def:h3vpu}

As far as the \textit{r.h.s.} of the continuity equation \eqref{continuity3} is concerned,
 the $x$-derivative of the second-order longitudinal velocity, $U_2$ of \eqref{def:U2}, is
\be
\begin{split}
U_{2,x}&= \overline{\Phi_{11}}(Y,Z)u_{1,xz}(x,y_0,0,t) +\left[\overline{\Phi_{21}}(Y,Z)+Y\,\overline{\Phi_{1}(Y,Z)}\right]u_{0,xyz}(x,y_0,0,t)+{} \\
&\overline{U_{22}}(Y,Z)p_{0,xx}(x,y_0,0,t)+\overline{U_{23}}(Y,Z)v_{0,xxz}(x,y_0,0,t) .
\end{split}
\ee{prepW36}
Each term in turn must be inserted into four problems similar to \eqref{P13}, one with \textit{r.h.s.} $-\overline{\Phi_{11}}(Y,Z)$ yielding the same $\overline{V_{23}}, \overline{W_{23}}, \overline{P_{13}}$ as \eqref{P13}, one with \textit{r.h.s.} $-\overline{\Phi_{21}}(Y,Z)-Y\overline{\Phi_{11}}(Y,Z)$ yielding \revised{a new triplet} $\overline{V_{35}}, \overline{W_{35}}, \overline{P_{25}}$, one with \textit{r.h.s.} $-\overline{U_{22}}(Y,Z)$ yielding \revised{a new triplet} $\overline{V_{36}}, \overline{W_{36}}, \overline{P_{26}}$  and one with \textit{r.h.s.} $-\overline{U_{23}}(Y,Z)$ yielding \revised{a new triplet} $\overline{V_{37}}, \overline{W_{37}}, \overline{P_{27}}$.

Just as with \eqref{P11CR}, the term $-\overline{\Phi_{21}}(Y,Z)-Y\overline{\Phi_{11}}(Y,Z)$ requires special care for the presence of a part proportional to $Y$. Inserting this binomial into a Stokes problem like \eqref{P13} gives
\begin{align*}
(\overline{V_{35}}+Y\overline{V_{23}})_{YY}+(\overline{V_{35}}+Y\overline{V_{23}})_{ZZ} - (\overline{P_{25}}+Y\overline{P_{13}})_{Y} &=0,\\
(\overline{W_{35}}+Y\overline{W_{23}})_{YY}+(\overline{W_{35}}+Y\overline{W_{23}})_{ZZ} - (\overline{P_{25}}+Y\overline{P_{13}})_{Z} &=0,\\
(\overline{V_{35}}+Y\overline{V_{23}})_{Y} + (\overline{W_{35}}+Y\overline{W_{23}})_{Z} &=-\overline{\Phi_{21}}-Y\overline{\Phi_{11}} ,
\end{align*}
\textit{i.e.}, upon remembering that $\overline{V_{23}},\overline{W_{23}},\overline{P_{13}}$ themselves satisfy  \eqref{P13},
\begin{subequations}
\begin{align}
\overline{V_{35}}_{,YY}+\overline{V_{35}}_{,ZZ} - \overline{P_{25}}_{,Y} &=\overline{P_{13}}-2\overline{V_{23}}_{,Y},\\
\overline{W_{35}}_{,YY}+\overline{W_{35}}_{,ZZ} - \overline{P_{25}}_{,Z} &=-2\overline{W_{23}}_{,Y},\\
\overline{V_{35}}_{,Y} + \overline{W_{35}}_{,Z} &=-\overline{\Phi_{21}}-\overline{V_{23}} .
\end{align}
\label{V35}
\end{subequations}
This is the system of equations to obtain $\overline{V_{35}},\overline{W_{35}},\overline{P_{25}}$ from.

There will result three new contributions to the $v$ and $w$ equivalent boundary conditions \eqref{nlbc3},
\[
\left\{ \begin{array}{l}
\eps^3 f_3^{(u_{xyz})} u_{xyz}+\eps^3 f_3^{(p_{xx})}p_{xx}+\eps^3 f_3^{(v_{xxz})}v_{xxz} \\
\eps^3 g_3^{(u_{xyz})} u_{xyz}+\eps^3 g_3^{(p_{xx})}p_{xx}+\eps^3 g_3^{(v_{xxz})}v_{xxz} \\
\end{array}\right.
\]
where
\be
\begin{split}
&f_3^{(u_{xyz})}=\lim_{Z\rightarrow\infty}[\overline{V_{35}}(Y,Z)+Z^3/6],\quad f_3^{(p_{xx})}=\lim_{Z\rightarrow\infty}\overline{V_{36}}(Y,Z),\\
&f_3^{(v_{xxz})}=\lim_{Z\rightarrow\infty}\overline{V_{37}}(Y,Z)
\end{split}
\ee{def:f3}
and
\be
\begin{split}
&g_3^{(u_{xyz})}=\lim_{Z\rightarrow\infty}[\overline{W_{35}}(Y,Z)+(h_2+f_2 ) Z], \\
&g_3^{(p_{xx})}=\lim_{Z\rightarrow\infty}[\overline{W_{36}}(Y,Z)+Z^3/6+h_2^{(p_x)}Z], \\
&g_3^{(v_{xxz})}=\lim_{Z\rightarrow\infty}[\overline{W_{37}}(Y,Z)+h_2^{(v_{xz})}Z].
\end {split}
\ee{def:g3}

\subsection{\revised{Fourth and higher order}}
\revised{
Although we shall not in this paper pursue the detailed calculation of fourth- and higher-order contributions, the road should now be clear. The next order in the expansion of \eqref{NS3eps}, for instance, is
\\[0.5\baselineskip]
Fourth order:
\begin{subequations}
\begin{align}
&U_{4,YY}+U_{4,ZZ} = P_{2,x}-U_{2,xx}+U_{2,t} +U_1U_{1,x}+ V_2U_{1,Y}+ V_1U_{2,Y} + W_2U_{1,Z} + W_1U_{2,Z},\\
&V_{4,YY}+V_{4,ZZ} - P_{3,Y} = -V_{2,xx}+V_{2,t} +U_1V_{1,x} + V_2V_{1,Y} + V_1V_{2,Y} + W_2V_{1,Z} + W_1V_{2,Z},\\
&W_{4,YY}+W_{4,ZZ} - P_{3,Z} = -W_{2,xx}+W_{2,t} +U_1W_{1,x} + V_2W_{1,Y} + V_1W_{2,Y} + W_2W_{1,Z} + W_1W_{2,Z},\\
&V_{4,Y} + W_{4,Z} = - U_{3,x} .
\end{align}
\label{NS3order4}
\end{subequations}
As can be seen, linear terms follow a similar pattern at every order, whereas nonlinear, advective terms produce a convolution of previous orders stemming from their product structure. When these systems are solved in a cascade they will produce a general $n^\text{th}$ order equivalent boundary condition for the velocity vector $(u,v,w)$ at the equivalent wall $z=0$ akin to \eqref{eq:bcn}-\eqref{eq:bcN}. Just as \eqref{eq:bcn}-\eqref{eq:bcN} contained higher and higher lateral derivatives of $\psi_{zz}$ and $\psi_{zzz}$, this boundary condition will contain partial derivatives of order up to $n-1$ with respect to $x,y$ and $t$ of the triplet $(u_z,v_z,p)$. As can be imagined the number of terms goes up rapidly with order, though not as rapidly as it might, because many of them will have zero coefficients. Rather than trying to provide a tensorial notation for the general formula, which would be cumbersome, we prefer to name each relevant term individually in the next section.
}

\section{Summing-up and numerical examples}\label{numerical}
Collecting the results of the last five sections together, the complete, up to third-order, equivalent boundary condition obtained through matched asymptotic expansions is
\be
\begin{split}
&\left[\begin{matrix}
u\\
v\\
w
\end{matrix}\right] =  \
\eps \left[\begin{matrix}
h_1 u_z\\
a_1 v_z\\
0
\end{matrix}\right] + \
\eps^2 \left[\begin{matrix}
\cg{h_2 u_{yz}} +{} \\
\cg{a_2 v_{yz}} +{} \\
{}
\end{matrix}\begin{matrix}
h_2^{(p_x)} p_x+\cg{h_2^{(v_{xz})} v_{xz}}\\
b_2 p_{y}+\cg{f_2  u_{xz}}\\
-c_2 v_{yz}+g_2  u_{xz}
\end{matrix}\right]+\eps^3 \left[\begin{matrix}
h_3 u_{yyz}+{}\\
a_3 v_{yyz} +{}\\
{}
\end{matrix}\right. \\
& \left.\begin{matrix}
\cg{h^{(p_{xy})}_3 p_{xy}}+ h_3^{(u_{zt})} (u_{zt}-u_{xxz})+\cg{h^{(nl)}_3 u_z v_z}+h_3^{(v_{xyz})} v_{xyz}+ h_3^{(u_{xxz})}u_{xxz}\\
\cg{b_3 p_{yy}}+e_3^{(v_{zt})}(v_{zt}-v_{xxz})+\cg{e_3^{(nl)}v_{z}^2}+f_3^{(u_{xyz})} u_{xyz}+\cg{f_3^{(p_{xx})}p_{xx}}+f_3^{(v_{xxz})}v_{xxz}\\
\cg{-c_3 v_{yyz}}-d_3  p_{yy}+\cg{g_3^{(u_{xyz})} u_{xyz}}+g_3^{(p_{xx})}p_{xx}+\cg{g_3^{(v_{xxz})}v_{xxz}}
\end{matrix}\right]
\end{split}
\ee{effectivebc}
where terms that are forbidden by left-right symmetry (when the wall texture has got such a symmetry) have been greyed out.

\begin{table}
  \begin{center}
  \def~{\hphantom{0}}
  \input{coeftable}
  \caption{Up to third-order \revised{protrusion coefficients} for some typical riblet shapes.}
  \label{tab:coefficients}
  \end{center}
\end{table}

The coefficients involved in this formula have been computed numerically for \revised{six} wall shapes, chosen for the sake of example but also for their objective interest:
\begin{enumerate}
\item Triangular riblets with an equilateral shape, tip angle $60^\circ$, and height $\sqrt{3/4} s^*$. A relatively common shape in both simulations and experiments \dc{\citep{Choi93direct,Bechert97experiments,Endrikat21influence,Modesti21dispersive,Wong24viscous}, coefficients also computed in table~4 of \citet{Ahmed25exploring}}.
\item \dc{Triangular riblets with tip angle $90^\circ$, and height $0.5 s^*$. These too considered in \citep{Choi93direct,Bechert97experiments,Endrikat21influence,Modesti21dispersive,Wong24viscous}. 2nd-order and partial 3rd-order coefficients for this geometry were computed by \citet{Bottaro20effective}.}
\item Rectangular riblets with height $0.5 s^*$ and width $0.2 s^*$. Not particularly effective for drag reduction, but easy to mesh for the purpose of numerical simulation and therefore useful for comparison with other authors \dc{\citep{Endrikat21influence,Modesti21dispersive,Wong24viscous}, coefficients also computed in table~4 of \citet{Ahmed25exploring}}. \dc{\citet{Garcia-Mayoral11hydrodynamic} considered a similar shape but with tip width $0.25s^*$.}
\item Trapezoidal riblets with height $0.5 s^*$ and tip angle $30^\circ$, close to what is empirically considered to provide the best drag reduction with a practical shape \dc{\citep{Endrikat21influence,Modesti21dispersive,Wong24viscous}, coefficients also computed in table~4 of \citet{Ahmed25exploring}}.
\item Asymmetric riblets with a sawtooth shape, height $0.5 s^*$, and tip angle such as to yield a width equal to the period $s^*$. Devised to test the effects of asymmetry on those coefficients of \eqref{effectivebc} that are mandated to be zero when riblets are left-right symmetric. Some flow in this geometry was \dc{first simulated} by \citet{Modesti21dispersive}\dc{, coefficients also computed in table~4 of \citet{Ahmed25exploring}}.
\item The \revised{sixth} pattern is not really an indented geometry, but a pattern of slip and no-slip stripes on an impermeable flat surface, with $50\%$--$50\%$ surface coverage. This configuration is the simplest idealized model of a superhydrophobic surface when used for drag reduction. It also possesses analytical solutions \citep{Philip}, which provide its two first-order protrusion heights as $h_1=\log(2)/(2\upi), a_1=h_1/2$.
\dc{These have been computed by \citet{Jelly14turbulence} and \citet{Turk14turbulent}}.
\end{enumerate}

\revised{Of course the calculation can be easily repeated for different-aspect-ratio or more convoluted riblet shapes; for instance, an alternating sequence of small and large riblets is nothing else than a periodic sequence with double the period. All the numerical codes required are available at \url{https://cplcode.net/Applications/article-CPLcodes/horbcs/}.}

For each of the coefficients in \eqref{effectivebc} a corresponding numerical viscous problem has been solved. This is less daunting a computational task than it may look like, because only two solvers have to be programmed, one for a 2D Poisson problem and one for a 2D Stokes problem, with a single set of boundary conditions each, and then applied multiple times to different right-hand sides of either the equations or boundary conditions or both. The equations for either problem have been discretized on a square grid using a second-order, finite-difference, immersed-boundary method \citep{Luchini25ibm}, enhanced with an analytical correction for the tip singularity. The resulting discrete linear system has been LU-decomposed by direct banded Gauss elimination, just once, and then applied to a battery of different right-hand sides. The values obtained for the various coefficients are collected in table \ref{tab:coefficients}.

\begin{figure}
\centerline{{\includegraphics[trim=112 291 113 120, clip]
{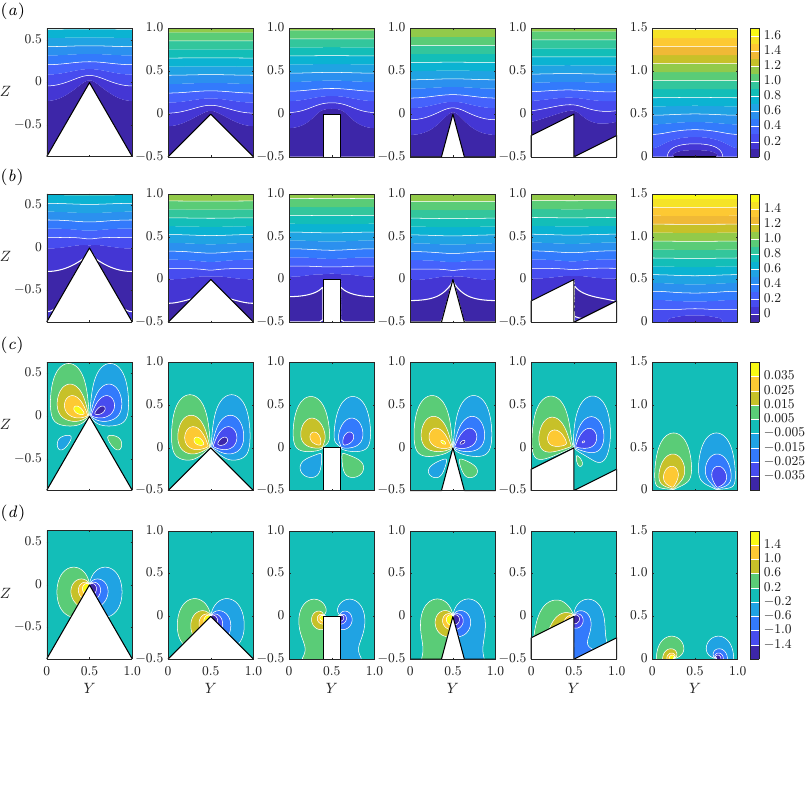}}}
\caption{(\textit{a}) First-order streamwise velocity \revised{$\overline{U_{11}}=\overline{\Phi_{11}}$ of \eqref{Phi1}, responsible for the $h_1$ longitudinal protrusion height}, (\textit{b}) spanwise velocity \revised{$\overline{V_{11}}=\overline{\Psi_{21}}_{,Z}$ of \eqref{Psi21}, responsible for the $a_1$ transverse protrusion height}, (\textit{c}) wall-normal velocity \revised{$\overline{W_{11}}=-\overline{\Psi_{21}}_{,Y}$}, and (\textit{d}) \revised{connected} zeroth-order pressure \revised{$\overline{P_{01}}$ of \eqref{CauchyRiemann}}, for the \revised{six} considered geometries. The \revised{corresponding} riblet profile is drawn on top of each figure.}
\label{fig:1storder}
\end{figure}

\begin{figure}
\centerline{{\includegraphics[trim=112 381 113 120, clip]
{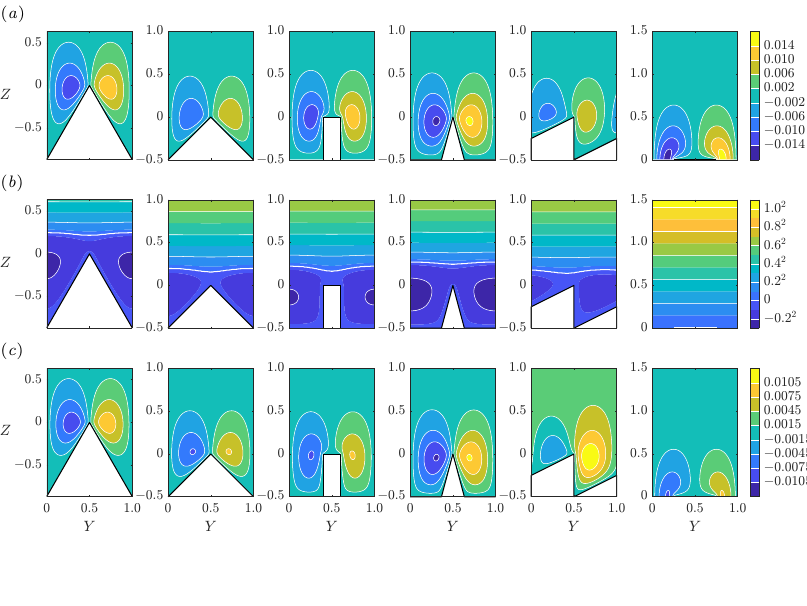}}}
\caption{Second-order contributions to the streamwise velocity. (\textit{a}) $\overline{\Phi_{21}}$ of \eqref{Phi20}, \revised{producing coefficient $h_2$}, (\textit{b}) $\overline{U_{22}}$ of \eqref{Upx}, \revised{producing coefficient $h_2^{(p_x)}$}, (\textit{c}) $\overline{U_{23}}$ of \eqref{U22}, \revised{producing coefficient $h_2^{(v_{xz})}$}.}
\label{fig:2ndorder_u}
\end{figure}

\begin{figure}
\centerline{{\includegraphics[trim=112 381 113 120, clip]
{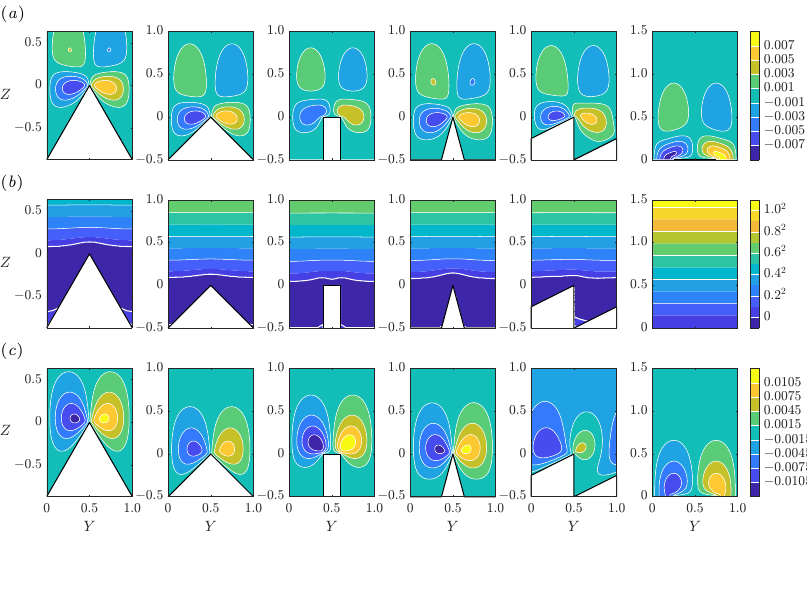}}}
\caption{Second-order contributions to the spanwise velocity. (\textit{a}) $\overline{\Psi_{31}}_{,Z}$ of \eqref{Psi31}, \revised{producing $a_2$ and $c_3$}, (\textit{b}) $\overline{\Psi_{32}}_{,Z}$ of \eqref{Psi32}, \revised{producing $b_2$ and $d_3$}, (\textit{c}) $\overline{V_{23}}$ of \eqref{V23}, \revised{producing $f_2$}.}
\label{fig:2ndorder_v}
\end{figure}
\begin{figure}
\centerline{{\includegraphics[trim=112 475 113 120, clip]
{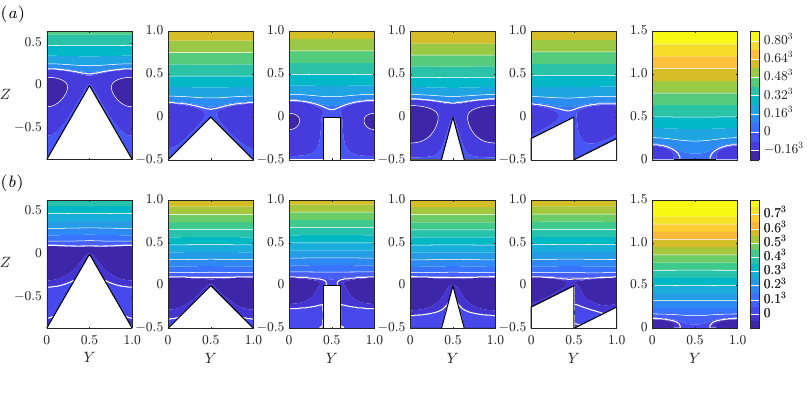}}}
\caption{Third-order velocities ingenerated by the time derivative. (\textit{a}) streamwise velocity $\overline{U_{33}}$ of \eqref{def:U33}, \revised{producing $h_3^{(u_{zt})}$}, (\textit{b}) spanwise velocity $\overline{\Psi_{43}}_{,Z}$ of \eqref{Psi43}, \revised{producing $e_3^{(v_{zt})}$}.}
\label{fig:3rdorder_unsteady}
\end{figure}

\begin{figure}
\centerline{{\includegraphics[trim=112 475 113 120, clip]
{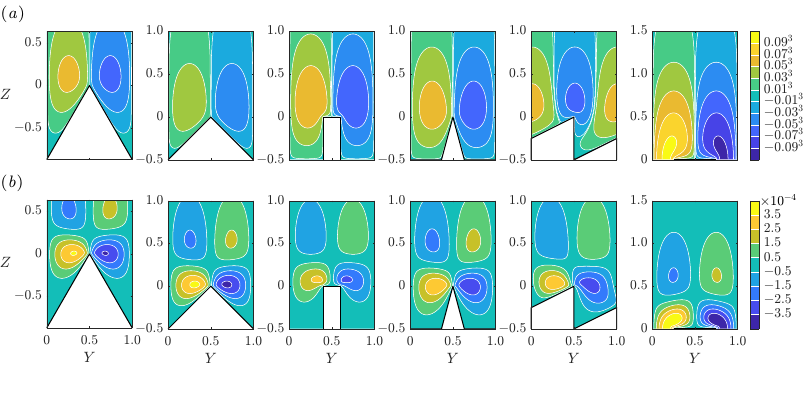}}}
\caption{Nonlinear third-order velocities. (\textit{a}) streamwise velocity $\overline{U_{34}}$ of \eqref{def:U34}, \revised{producing $h_3^{(nl)}$}, (\textit{b}) spanwise velocity $\overline{\Psi_{44}}_{,Z}$ of \eqref{Psi44}, \revised{producing $e_3^{(nl)}$}.}
\label{fig:3rdorder_nl}
\end{figure}

\begin{figure}
\centerline{{\includegraphics[trim=112 571 113 120, clip]
{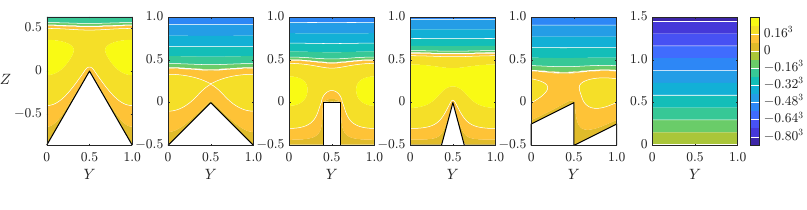}}}
\caption{Wall-normal velocity corresponding to the largest wall-normal third-order correction, $\overline{W_{36}}$ defined just below \eqref{prepW36}, \revised{producing $g_3^{(p_{xx})}$}.}
\label{fig:3rdorder_w36}
\end{figure}

A selection of corresponding velocity fields are collected in figures \ref{fig:1storder}--\ref{fig:3rdorder_w36}. Each plot covers a $1\times 1.5$ patch of a taller $1\times 3$ computational domain covered with $241\times 721$ grid points ($121\times 181$ displayed). The computational domain can be kept relatively short because each solution converges exponentially to its asymptotic behaviour in $Z$. 

We can remark that, of the 2 first-order, 8 second-order, and 18 third-order potential coefficients, several are zero. 
For left-right symmetric riblets, which means for the first three columns of table \ref{tab:coefficients}, 2 first-order, 4 second-order and 10 third-order coefficients are allowed to be non-zero, as the rest are forbidden by symmetry (as brought out in table \ref{tab:coefficients}). In addition, two pairs of second-order coefficients are tied to each other ($b_2=-c_2, h_2^{(p_x)}=g_2$), thus making the number of independent second-order coefficients just $2$ \revised{for symmetric riblets}. One can observe in figures \ref{fig:2ndorder_u} and \ref{fig:2ndorder_v} that for each forbidden coefficient the corresponding velocity field is of odd symmetry, thus making its $Y$-average zero.

But even for asymmetric riblets, in the \revised{fifth} column, astoundingly enough 2 second-order coefficients and 2 third-order coefficients are zero, thus making the non-zero totals 2, 6 and 16. This is not just a coincidence or a numerical approximation: as shown in appendix \ref{zeroproof}, these coefficients can actually be proved to be zero for any wall shape. They are also not just any of the terms: $h_2$ and $a_2$ are the first higher-order \revised{protrusion coefficients} that came about in the derivation (\eqref{def:h2} and \eqref{def:a2b2} respectively), and even more importantly $e_3^{(nl)}$ and $h_3^{(nl)}$ are the only third-order terms that involve nonlinearities of the Navier-Stokes equations (\eqref{def:e3} and \eqref{def:h3nl} respectively). Therefore nonlinearities, which were already barred by symmetry for left-right symmetric wall textures, are actually absent, up to third order, for any textures.

One can also notice from the table that not all coefficients are independent, since $b_2=-c_2$, $b_3=c_3$, $f_2=-h_2^{(v_{xz})}$, $g_2 =h_2^{(p_x)}$, $h_3^{(p_{xy})}=-g_3^{(u_{xyz})}$, and $f_3^{(p_{xx})}=-g_3^{(v_{xxz})}$. Theoretical justification for some of these identities is provided in appendix \ref{symmetrysec}.

An even larger number of terms are zero in the boundary conditions for the slip-no-slip surface, for the reason that some of the polynomial behaviours required at $Z\rightarrow\infty$ already have $u$ and its derivative $u_z$ (or $v$ and $v_z$) simultaneously vanish at the wall, and therefore satisfy the boundary conditions on both the slip and no-slip sections of the wall without any $Y$-dependence required. Such polynomial solutions do not entail any corrections and produce zero coefficients. In fact, as can be seen from figures \ref{fig:2ndorder_u} and \ref{fig:2ndorder_v}, all the second-order velocity fields in the last column are either constant with $Y$ or of odd parity. Astoundingly enough, \emph{no} second-order \revised{protrusion coefficients} and only 8 third-order \revised{protrusion coefficients} have non-zero values in the slip-no-slip case of table \ref{tab:coefficients}. Among these, all coefficients involved in the $w(x,y,0,t)$ boundary condition (bottom row of \eqref{effectivebc}) are zero, for the trivial reason that the boundary condition $w(x,y,0,t)=0$ is exact.

\begin{figure}
\centerline{\input{testerr.tex}}
\caption{Homogenisation error of the equivalent boundary conditions \eqref{effectivebc}, calculated for sawtooth riblets with $U_Z(0)=0.480125-0.754587\im, V_Z(0)=0.268092-1.27046\im, P(0)=0.592799-0.367651\im, \alpha=0.835223, \beta=0.777775, \omega=1.26823$ (see Appendix \ref{Testing} for the precise definition of these quantities).}\label{testerr}
\end{figure}
\revised{A word is also needed about the accuracy of formula \eqref{effectivebc} and of the coefficient values in Table \ref{tab:coefficients}, not just intended as numerical accuracy but as freedom from formulation or programming mistakes, since the formulation itself is complicated enough that to rely on one's attention only to exclude inadvertent mistakes is insufficient. For this purpose a test case has been developed, consisting of a laminar flow past riblets with an imposed sinusoidal stress excitation. (Sinusoidal in all of the $x$, $y$, $t$ directions, and in all three surface stress components, so as to excite all possible degrees of freedom. For the same reason, of exercising all possible coefficients in \eqref{effectivebc}, the test was conducted upon asymmetric sawtooth riblets. Needless to say, the error plot of symmetric riblets turns out similar.) A predefined sinusoidal stress vector was imposed on a plane at a certain distance (3 riblet periods) from the wall, and the resulting velocity vector on the same plane was measured. The 3D, unsteady Stokes problem for this flow was solved both exactly and in homogenised form, and the absolute value of the difference between the two resulting complex velocity vectors, $\left(\left|U_\text{out,hom.}-U_\text{out,ex.}\right|^2+\left|V_\text{out,hom.}-V_\text{out,ex.}\right|^2+\left|W_\text{out,hom.}-W_\text{out,ex.}\right|^2\right)^{1/2}$, was chosen to represent the homogenisation error. Nontrivial difficulties had to be overcome, related to the possible interference of discretization error with the error of the expansion we are trying to measure, the details of which are discussed in Appendix \ref{Testing}. The results of this test are displayed in Figure \ref{testerr}.}

\revised{It can be visually verified that the slope of this bilogarithmic plot, to a very good accuracy, is $3$ for the third-order equivalent boundary condition \eqref{effectivebc}, $2$ for the second-order equivalent boundary condition obtained by truncating \eqref{effectivebc} to second order, and $1$ for the classical first-order equivalent boundary condition containing the two original protrusion heights only. (As explained in Appendix \ref{Testing}, this is relative error on top of a velocity which is O$(\eps)$ itself, thus the O$(\eps^4)$ absolute error of an expansion up to third order appears as an O$(\eps^3)$ relative error.) Although the very value of velocity is obviously not accurate to $10^{-12}$, nor it needs be in practice, numerically observing the expected convergence rate of its homogenisation error all the way down to machine error gives us confidence, since this convergence would be highly unlikely to occur by chance if the computational procedure, and by extension the theory from which the procedure was derived, contained any oversights.}

\revised{Figure \ref{testerr} also gives the interesting indication that the present expansion can be practically useful in a more and more extended range of values of $\eps$ with higher and higher order, as it can be seen that the error curve for third order is consistently below the one for second order, and the error curve for second order is consistently below the one for first order, over the whole range of $\eps$. This was not a priori granted: it is not uncommon for an asymptotic expansion to exhibit a shorter validity range with increasing order, but evidently the present case has favourable behaviour.}

\dc{\subsection{Comparison with previous literature}}

\dc{Up to the second order, the asymptotic expansion (\ref{effectivebc})
is consistent with the two-dimensional $(y,z)$, two-component $(v,w)$ rough-wall
equations (7) and (8) and table~5 of \citet{Sudhakar21higher},
written here in the present notation and wall scaling (but without changing
the numerical subscripts of the coefficients):
\begin{equation}\label{eq:Sudhakar21}
v = \varepsilon[\mathcal{L}_{11}(v_z+w_y)]
  + \varepsilon^2[\mathcal{K}_{11} (-p_y + 2 w_{zy})], \quad
w = \varepsilon^2 [\mathcal{M}_{211}(v_{zy} + w_{yy})].
\end{equation}
Following Appendix~A of \citet{Ahmed25exploring}, we note that
$w = \mathrm{O}(\varepsilon^2)$ to neglect the $y$-derivatives
$w_y$, $w_{zy}$ and $w_{yy}$ on the right-hand sides, and then identify,
by matching with (\ref{effectivebc}) presently, that
$\mathcal{L}_{11} = a_1$, $-\mathcal{K}_{11} = b_2$,
$\mathcal{M}_{211} = -c_2$.
Noting that $a_2= 0$ (Appendix~\ref{zeroproof}) and
ignoring the third component $u$ or third dimension $x$ in (\ref{effectivebc}),
it can be seen that (\ref{eq:Sudhakar21})
and (\ref{effectivebc}) are identical up to the second order.
}

\dc{Likewise, the present (\ref{effectivebc}) is consistent with
the two-dimensional $(y,z)$, two-component $(v,w)$
third-order effective boundary conditions in equations (54) and (55) of
\citet{Bottaro20effective}, written in the present notation and wall scaling
(without changing the numerical subscripts of the coefficients
and dropping the superscripts of the coefficients):
\begin{align*}
&v = \varepsilon [\lambda (v_z + w_y)]
   + \varepsilon^2[m_{12} (-p_y + 2w_{zy})]
      \nonumber\\
&\quad+ \varepsilon^3[\theta (v_{zyy} + w_{yyy}) + q_{12} (v_{zt} + w_{yt})], \\
&w = \varepsilon^2[m_{21}(v_{zy}+w_{yy})]
    + \varepsilon^3[q_{21}(-p_{yy} + 2w_{zyy})].
\end{align*}
First, up to second order, it can be verified that this is
consistent with (\ref{eq:Sudhakar21}) above, i.e.\ the equation
of \citet{Sudhakar21higher}, with $\lambda=\mathcal{L}_{11}$,
$m_{12} = \mathcal{K}_{11}$ and $m_{21} = \mathcal{M}_{211}$ and,
therefore, also consistent with (\ref{effectivebc}).
Further, it has been recognised, e.g.\ table~3 of \citet{Bottaro20effective},
that $m_{21} = -m_{12}$, i.e.\ $\mathcal{K}_{11} = -\mathcal{M}_{211}$,
consistent with the numerical values in table~5 of \citet{Sudhakar21higher},
and also, in the present notation, $-b_2 = c_2$, also verified in the
present table~\ref{tab:coefficients} and now proved in
Appendix~\ref{symmetrysec}.
Focussing now on the third-order terms,
noticing again that $w = \mathrm{O}(\varepsilon^2)$, such
that the $y$-derivatives $w_{yyy}$, $w_{yt}$ and $w_{zyy}$ can be neglected,
we then identify by comparing with (\ref{effectivebc}) that
$\theta = a_3$, $q_{12} = e_3^{(v_{zt})}$ and $q_{21} = d_3$
in addition to $\lambda = a_1$, $m_{12} = -b_2$ and $m_{21} =  -c_2$
already established at second order.
The remaining unmatched, extra terms in (\ref{effectivebc}) are either
zero for symmetric shapes (indicated in grey) or involve the third
dimension $x$ or the third component $u$.
The third-order coefficient $e_3^{(nl)}$ is
zero (Appendix~\ref{zeroproof}), but
the third-order coefficient $b_3 = c_3$
(cf.\ Appendix~\ref{symmetrysec}), multiplying $p_{yy}$ and $v_{yyz}$,
is non-zero for asymmetric shapes (table~\ref{tab:coefficients}),
but not previously discussed, e.g.\ by \citet{Bottaro20effective}.
The claim that $q_{12} = q_{21}$ \citep{Bottaro20effective},
i.e.\ that $e_3^{(v_{zt})}$ and $d_3$ are equal is not
supported by the present data (table~\ref{tab:coefficients}).}

\dc{So far, no coefficients have been compared numerically with
previous literature.
We address this now through the three-dimensional second-order
boundary condition for a rough wall, (2.21) of
\citet{Ahmed25exploring} written with the present notation and wall scaling:
\begin{align*}
&u = \varepsilon [\lambda_x (u_z+w_x)]
   + \varepsilon^2 [\mathcal{K}_{xz}^{itf}(-p_x+2w_{zx})],\\
&v = \varepsilon [\lambda_y (v_z+w_y)]
   + \varepsilon^2 [\mathcal{K}_{yz}^{itf}(-p_y+2w_{zy})],\\
&w = \varepsilon^2 [ -\mathcal{K}_{xz}^{itf}(u_{zx}+w_{xx})
                     -\mathcal{K}_{yz}^{itf}(v_{zy}+w_{yy})],
\end{align*}
cf.\ (68) and (69) of \citet{Naqvi21interfacial}.
For a rough wall, $\mathcal{K}_{zz} = 0$ \citep[][Appendix~C]{Ahmed25exploring}.
These 4 coefficients
$(\lambda_x, \lambda_y, \mathcal{K}_{xz}^{itf},\mathcal{K}_{yz}^{itf})$ are
to be compared with
$(h_1, a_1, h_2^{(p_x)}=g_2, b_2=-c_2)$ of (\ref{effectivebc}),
numerically tabulated presently in table~\ref{tab:coefficients}
($60^\circ$-triangular, rectangular, trapezoidal, sawtooth riblets)
corresponding respectively to table~4
of \citet{Ahmed25exploring}
(equilateral triangle, thick blade, trapezoidal, right asymmetric triangle),
showing that the numbers agree.
At second order in three dimensions and three components, the new
coefficient is $f_2 = -h_2^{(v_{xz})}$, which is only non-zero for
asymmetric shapes, multiplying $u_{xz}$ and $-v_{xz}$,
previously undocumented by either \citet{Ahmed25exploring}
or \citet{Sudhakar21higher}}.

\dc{In summary, compared to previous literature,
(\ref{effectivebc}) highlights missing coefficients
for asymmetric shapes.
In particular, $f_2 = -h_2^{(v_{xz})}$ was missing at second order
for three-dimensional and three-component flow, and
$b_3 = c_3$ was missing at third order
for two-dimensional and two-component flow.
Moreover, a third-order boundary condition for three-dimensional and
three-component flow had not been attempted, now addressed by
(\ref{effectivebc}), revealing an additional seven new coefficients $h_3$,
$h_3^{(u_{zt})}$, $h_3^{(v_{xyz})}$, $h_3^{(u_{xxz})}$,
$f_3^{(u_{xyz})}$, $f_3^{(v_{xxz})}$ and $g_3^{(p_{xx})}$
for symmetric shapes plus a further two,
$h_3^{(p_{xy})} = -g_3^{(u_{xyz})}$ and
$f_3^{(p_{xx})}= -g_3^{(v_{xxz})}$
(Appendix~\ref{symmetrysec}) for asymmetric shapes.
In (\ref{effectivebc}), regardless of shape symmetry,
$h_2=a_2=0$ and surprisingly $h_3^{(nl)} = e_3^{(nl)} = 0$
(Appendix~\ref{zeroproof}),
i.e.\ nonlinear terms are absent at third order.}
\\
\\

\section{Conclusions}
As discussed in the introduction, the original concept of longitudinal and transverse protrusion heights, and the connection of their difference with the initial slope of the drag-reduction curve of riblets, have been extended by several authors over the years, both in order to further inform the design of drag-affecting surfaces by predicting their performance beyond the linear regime, and as equivalent slip-length boundary conditions to simplify numerical simulations of the same. This paper is an attempt to systematize such extensions by continuing the asymptotic expansion, of which the protrusion heights represent the first order, to higher orders, and in the while to discover the role of nonlinearities. The appropriate expansion parameter is unequivocally $\eps=s^+$, the riblet period in wall units\revised{, as discussed by nondimensionalization arguments at the beginning of \S\ref{2DNavierStokes}}. The strength and at the same time the weakness of this approach is that it is a formal asymptotic expansion in the limit of $\eps\rightarrow 0$, so on the positive side every term is strictly identified and there is no ambiguity as to what is negligible compared to what, on the negative side there is no guarantee that the approximation will be useful for practically interesting values of $\eps$ (\textit{i.e.}, $s^+$). Nonetheless, there also is no guarantee for empirical approximations not based on a formal expansion, and at the very least the availability of approximations of multiple orders allows one to estimate a posteriori how close we are to reality by comparing successive orders. 

The main operational results, at a first reading that skips the mathematics involved, are collected in the third-order equivalent boundary condition \eqref{effectivebc} and the table of its coefficients for \revised{six} typical wall textures, table \ref{tab:coefficients}. We recall that the purpose of such an equivalent boundary condition is to replace the actual fine wall geometry, in the \emph{small-roughness} limit where all of its dimensions tend to zero simultaneously, by a virtual flat wall where an approximate boundary condition is imposed. The advantage for numerical simulations, which otherwise become increasingly difficult as the texture becomes smaller and smaller in size, should be obvious.

Already deciding the appropriate, most general form of such an equivalent boundary condition was not evident before we started the present work. The practical answer is contained in the Cauchy expansion of an initial-value problem: a partial differential equation, or a system of those, can at least formally be specified as an initial-value problem in the wall-normal coordinate which for the Navier-Stokes equations involves six quantities, three components of the velocity vector and three components of the normal-stress vector, and their surface-parallel and time derivatives. The most general boundary condition that transforms this into an elliptic boundary-value problem involves three constraints among these six quantities. Most convenient is to write these three constraints in explicit form by assigning the three velocity components as functions of the three stress components and their wall-parallel derivatives, so that the solid-wall condition of zero velocity is naturally recovered when texture size $s^+$ tends to zero. This is what \eqref{effectivebc} does. One must note that, whereas wall-normal derivatives of stresses (equivalently replaced by $u_z$, $v_z$ and $p$) could in principle also appear in this formula, they can always be algebraically eliminated by using the differential equations themselves and their derivatives. Therefore a canonical form where wall-normal derivatives of $u_z$, $v_z$ and $p$ never appear can and should be used to eliminate \revised{redundance}. This is notably also true of the wall-normal derivative $w_z$ of the wall-normal velocity, which can always be eliminated through the continuity equation and should not appear in the equivalent boundary condition at all.

The analysis proceeded through several intermediate problems, starting with the 2D Laplace and the 2D Stokes problem (which, not by accident, were the basic ingredients of the 1991 protrusion-height theory as well), and discovering step by step that these are also the building blocks of both 2D and 3D Navier-Stokes problems, when the wall geometry itself is 2D (as it is in classical riblets but not, for instance, in recent attempts at wavy riblets, or in general roughness). \revised{Our final result, the equivalent boundary condition \eqref{effectivebc} and the table of its coefficients, was confirmed by a numerical test case showing that the homogenisation error actually exhibits its theoretical convergence rate. Nontrivially, the homogenisation error of the higher-order condition is found to be lower not just asymptotically but in the whole range of $\eps$ (and, implicitly, wavenumbers) examined, thus providing a uniformly better approximation.}

\revised{One may wonder how the present analysis of riblets, which are a very specific form of roughness, relates to more general 3D roughness.} The limitation of the present article to \revised{spatially periodic} 2D geometries, which anyway include a large number of practically interesting cases, is intentional as it gives the theory interesting distinctive features, but is not insurmountable. Effective boundary conditions for more general 3D textures were already introduced at first order by \citet{Sarkar96} and by \citet{Luchini13linearised}, in the small-roughness limit, and at second order in the shallow-roughness limit by \citet{Kamrin10}, and could be generalized to higher orders in the future. \revised{We note, however, that the peculiarity of riblets, that the inner solution is 2D even when the outer flow is 3D, would be lost.}

One outcome was to clarify, if there was any need, why only one longitudinal and one transverse protrusion heights were included in the \citep{Luchini91resistance} original theory, despite the presence of two wall boundary conditions in the cross-flow problem. A formalization that specifies the most general relationship between the zeroth and first derivatives of the streamfunction $\psi$ on one side and its second and third derivatives on the other, through a matrix of four $a,b,c,d$ coefficients \eqref{eq:bc}, shows that these coefficients are of different orders in the scale ratio $\eps$, and $a$ alone has a role at first order. The $b,c,d$ coefficients only appear at higher orders, and were appropriately omitted at first order by \citet{Luchini91resistance}; they do play a role here, on the other hand.

Among the qualitative conclusions that can be drawn of the present analysis, even more interesting than what is included in the effective boundary condition \eqref{effectivebc} is probably what is not there. In particular, symmetry considerations forbid about half of possible couplings when riblets are left-right symmetric, because these coefficients would mix incompatible symmetries if present. We already mentioned that normal derivatives different from the basic shear components $u_z$, $v_z$ have no role in it, not so much because they cannot be there but because they need not be there; the Cauchy expansion process can always replace them with other quantities. In addition, undifferentiated pressure never appears (not so surprisingly because in incompressible flow pressure is defined up to an additive arbitrary constant, and having undifferentiated pressure in the boundary condition would make this constant no longer arbitrary); this is another reason why the first-order equivalent boundary condition, in the first block of \eqref{effectivebc}, only contains two protrusion heights.

At second order we see the appearance, still in the first line of \eqref{effectivebc}, of an effect of the longitudinal and transverse pressure gradients on the respective wall-parallel velocity components, but also effects of shear-stress gradients $v_{yz}$ and $u_{xz}$ on the wall-normal velocity. All these effects have coefficients with the dimensions of an area, just because being of second order in the expansion they are proportional to the second power $s^2$ of the texture period. The second derivative of wall normal velocity $w_{zz}$ need not appear because, from the continuity equation, it can always be replaced by $-u_{xz}-v_{yz}$. In addition, an unforeseen relationship was discovered between the effect of each pressure gradient on the corresponding velocity and the effect of the gradient of shear stress on normal velocity, in that their coefficients are necessarily equal, as evidenced in table \ref{tab:coefficients} and justified in appendix \ref{symmetrysec}. The number of independent second-order \revised{protrusion coefficients} for left-right symmetric riblets thus turns out to be just two.  Two more second-order terms, only present for asymmetric riblets, involve cross effects of $v_{xz}$ upon longitudinal velocity $u$ and of $u_{xz}$ upon transverse velocity $v$, also with equal (and opposite in sign) coefficients. Two more coefficients, $h_2$ tying $u$ to $u_{yz}$, and $a_2$ tying $v$ to $v_{yz}$, will never be present for either symmetric or asymmetric riblets (see appendix \ref{zeroproof}), despite these being the first that naturally turned up as ``second-order \revised{protrusion coefficients}" in the theory.

Concerning the slip-no-slip striped surface that is the simplest idealized model of a superhydrophobic surface, we found an unexpected important simplification in that \emph{all} second-order effects are zero and the first non-zero corrections to the standard protrusion heights are of the third order. With hindsight this lack of second-order effects can be traced back for the normal velocity to its lack of any corrections because the impermeability condition is satisfied exactly, and for the pressure gradients to the above property which states that the second-order effects of pressure gradients have the same weights as the effects upon normal velocity.

But our perhaps most striking conclusion is that, among the numerous third-order effects that are listed in table \ref{tab:coefficients} and will not be repeated here, nonlinear terms play no role. Already that nonlinearities of the Navier-Stokes equations should not appear before the third order only becomes obvious after the present formalization of the matched asymptotic expansion; but even at third order, when nonlinearities do produce nontrivial inner velocity perturbations, their contributions to the equivalent boundary conditions vanish, for symmetric riblets because they are forbidden by symmetry, and more generally for any geometry as shown in appendix \ref{zeroproof}. Therefore nonlinearity is repelled to fourth order (at least) in $s^+$, and linear equivalent boundary conditions can work unencumbered up to third order.

\backsection[Acknowledgements]{
P. Luchini wishes to thank the Department of Mechanical Engineering of the University of Melbourne for a two-month invitation (March-April 2024) during which the core of this work was conceived.
}

\backsection[Funding]{
D. Chung acknowledges support by the Air Force Office of Scientific Research (AFOSR) under award number FA2386-23-1-4071 (AOARD program manager: David Newell).
}

\renewcommand{\theHsection}{A\arabic{section}}
\appendix
\section{Symmetry of the boundary-condition matrix\label{symmetrysec}}
The Stokes equations satisfy a variational principle which gives them special symmetry properties. There is actually more than one version of this variational principle, but for the purposes of the biharmonic equation \eqref{biharmonic} its most convenient formulation uses the quantity
\be
J = \int_\mathcal{D} \frac{(\Delta_2\psi)^2}{2} \d S ,
\ee{variational}
$\mathcal{D}$ being the domain $0\le Y \le 1$, $F(Y)\le Z \le L$.
Calculating the variation $\delta J$ gives
\[
\delta J = \int_\mathcal{D}\Delta_2\delta \psi \Delta_2\psi \d S = \int_\mathcal{\partial D} \left( \de{\delta\psi}{n} \Delta_2 \psi - \delta\psi \de{\Delta_2 \psi}{n}\right) \d l + \int_\mathcal{D}\delta\psi \underbrace{\Delta_2 \Delta_2\psi}_{{}=\, 0} \d S .
\]
This is usually interpreted to say that, if boundary conditions are such as to nullify the boundary integral (for instance, if $\psi$ and $\partial\psi/\partial n$ are assigned on the whole boundary, so that their variations are zero), a solution of the biharmonic equation minimizes \eqref{variational}. However, more generally, this shows that if conditions are zero on part of the boundary and constant on the rest,
\[
\psi = \de{\psi}{n} = 0 \text{ on } \partial\mathcal{D}_1;\quad \de{\psi}{n} = r_1, \; \psi=r_2 \text{ on } \partial\mathcal{D}_2
\]
one shall have
\[
\de{J}{r_1}=\Delta_2 \psi= s_1,\quad \de{J}{r_2}=- \de{\Delta_2 \psi}{n} =s_2 .
\]
If now the solution of the equation is seen as a system with inputs $r_j$ and outputs $s_i$, $s_i$ will be linear functions of $r_j$, $s_i=A_{ij} r_j$. But
\[
A_{ij}=\de{^2J}{r_i \partial r_j} ,
\]
and therefore $A_{ij}$ is a symmetric matrix, $A_{21}=A_{12}$. Similarly, if $s_i$ are assigned inputs and $r_j$ are outputs, the matrix tying one to the other, $A^{-1}$, will be symmetric.
In the periodic flow problem with a riblet wall near $z=0$ and an opposite flat wall at distance $L\gg s$, the boundary integral vanishes on the riblet wall and the periodic boundaries, and $\partial\mathcal{D}_2$ is just the flat wall. In addition, on the flat wall conditions become asymptotically constant with $y$, and $\Delta_2\psi$ becomes just $\psi_{zz}$, as $L$ grows larger and larger. Therefore the relationship between $r_1=\partial \psi/\partial z, r_2=\psi, s_1=\psi_{zz}$, and $s_2=-\psi_{zzz}$ is symmetric. If one writes
\[
\left[\begin{matrix}
\psi_z(y,L) \\
\psi(y,L)
\end{matrix}\right]
=\left[\begin{matrix}
a_L & -b_L \\
c_L & -d_L \\
\end{matrix}\right]
\left[\begin{matrix}
\psi_{zz}(y,L) \\
-\psi_{zzz}(y,L)
\end{matrix}\right] ,
\]
at distance $L$ \revised{(notice the sign change of $b_L$ and $d_L$, with respect to \eqref{abcd}, to compensate for the presence of $-\psi_{zzz}$)}, the $[[a_L,-b_L][c_L,-d_L]]$ matrix will be symmetric. If this matrix is then relayed back to $z=0$ using \eqref{transfer} \revised{to produce $a_1,-b_2,c_2,-d_3$ of \eqref{abcd}}, it can be verified that it will still remain symmetric\revised{, in particular we shall have $-b_2=c_2$}.

Another identity can be proved as follows. By integrating both sides of \eqref{P133} over $\mathcal{D}$ and applying the divergence theorem one gets
\[
 -\int_\mathcal{D}\overline{\Phi_{11}}\d Y\d Z = \int_\mathcal{D}\left(\overline{V_{23}}_{,Y} + \overline{W_{23}}_{,Z}\right) \d Y\d Z = \int_0^1\overline{W_{23}}(Y,L)\d Y ;
\]
\textit{i.e.}, on account of \eqref{def:f2g2}, an alternate expression of $g_2$:
\be
g_2=\lim_{L\rightarrow\infty}\left( -\int_\mathcal{D}\overline{\Phi_{11}}\d Y\d Z +h_1 L+L^2/2\right).
\ee{g2id}

On the other hand, multiplying \eqref{Upx} by $\overline{\Phi_{11}}$ and integrating gives
\[
\int_\mathcal{D}\overline{\Phi_{11}}\d Y\d Z =\int_\mathcal{D}\overline{\Phi_{11}}\Delta_2 \overline{U_{22}}\d Y\d Z =\int_{\partial \mathcal{D}}\left(\overline{\Phi_{11}}\de{ \overline{U_{22}}}{n}-\de{\overline{\Phi_{11}}}{n} \overline{U_{22}}\right)\d l + \int_\mathcal{D}\underbrace{\Delta_2\overline{\Phi_{11}}}_{=\,0} \overline{U_{22}}\d Y\d Z,
\]
\textit{i.e.}, since $\overline{U_{22}}\sim Z^2/2+h_2^{(p_x)}$, according to \eqref{def:h2px}, and $\overline{\Phi_{11}}\sim Z+h_1$, according to \eqref{def:h1},
\be
\int_\mathcal{D}\overline{\Phi_{11}}\d Y\d Z \sim (L+h_1)L-(L^2/2+h_2^{(p_x)})=L^2/2+h_1L- h_2^{(p_x)} .
\ee{h2pxid}
Comparing \eqref{g2id} with \eqref{h2pxid} yields $g_2=h_2^{(p_x)}$, consistently with what \revised{is} observed in table \ref{tab:coefficients}. 

Four more identities were detected in table \ref{tab:coefficients},
\[
f_2=-h_2^{(v_{xz})},\quad b_3=c_3,\quad h_3^{(p_{xy})}=-g_3^{(u_{xyz})},\quad f_3^{(p_{xx})}=-g_3^{(v_{xxz})},
\]
all of which are only relevant to asymmetric riblets (the coefficients involved being zero for symmetric ones) and will not be formally proved here. They have, nonetheless, been subjected to an additional numerical test using a \revised{sawtooth} riblet with smoothed edge, where numerical accuracy is much higher and has allowed them to be verified to several significant digits.

\section{Proof that some coefficients turn out to be zero for any geometry\label{zeroproof}}
It was observed in \S\ref{numerical} that four coefficients of the effective boundary condition \eqref{effectivebc}, among those that are expected to be zero for left-right symmetric wall geometries, are numerically zero even for an instance of asymmetric geometry. In fact these coefficients can be proved to be zero for any geometry, as follows.

\subsection{$h_2$ \eqref{def:h2}}\label{apph2}
The adjoint equation of \eqref{eq_200} is
\be
\Delta_2\overline{\Phi_{21}}^\dagger = 0
\ee{adjeq_200}
with boundary conditions
 $\overline{\Phi_{21}}^\dagger[Y,F(Y)]=0$, $\overline{\Phi_{21}}^\dagger_{,Z}(Y,\infty) = 1$, and allows $h_2$ to be calculated from the \textit{r.h.s} of \eqref{eq_200} as
 \[
 h_2=-\frac{1}{2\upi}\int_\mathcal{D}\overline{\Phi_{21}}^\dagger\,2\,\overline{\Phi_{11}}_{,Y} \d Y \d Z,
 \]
$\mathcal{D}$ being the domain $0\le Y \le 1$, $F(Y)\le Z <\infty$.

 But equation \eqref{adjeq_200} and its boundary conditions are the same as \eqref{Phi1}, and therefore $\overline{\Phi_{21}}^\dagger=\overline{\Phi_{11}}$. It follows that
 \[
 h_2=-\frac{1}{2\upi}\int_\mathcal{D}\de{}{Y}\left(\overline{\Phi_{11}}^2\right) \d Y \d Z
 \]
and the latter integral is zero, because $\overline{\Phi_{11}}$ (and thus also $\overline{\Phi_{11}}^2$) is periodic in $Y$ of period $1$, and the integral of the derivative of a periodic function is always zero.

\subsection{$a_2$ \eqref{def:a2b2}}
The adjoint equation of \eqref{Psi31} is
\be
\Delta_2 \Delta_2\overline{\Psi_{31}}^\dagger=0
\ee{adjeq_Psi31}
with boundary conditions
\[
\overline{\Psi_{31}}^\dagger[Y,F(Y)]=\overline{\Psi_{31}}^\dagger_{,N}[Y,F(Y)]=0,\quad\overline{\Psi_{31}}^\dagger_{,ZZ}[Y,\infty]=1,\quad \overline{\Psi_{31}}^\dagger_{,ZZZ}[Y,\infty] =0 ,
\]
and allows $a_2$ to be calculated from the \textit{r.h.s} of \eqref{Psi31} as
 \[
 a_2=-\frac{1}{2\upi}\int_\mathcal{D}\overline{\Psi_{31}}^\dagger\,4\, \Delta_2\overline{\Psi_{21}}_{,Y} \d Y \d Z.
 \]
 But equation \eqref{adjeq_Psi31} and its boundary conditions are the same as \eqref{Psi21}, and therefore $\overline{\Psi_{31}}^\dagger=\overline{\Psi_{21}}$. It follows that
 \[
 a_2=-\frac{1}{2\upi}\int_\mathcal{D}\overline{\Psi_{21}}\,4\, \Delta_2\overline{\Psi_{21}}_{,Y} \d Y \d Z = \frac{2}{\upi}\int_\mathcal{D}\bnabla\overline{\Psi_{21}}\boldsymbol{\cdot}\bnabla \overline{\Psi_{21}}_{,Y} \d Y \d Z =\frac{1}{\upi}\int_\mathcal{D}\de{}{Y}\left(\bnabla\overline{\Psi_{21}}\right)^2 \d Y \d Z, 
 \]
and the latter integral is zero for the same reason as in \S\ref{apph2}.

\subsection{$h_3^{(nl)}$ \eqref{def:h3nl}}
The adjoint equation of \eqref{def:U34} is the same as \eqref{adjeq_200}, and therefore its solution $\overline{U_{34}}^\dagger$ also coincides with $\overline{\Phi_{21}}^\dagger$ and with $\overline{\Phi_{11}}$. It follows that \eqref{def:h3nl} can be alternately calculated from the \textit{r.h.s} of \eqref{def:U34} as
\[
\begin{split}
&h_3^{(nl)}=-\frac{1}{2\upi}\int_\mathcal{D} \left(
 \overline{\Psi_{21}}_{,Z}\overline{\Phi_{11}}_{,Y}-\overline{\Psi_{21}}_{,Y}\overline{\Phi_{11}}_{,Z}\right)\overline{\Phi_{11}} \d Y \d Z ={}\\
& -\frac{1}{\upi}\int_\mathcal{D} \left[
 \overline{\Psi_{21}}_{,Z}(\overline{\Phi_{11}}^2)_{Y}-\overline{\Psi_{21}}_{,Y}(\overline{\Phi_{11}}^2)_{Z}\right] \d Y \d Z ={}\\
& -\frac{1}{\upi}\int_\mathcal{D} \left( \overline{\Psi_{21}}_{,Z}\overline{\Phi_{11}}^2\right)_{Y}-\left(\overline{\Psi_{21}}_{,Y}\overline{\Phi_{11}}^2\right)_{Z} \d Y \d Z = 0,
\end{split}
\]
the last step coming from the divergence theorem and $\overline{\Phi_{11}}^2[Y,F(Y)]=0$, $\overline{\Psi_{21}}_{,Y}(Y,\infty)=0$.

\subsection{$e_3^{(nl)}$ \eqref{def:e3}}
The adjoint equation of \eqref{Psi44} is the same as \eqref{adjeq_Psi31}, and therefore its solution $\overline{\Psi_{44}}^\dagger$ also coincides with $\overline{\Psi_{31}}^\dagger$ and with $\overline{\Psi_{21}}$. It follows that the second of \eqref{def:e3} can be alternately calculated from the \textit{r.h.s} of \eqref{Psi44} as
\[
\begin{split}
&e_3^{(nl)} =-\frac{1}{2\upi}\int_\mathcal{D} \left( \overline{\Psi_{21}}_{,Z} \Delta_2\overline{\Psi_{21}}_{,Y}- \overline{\Psi_{21}}_{,Y} \Delta_2\overline{\Psi_{21}}_{,Z}\right)\overline{\Psi_{21}} \d Y \d Z ={} \\
& -\frac{1}{\upi}\int_\mathcal{D}
 \left[ (\overline{\Psi_{21}}^2)_{Z} \Delta_2\overline{\Psi_{21}}_{,Y}- (\overline{\Psi_{21}}^2)_{Y} \Delta_2\overline{\Psi_{21}}_{,Z}\right]  \d Y \d Z
 ={}\\
 &-\frac{1}{\upi}\int_\mathcal{D}
  \left[( \overline{\Psi_{21}}^2)_{Z} \Delta_2\overline{\Psi_{21}}\right]_{Y}- \left[(\overline{\Psi_{21}}^2)_{Y} \Delta_2\overline{\Psi_{21}}\right]_{Z}  \d Y \d Z = 0,
\end{split}
\]
the last step coming from the divergence theorem and $(\overline{\Psi_{21}}^2)_{Y}[Y,F(Y)]=(\overline{\Psi_{21}}^2)_{Z}[Y,F(Y)]=0$, $\Delta_2\overline{\Psi_{21}}(Y,\infty)=0$.

\section{Testing a third-order expansion through a second-order discretization\label{Testing}}
\input{testing}

\bibliographystyle{agsm}
\bibliography{intro}

\end{document}

%% file: intro.tex
Surfaces that are not smooth alter the flow-induced transfer of
momentum, heat and mass \citep{Jimenez04turbulent,Bottaro19flow,Garcia-Mayoral19control,
Chung21predicting}. Examples include riblets \citep{Garcia-Mayoral11drag},
shark denticles \citep{Lang08bristled,Savino24thrust}, seal fur \citep{Itoh06turbulent,Gomez-de-Segura19turbulent},
superhydrophobic or liquid-infused surfaces \dc{\citep{VanBuren17substantial,Garcia-Cartagena19dependence}},
roughness \citep{Busse17reynolds,Thakkar18direct,Flack10review,Flack14roughness}
and wavy surfaces \citep{Charru13sand,Thorsness78comparison}.
Understanding the relationship between form (shape, topography,
morphology) and function (alteration of transfer properties) of these
surfaces is not trivial, generally requiring challenging laboratory
experiments \citep{Abrams85relaxation,Bechert97experiments,Schultz13reynolds} or numerical simulations 
\citep{Busse17reynolds,Garcia-Mayoral11hydrodynamic}.
However, in regimes where a small
parameter exists, inroads can be made with asymptotic analyses.

As discussed in \citet{Luchini13linearised}, surface texture characterized by a small
scale can be analysed in two separate distinguished limits, one of which is a
shallow-roughness limit where height of the texture tends to zero while
its horizontal scales stay fixed. These are shallow or wavy surfaces
\citep{Luchini19large,Abrams85relaxation}. The other limit is the
small-roughness limit where all three dimensions tend to zero
proportionally to each other, i.e. the texture stays geometrically
self-similar, e.g. riblets \citep{Garcia-Mayoral11hydrodynamic,Endrikat21influence,
Modesti21dispersive}
and rough surfaces \citep{Thakkar18direct,Sharma20turbulent}
with small surface features compared to
flow features. In this paper we shall be concerned with the
small-roughness limit.  In particular, we will develop a higher-order
matched asymptotic expansion that encapsulates the relationship between
flow alteration and surface details via equivalent (homogenised)
boundary conditions. Apart from their direct use for informing design
\citep[e.g.][]{Garcia-Mayoral11drag}, \dc{it is hoped that} such boundary conditions
circumvent the need to resolve the small surface features in flow
simulations, easing a bottleneck\dc{, which then} enables routine computations and
subsequent analysis towards overall drag savings, for example
\citep{Lacis20transfer,Ahmed25exploring}.

In the context of riblets, \citet{Bechert89viscous} first
developed the idea of the longitudinal protrusion height, which can also
be interpreted as a slip length \citep[cf.][]{Bottaro19flow}. When applied in the
turbulent-flow regime, it predicts the wall-normal distance below a
reference plane (often taken as the plane of the riblet tips) where the
mean (longitudinal) flow extrapolates \revised{linearly} to zero. In this sense, the theory
predicts that the mean flow is shifted by riblets, an idea which has
been confirmed in numerical simulations \citep{Garcia-Mayoral11hydrodynamic}.
In the presence of three-dimensional turbulence, it was also
necessary to consider the other (non-longitudinal) flow components\revised{, and their own shift analogous to the one of the mean flow}.
This was explored by \citet{Luchini91resistance}, who identified the
shift in \revised{turbulent fluctuations} to be the lateral- or cross-flow (spanwise to the
mean) protrusion height. \revised{Only after both shifts are identified} can the drag change that modifies
the inner layer of wall turbulence be explained and extrapolated to full
scale \citep{Luchini96reducing}.  Specifically, the drag change\revised{, in the linearized limit,} is proportional to
the difference between the longitudinal and cross-flow protrusion
heights, which is independent of the reference plane from which the
protrusion heights are measured.

The protrusion heights \dc{\citep{Luchini91resistance}} describe well the \dc{linear} asymptotic change in drag for
vanishingly small roughness in \dc{turbulent} flow \citep{Wong24viscous},
however, the effect of wall transpiration, i.e. wall-normal component of
motion, needs to also be accounted for where small roughness features
are no longer vanishingly small \dc{\citep{Gomez18manipulation,Lacis20transfer,Ibrahim21smooth,Sudhakar21higher}}.
\dc{Transpiration is akin to a wall-normal shift in ``virtual origin" of
the wall-normal velocity component \citep{Ibrahim21smooth}. In cases even beyond the
linear asymptotic regime, turbulence is often unmodified, except
for a wall-normal shift, itself a
nonlinear combination of shifts of both spanwise and wall-normal
velocities \citep{Ibrahim21smooth}.}
For riblets that are comparable in size to the quasi-streamwise vortices
of turbulence \dc{(i.e.\ no longer vanishingly small)}, a physically based so-called viscous vortex model was
developed by \citet{Wong24viscous} to allow \dc{prediction for the
shift in turbulence, consistent with non-zero} wall-normal \dc{(transpiration)} component.

Another approach is the transpiration-resistance model of \citet{Lacis20transfer,Khorasani22near},
which are essentially homogenised
boundary conditions for the macroscopic flow, with coefficients obtained
from microscopic flow calculations performed in a single repeating unit
of surface texture.  
A similar \dc{homogenisation} approach is also pursued by \citet{Zampogna19generalized}.
\dc{As shown by \citet{Bottaro20effective} and \citet{Sudhakar21higher}, the transpiration part appears at the second order of an asymptotic expansion.}
 \revised{(The word ``order'', here and everywhere in this paper, denoting the power of the expansion parameter $\eps$ with which various quantities approach the limit of $\eps\rightarrow 0$.)}
\dc{Third-order boundary conditions are developed by \citet{Bottaro20effective}.
However, the focus of \citet{Bottaro20effective} and
\citet{Sudhakar21higher} are both on two-dimensional, two-component flow;
a three-dimensional, three-component third-order boundary condition
has not yet been developed.}
\dc{In terms of range of applicability, a limitation is the absence of advection
at lower orders, and so an innovation is a linear Oseen-like
approximation \citep{Ahmed24laminar,Ahmed25exploring},
a technique already used for porous-medium homogenisation \citep{Zampogna16fluid,Wittkowski24quasi}.
Another approach in this vein
is the slip-transpiration-vortex model \citep{Bottaro25slip}
that approximates the effect of advection.}

Although
the effect of wall transpiration has been recognised as a second-order
effect and modelled to varying degrees \citep{Lacis20transfer,Ibrahim21smooth,Wong24viscous,Ahmed25exploring},
it is unclear \dc{at which orders do unsteadiness, nonlinear advection, etc.\ appear and, conversely, what effects appear at the next (third) order.}
  This is one of
the questions this paper will address.  
\dc{We will also see that the effect of shape asymmetry appears to have been
missed.}
Another way to view the effect
of wall transpiration is the influence of boundary conditions that vary
in the directions parallel to the wall. The first-order asymptotics as
developed by \citet{Luchini91resistance} effectively assumed that the boundary
conditions are spatially uniform, and so the wall-normal velocity must
be zero at this level of approximation for an impermeable wall at the
macroscopic flow scale. However, wall-parallel spatial variations in
boundary conditions will allow for effective wall transpiration at the
flow scale \citep{Lacis20transfer,Wong24viscous,Hao25turbulent}.
Further, we will also see that order-by-order matching in
physical space does not require linearity between the macroscopic
forcing and the microscopic response, with no ad-hoc modelling nor
tuning parameters.

\revised{The approach we adopt here will be a rational asymptotic expansion in powers of a length ratio $\eps$. Any asymptotic expansion critically depends on the quantities that are kept constant during the limiting process of $\eps\rightarrow 0$, \textit{i.e.} on the scaling of quantities that are not explicitly involved in the process (a bit like partial derivatives depend on what other variables are implicitly kept constant during the differentiation). This is what is meant by ``distinguished limit'', and it has to be specified carefully. Flow past a riblet surface can be considered as having three characteristic lengths: the riblet size $s^*$ (chosen to be represented by their period), the distance to the opposing wall $L^*$ and a dynamic length which will be taken to be the viscous length $\ell^* = \nu^*/u^*_\tau$. These three lengths will combine in various ways to determine the distinguished limit, according to their ordering. Another defining feature of the distinguished limit that we adopt here is that all three components of velocity are assumed to scale proportionally to each other, \textit{i.e.} share the same reference velocity $u^*_R=u^*_\tau$. This is ordinarily the case in a near-wall turbulent flow but not, for instance, in a laminar boundary layer.}

The exercise recovers the first-order protrusion
heights \citep{Bechert89viscous,Luchini91resistance} and then
extends them to higher orders, which contain nonlinear terms that reflect
the nonlinearity of Navier--Stokes equations. The general idea that made this possible is to apply matched inner and outer asymptotic expansions in the classical sense of \citet{vanDyke} but then, instead of solving the inner and outer equations in a back-and-forth cascade, represent the outer solution by a formal Cauchy series and only (numerically) solve the inner equations. This allows an equivalent boundary condition to be formulated, at any desired order of accuracy, that completely replaces the inner dynamics by its projection on a flat virtual wall. First we start with  the
2D Laplace equation to demonstrate the mechanics of the process, and then
apply it with increasing complexity to Stokes and then the
3D Navier--Stokes equations. \revised{The process will gradually lead to our} main operational result \revised{exhibited in \S\ref{numerical}}, the third-order equivalent boundary condition \eqref{effectivebc} and the table of its coefficients for six typical wall textures, table \ref{tab:coefficients}.

%% file: coeftable.tex
\begin{tabular}{llcccccc}%
\multicolumn{2}{l}{Riblet shape:} & $60^\circ$-triangular & $90^\circ$-triangular & rectangular & $30^\circ$-trapezoidal & sawtooth & slip-no-slip \\
\\
$h_1$ & \eqref{def:h1}               & ~0.170741 & ~0.139685 & ~0.114319 & ~0.191366 & ~0.127310 & ~0.110325 \\
$a_1$ & \eqref{def:a1}               & ~0.080237 & ~0.077773 & ~0.048021 & ~0.081519 & ~0.073446 & ~0.055162 \\
\\                                                           
$h_2$ & \eqref{def:h2}               & ~0.~~~~~~ & ~0.~~~~~~ & ~0.~~~~~~ & ~0.~~~~~~ & ~0.~~~~~~ & ~0.~~~~~~ \\
$h_2^{(p_x)}$ & \eqref{def:h2px}     & -0.028213 & -0.016816 & -0.021010 & -0.034825 & -0.014106 & ~0.~~~~~~ \\
$h_2^{(v_{xz})}$ & \eqref{def:h2vxz} & ~0.~~~~~~ & ~0.~~~~~~ & ~0.~~~~~~ & ~0.~~~~~~ & ~0.003393 & ~0.~~~~~~ \\
$a_2$ & \eqref{def:a2b2}             & ~0.~~~~~~ & ~0.~~~~~~ & ~0.~~~~~~ & ~0.~~~~~~ & ~0.~~~~~~ & ~0.~~~~~~ \\
$b_2$ & \eqref{def:a2b2}             & -0.005822 & -0.005800 & -0.002077 & -0.005398 & -0.004789 & ~0.~~~~~~ \\
$c_2$ & \eqref{def:c2}               & ~0.005820 & ~0.005798 & ~0.002074 & ~0.005396 & ~0.004787 & ~0.~~~~~~ \\
$f_2 $ & \eqref{def:f2g2}            & ~0.~~~~~~ & ~0.~~~~~~ & ~0.~~~~~~ & ~0.~~~~~~ & -0.003393 & ~0.~~~~~~ \\
$g_2 $ & \eqref{def:f2g2}            & -0.028211 & -0.016814 & -0.021008 & -0.034823 & -0.014103 & ~0.~~~~~~ \\
\\                                                           
$h_3$ & \eqref{def:h3}               & ~0.001658 & ~0.000907 & ~0.000497 & ~0.002334 & ~0.000686 & -0.000767 \\
$h_3^{(u_{zt})}$ & \eqref{def:h3uzt} & -0.003593 & -0.001960 & -0.002356 & -0.004819 & -0.001505 & -0.001050 \\
$h_3^{(nl)}$ & \eqref{def:h3nl}      & ~0.~~~~~~ & ~0.~~~~~~ & ~0.~~~~~~ & ~0.~~~~~~ & ~0.~~~~~~ & ~0.~~~~~~ \\
$h_3^{(v_{xyz})}$& \eqref{def:h3vpu} & -0.002620 & -0.001222 & -0.001331 & -0.003638 & -0.000977 & -0.000758 \\
$h_3^{(p_{xy})}$ & \eqref{def:h3vpu} & ~0.~~~~~~ & ~0.~~~~~~ & ~0.~~~~~~ & ~0.~~~~~~ & -0.000213 & ~0.~~~~~~ \\
$h_3^{(u_{xxz})}$& \eqref{def:h3vpu} & ~0.001897 & ~0.001015 & ~0.004648 & ~0.002038 & ~0.000994 & ~0.001729 \\
$a_3$ & \eqref{def:a3b3}             & -0.000589 & -0.000420 & -0.000131 & -0.000628 & -0.000412 & -0.000377 \\
$b_3$ & \eqref{def:a3b3}             & ~0.~~~~~~ & ~0.~~~~~~ & ~0.~~~~~~ & ~0.~~~~~~ & ~0.000196 & ~0.~~~~~~ \\
$c_3$ & \eqref{def:c3}               & ~0.~~~~~~ & ~0.~~~~~~ & ~0.~~~~~~ & ~0.~~~~~~ & ~0.000196 & ~0.~~~~~~ \\
$d_3$ & \eqref{def:d3}               & -0.001664 & -0.001140 & -0.000981 & -0.001974 & -0.000836 & ~0.~~~~~~ \\
$e_3^{(v_{zt})}$ & \eqref{def:e3}    & -0.000575 & -0.000540 & -0.000295 & -0.000572 & -0.000442 & -0.000262 \\
$e_3^{(nl)}$ & \eqref{def:e3}        & ~0.~~~~~~ & ~0.~~~~~~ & ~0.~~~~~~ & ~0.~~~~~~ & ~0.~~~~~~ & ~0.~~~~~~ \\
$f_3^{(u_{xyz})}$& \eqref{def:f3}    & -0.001351 & -0.000724 & -0.000559 & -0.001832 & -0.000551 & -0.000758 \\
$f_3^{(p_{xx})}$ & \eqref{def:f3}    & ~0.~~~~~~ & ~0.~~~~~~ & ~0.~~~~~~ & ~0.~~~~~~ & ~0.001071 & ~0.~~~~~~ \\
$f_3^{(v_{xxz})}$& \eqref{def:f3}    & -0.000998 & -0.000694 & -0.000770 & -0.001394 & -0.001029 & -0.000454 \\
$g_3^{(u_{xyz})}$& \eqref{def:g3}    & ~0.~~~~~~ & ~0.~~~~~~ & ~0.~~~~~~ & ~0.~~~~~~ & ~0.000213 & ~0.~~~~~~ \\
$g_3^{(p_{xx})}$ & \eqref{def:g3}    & ~0.010247 & ~0.003754 & ~0.009021 & ~0.012171 & ~0.003064 & ~0.~~~~~~ \\
$g_3^{(v_{xxz})}$& \eqref{def:g3}    & ~0.~~~~~~ & ~0.~~~~~~ & ~0.~~~~~~ & ~0.~~~~~~ & -0.001071 & ~0.~~~~~~ \\
\end{tabular}

%% file: testerr.tex
\setlength{\unitlength}{0.120450pt}
\ifx\plotpoint\undefined\newsavebox{\plotpoint}\fi
\ifx\transparent\undefined%
    \providecommand{\gpopaque}{}%
    \providecommand{\gptransparent}[2]{\color{.!#2}}%
\else%
    \providecommand{\gpopaque}{\transparent{1.0}}%
    \providecommand{\gptransparent}[2]{\transparent{#1}}%
\fi%
\begin{picture}(3000,1800)(0,0)
\miterjoin\buttcap
\color{black}
\sbox{\plotpoint}{\rule[-0.200pt]{0.400pt}{0.400pt}}%
\linethickness{0.4pt}%
\multiput(472,265)(24.907,0.000){95}{\usebox{\plotpoint}}
\put(2835,265){\usebox{\plotpoint}}
\Line(472,265)(513,265)
\Line(2835,265)(2794,265)
\put(431,265){\makebox(0,0)[r]{$1\times10^{-12}$}}
\multiput(472,265)(24.907,0.000){95}{\usebox{\plotpoint}}
\put(2835,265){\usebox{\plotpoint}}
\Line(472,265)(492,265)
\Line(2835,265)(2815,265)
\multiput(472,410)(24.907,0.000){66}{\usebox{\plotpoint}}
\put(2097,410){\usebox{\plotpoint}}
\multiput(2794,410)(24.907,0.000){2}{\usebox{\plotpoint}}
\put(2835,410){\usebox{\plotpoint}}
\Line(472,410)(492,410)
\Line(2835,410)(2815,410)
\multiput(472,555)(24.907,0.000){95}{\usebox{\plotpoint}}
\put(2835,555){\usebox{\plotpoint}}
\Line(472,555)(513,555)
\Line(2835,555)(2794,555)
\put(431,555){\makebox(0,0)[r]{$1\times10^{-10}$}}
\multiput(472,555)(24.907,0.000){95}{\usebox{\plotpoint}}
\put(2835,555){\usebox{\plotpoint}}
\Line(472,555)(492,555)
\Line(2835,555)(2815,555)
\multiput(472,700)(24.907,0.000){95}{\usebox{\plotpoint}}
\put(2835,700){\usebox{\plotpoint}}
\Line(472,700)(492,700)
\Line(2835,700)(2815,700)
\multiput(472,845)(24.907,0.000){95}{\usebox{\plotpoint}}
\put(2835,845){\usebox{\plotpoint}}
\Line(472,845)(513,845)
\Line(2835,845)(2794,845)
\put(431,845){\makebox(0,0)[r]{$1\times10^{-8}$}}
\multiput(472,845)(24.907,0.000){95}{\usebox{\plotpoint}}
\put(2835,845){\usebox{\plotpoint}}
\Line(472,845)(492,845)
\Line(2835,845)(2815,845)
\multiput(472,991)(24.907,0.000){95}{\usebox{\plotpoint}}
\put(2835,991){\usebox{\plotpoint}}
\Line(472,991)(492,991)
\Line(2835,991)(2815,991)
\multiput(472,1136)(24.907,0.000){95}{\usebox{\plotpoint}}
\put(2835,1136){\usebox{\plotpoint}}
\Line(472,1136)(513,1136)
\Line(2835,1136)(2794,1136)
\put(431,1136){\makebox(0,0)[r]{$1\times10^{-6}$}}
\multiput(472,1136)(24.907,0.000){95}{\usebox{\plotpoint}}
\put(2835,1136){\usebox{\plotpoint}}
\Line(472,1136)(492,1136)
\Line(2835,1136)(2815,1136)
\multiput(472,1281)(24.907,0.000){95}{\usebox{\plotpoint}}
\put(2835,1281){\usebox{\plotpoint}}
\Line(472,1281)(492,1281)
\Line(2835,1281)(2815,1281)
\multiput(472,1426)(24.907,0.000){95}{\usebox{\plotpoint}}
\put(2835,1426){\usebox{\plotpoint}}
\Line(472,1426)(513,1426)
\Line(2835,1426)(2794,1426)
\put(431,1426){\makebox(0,0)[r]{$0.0001$}}
\multiput(472,1426)(24.907,0.000){95}{\usebox{\plotpoint}}
\put(2835,1426){\usebox{\plotpoint}}
\Line(472,1426)(492,1426)
\Line(2835,1426)(2815,1426)
\multiput(472,1571)(24.907,0.000){95}{\usebox{\plotpoint}}
\put(2835,1571){\usebox{\plotpoint}}
\Line(472,1571)(492,1571)
\Line(2835,1571)(2815,1571)
\multiput(472,1716)(24.907,0.000){95}{\usebox{\plotpoint}}
\put(2835,1716){\usebox{\plotpoint}}
\Line(472,1716)(513,1716)
\Line(2835,1716)(2794,1716)
\put(431,1716){\makebox(0,0)[r]{$0.01$}}
\multiput(472,1716)(24.907,0.000){95}{\usebox{\plotpoint}}
\put(2835,1716){\usebox{\plotpoint}}
\Line(472,1716)(492,1716)
\Line(2835,1716)(2815,1716)
\multiput(472,265)(0.000,24.907){59}{\usebox{\plotpoint}}
\put(472,1716){\usebox{\plotpoint}}
\Line(472,265)(472,306)
\Line(472,1716)(472,1675)
\put(472,182){\makebox(0,0){$0.0001$}}
\Line(650,265)(650,285)
\Line(650,1716)(650,1696)
\Line(754,265)(754,285)
\Line(754,1716)(754,1696)
\Line(828,265)(828,285)
\Line(828,1716)(828,1696)
\Line(885,265)(885,285)
\Line(885,1716)(885,1696)
\Line(932,265)(932,285)
\Line(932,1716)(932,1696)
\Line(971,265)(971,285)
\Line(971,1716)(971,1696)
\Line(1006,265)(1006,285)
\Line(1006,1716)(1006,1696)
\Line(1036,265)(1036,285)
\Line(1036,1716)(1036,1696)
\multiput(1063,265)(0.000,24.907){59}{\usebox{\plotpoint}}
\put(1063,1716){\usebox{\plotpoint}}
\Line(1063,265)(1063,306)
\Line(1063,1716)(1063,1675)
\put(1063,182){\makebox(0,0){$0.001$}}
\Line(1241,265)(1241,285)
\Line(1241,1716)(1241,1696)
\Line(1345,265)(1345,285)
\Line(1345,1716)(1345,1696)
\Line(1418,265)(1418,285)
\Line(1418,1716)(1418,1696)
\Line(1476,265)(1476,285)
\Line(1476,1716)(1476,1696)
\Line(1522,265)(1522,285)
\Line(1522,1716)(1522,1696)
\Line(1562,265)(1562,285)
\Line(1562,1716)(1562,1696)
\Line(1596,265)(1596,285)
\Line(1596,1716)(1596,1696)
\Line(1626,265)(1626,285)
\Line(1626,1716)(1626,1696)
\multiput(1654,265)(0.000,24.907){59}{\usebox{\plotpoint}}
\put(1654,1716){\usebox{\plotpoint}}
\Line(1654,265)(1654,306)
\Line(1654,1716)(1654,1675)
\put(1654,182){\makebox(0,0){$0.01$}}
\Line(1831,265)(1831,285)
\Line(1831,1716)(1831,1696)
\Line(1935,265)(1935,285)
\Line(1935,1716)(1935,1696)
\Line(2009,265)(2009,285)
\Line(2009,1716)(2009,1696)
\Line(2066,265)(2066,285)
\Line(2066,1716)(2066,1696)
\Line(2113,265)(2113,285)
\Line(2113,1716)(2113,1696)
\Line(2153,265)(2153,285)
\Line(2153,1716)(2153,1696)
\Line(2187,265)(2187,285)
\Line(2187,1716)(2187,1696)
\Line(2217,265)(2217,285)
\Line(2217,1716)(2217,1696)
\multiput(2244,265)(0.000,24.907){2}{\usebox{\plotpoint}}
\put(2244,306){\usebox{\plotpoint}}
\multiput(2244,555)(0.000,24.907){47}{\usebox{\plotpoint}}
\put(2244,1716){\usebox{\plotpoint}}
\Line(2244,265)(2244,306)
\Line(2244,1716)(2244,1675)
\put(2244,182){\makebox(0,0){$0.1$}}
\Line(2422,265)(2422,285)
\Line(2422,1716)(2422,1696)
\Line(2526,265)(2526,285)
\Line(2526,1716)(2526,1696)
\Line(2600,265)(2600,285)
\Line(2600,1716)(2600,1696)
\Line(2657,265)(2657,285)
\Line(2657,1716)(2657,1696)
\Line(2704,265)(2704,285)
\Line(2704,1716)(2704,1696)
\Line(2743,265)(2743,285)
\Line(2743,1716)(2743,1696)
\Line(2778,265)(2778,285)
\Line(2778,1716)(2778,1696)
\Line(2808,265)(2808,285)
\Line(2808,1716)(2808,1696)
\multiput(2835,265)(0.000,24.907){59}{\usebox{\plotpoint}}
\put(2835,1716){\usebox{\plotpoint}}
\Line(2835,265)(2835,306)
\Line(2835,1716)(2835,1675)
\put(2835,182){\makebox(0,0){$1$}}
\polygon(472,1716)(472,265)(2835,265)(2835,1716)
\put(2507,513){\makebox(0,0)[r]{$3^\text{rd}$ order}}
\color[rgb]{0.58,0.00,0.83}
\Line(2548,513)(2753,513)
\polyline(472,379)(472,379)(590,384)(708,312)
\Line(708,312)(786,265)
\polyline(844,265)(945,394)(1063,352)(1181,330)(1299,419)(1417,511)(1535,597)(1654,685)(1772,772)(1890,859)(2008,945)(2126,1032)(2244,1117)(2362,1202)(2481,1284)(2599,1360)(2717,1420)(2835,1458)
\put(472,379){\makebox(0,0){$+$}}
\put(590,384){\makebox(0,0){$+$}}
\put(708,312){\makebox(0,0){$+$}}
\put(945,394){\makebox(0,0){$+$}}
\put(1063,352){\makebox(0,0){$+$}}
\put(1181,330){\makebox(0,0){$+$}}
\put(1299,419){\makebox(0,0){$+$}}
\put(1417,511){\makebox(0,0){$+$}}
\put(1535,597){\makebox(0,0){$+$}}
\put(1654,685){\makebox(0,0){$+$}}
\put(1772,772){\makebox(0,0){$+$}}
\put(1890,859){\makebox(0,0){$+$}}
\put(2008,945){\makebox(0,0){$+$}}
\put(2126,1032){\makebox(0,0){$+$}}
\put(2244,1117){\makebox(0,0){$+$}}
\put(2362,1202){\makebox(0,0){$+$}}
\put(2481,1284){\makebox(0,0){$+$}}
\put(2599,1360){\makebox(0,0){$+$}}
\put(2717,1420){\makebox(0,0){$+$}}
\put(2835,1458){\makebox(0,0){$+$}}
\put(2650,513){\makebox(0,0){$+$}}
\color{black}
\put(2507,430){\makebox(0,0)[r]{$2^\text{nd}$ order}}
\color[rgb]{0.00,0.62,0.45}
\Line(2548,430)(2753,430)
\polyline(472,490)(472,490)(590,544)(708,606)(826,663)(945,721)(1063,779)(1181,837)(1299,895)(1417,953)(1535,1011)(1654,1069)(1772,1127)(1890,1184)(2008,1242)(2126,1298)(2244,1354)(2362,1408)(2481,1458)(2599,1498)(2717,1518)(2835,1515)
\put(472,490){\makebox(0,0){$\times$}}
\put(590,544){\makebox(0,0){$\times$}}
\put(708,606){\makebox(0,0){$\times$}}
\put(826,663){\makebox(0,0){$\times$}}
\put(945,721){\makebox(0,0){$\times$}}
\put(1063,779){\makebox(0,0){$\times$}}
\put(1181,837){\makebox(0,0){$\times$}}
\put(1299,895){\makebox(0,0){$\times$}}
\put(1417,953){\makebox(0,0){$\times$}}
\put(1535,1011){\makebox(0,0){$\times$}}
\put(1654,1069){\makebox(0,0){$\times$}}
\put(1772,1127){\makebox(0,0){$\times$}}
\put(1890,1184){\makebox(0,0){$\times$}}
\put(2008,1242){\makebox(0,0){$\times$}}
\put(2126,1298){\makebox(0,0){$\times$}}
\put(2244,1354){\makebox(0,0){$\times$}}
\put(2362,1408){\makebox(0,0){$\times$}}
\put(2481,1458){\makebox(0,0){$\times$}}
\put(2599,1498){\makebox(0,0){$\times$}}
\put(2717,1518){\makebox(0,0){$\times$}}
\put(2835,1515){\makebox(0,0){$\times$}}
\put(2650,430){\makebox(0,0){$\times$}}
\color{black}
\put(2507,347){\makebox(0,0)[r]{$1^\text{st}$ order}}
\color[rgb]{0.34,0.71,0.91}
\Line(2548,347)(2753,347)
\polyline(472,1169)(472,1169)(590,1198)(708,1227)(826,1256)(945,1285)(1063,1314)(1181,1343)(1299,1372)(1417,1401)(1535,1430)(1654,1459)(1772,1487)(1890,1516)(2008,1544)(2126,1572)(2244,1599)(2362,1625)(2481,1649)(2599,1667)(2717,1671)(2835,1649)
\put(472,1169){\makebox(0,0){$\ast$}}
\put(590,1198){\makebox(0,0){$\ast$}}
\put(708,1227){\makebox(0,0){$\ast$}}
\put(826,1256){\makebox(0,0){$\ast$}}
\put(945,1285){\makebox(0,0){$\ast$}}
\put(1063,1314){\makebox(0,0){$\ast$}}
\put(1181,1343){\makebox(0,0){$\ast$}}
\put(1299,1372){\makebox(0,0){$\ast$}}
\put(1417,1401){\makebox(0,0){$\ast$}}
\put(1535,1430){\makebox(0,0){$\ast$}}
\put(1654,1459){\makebox(0,0){$\ast$}}
\put(1772,1487){\makebox(0,0){$\ast$}}
\put(1890,1516){\makebox(0,0){$\ast$}}
\put(2008,1544){\makebox(0,0){$\ast$}}
\put(2126,1572){\makebox(0,0){$\ast$}}
\put(2244,1599){\makebox(0,0){$\ast$}}
\put(2362,1625){\makebox(0,0){$\ast$}}
\put(2481,1649){\makebox(0,0){$\ast$}}
\put(2599,1667){\makebox(0,0){$\ast$}}
\put(2717,1671){\makebox(0,0){$\ast$}}
\put(2835,1649){\makebox(0,0){$\ast$}}
\put(2650,347){\makebox(0,0){$\ast$}}
\color{black}
\polygon(472,1716)(472,265)(2835,265)(2835,1716)
\put(72,990){\rotatebox{-270}{\makebox(0,0){homogenisation error}}}
\put(1653,58){\makebox(0,0){$\eps$}}
\end{picture}

%% file: testing.tex
In order to be convinced that the $\eps$-expansion described in this paper is accurate to third order (or to any other desired order) a numerical test is advisable, if not else because even in the presence of a totally correct theory, unseen programming bugs may infiltrate the computer program written to evaluate the considerable number of independent coefficients involved. The first ingredient of such a test is to identify a problem which can be solved numerically with relative ease for a sequence of values of $\eps$, and a global output of it that can be graphically compared to its series expansion. If the empirically determined residual of the expansion decays with the appropriate power of $\eps$ when $\eps\rightarrow 0$, we can be probabilistically reassured both that the theory is correct and that the computer program is bug-free, in the sense that, although not impossible, it is highly unlikely that this convergence rate be achieved by chance in the presence of a mistake. Or at least this is so if appropriate precautions are taken, in that the chosen test exercises all of the coefficients in the expansion with comparable weights. It is pretty evident that if the test were too simplistic and some coefficients were multiplied by zero, an error in those coefficients would go undetected. (We don't say this lightly: a couple of nearly invisible programming bugs were actually discovered using this procedure, and only so after the test design became careful enough.)

\subsection{Problem setup}
We are helped by two occurrences: since we have proven (in Appendix \ref{zeroproof}) that nonlinearities are irrelevant up to third order, a linear test of the 3D, unsteady Stokes problem is sufficient; and in a linear, space- and time-invariant, problem sinusoidal inputs produce sinusoidal solutions. Therefore a Stokes problem forced by complex exponential boundary conditions, with single but nontrivial frequency and wavenumbers, fits our description of a useful test.
The 3D, unsteady, dimensionless Stokes equations
\begin{subequations}
\begin{align}
u_t +p_x &= u_{xx}+u_{yy}+u_{zz} \\
v_t +p_y &= v_{xx}+v_{yy}+v_{zz} \\
w_t +p_z &= w_{xx}+w_{yy}+w_{zz} \\
u_x + v_y + w_z &= 0 
\end{align}\label{3DStokes}
\end{subequations}
between a virtual flat wall $z=0$ with the general boundary conditions \eqref{effectivebc}, and an outer flat boundary $z=z_\text{test}$ with complex-exponential excitation
\be
\begin{split}
u_z(x,y,z_\text{test},t)&=u_{z,\text{test}}e^{\im (\omega t - \alpha x -\beta y)},\quad v_z(x,y,z_\text{test},t)=v_{z,\text{test}}e^{\im (\omega t - \alpha x -\beta y)},\\
p(x,y,z_\text{test},t)&=p_{\text{test}}e^{\im (\omega t - \alpha x -\beta y)},
\end{split}
\ee{3DSouter}
reduce to the 1D problem:
\begin{subequations}
\begin{align}
u_{zz} &= (\im\omega + \alpha^2+\beta^2) u -\im\alpha p \\
v_{zz} &= (\im\omega + \alpha^2+\beta^2) v -\im\beta p \\
w_z &= \im\alpha u + \im\beta v\\
p_z &= w_{zz} - (\im\omega + \alpha^2+\beta^2) w 
\end{align}\label{h3DStokes}
\end{subequations}
with boundary conditions:
\be
\begin{split}
&\left[\begin{matrix}
u(0)\\
v(0)\\
w(0)
\end{matrix}\right] =  \
\eps \left[\begin{matrix}
h_1 u_z(0)\\
a_1 v_z(0)\\
0
\end{matrix}\right] + \
\eps^2 \left[\begin{matrix}
-\im\alpha h_2^{(v_{xz})} v_{z}(0)-\im\beta h_2^{(p_x)} p(0)\\
-\im\alpha f_2  u_{z}(0)-\im\beta b_2 p(0)\\
-\im\alpha g_2  u_{z}(0)+\im\beta c_2 v_{z}(0)
\end{matrix}\right]+{} \\
&\eps^3 \left[\begin{matrix}
[ (\im\omega-\alpha^2)h_3^{(u_{zt})}-\alpha^2 h_3^{(u_{xxz})}-\beta^2 h_3] u_{z}(0)-\alpha\beta h_3^{(v_{xyz})} v_{z}(0)-\alpha\beta h^{(p_{xy})}_3 p(0)\\
-\alpha\beta f_3^{(u_{xyz})} u_{z}(0)+[(\im\omega-\alpha^2)e_3^{(v_{zt})}-\beta^2 a_3-\alpha^2 f_3^{(v_{xxz})}]v_{z}(0)-[\alpha^2 f_3^{(p_{xx})}+\beta^2 b_3] p(0)\\
-\alpha\beta g_3^{(u_{xyz})} u_{z}(0)+[\beta^2 c_3 -\alpha^2 g_3^{(v_{xxz})}]v_{z}(0)+[\beta^2 d_3 -\alpha^2 g_3^{(p_{xx})}]p(0)
\end{matrix}\right]
\end{split}
\ee{harmeffectivebc}

\[
u_z(z_\text{test})=u_{z,\text{test}},\quad v_z(z_\text{test})=v_{z,\text{test}},\quad p(z_\text{test})=p_{\text{test}}.
\]

The same equations \eqref{3DStokes} with the same outer excitation \eqref{3DSouter}, in inner variables over the actual riblet geometry, admit a Floquet solution of the form
\be
\begin{split}
u=\eps U(Y,Z)e^{\im (\omega t - \alpha x -\beta y)},\quad &v=\eps V(Y,Z)e^{\im (\omega t - \alpha x -\beta y)},\\
 w=\eps W(Y,Z)e^{\im (\omega t - \alpha x -\beta y)},\quad &p=P(Y,Z)e^{\im (\omega t - \alpha x -\beta y)},\quad 
\end{split}
\ee{Floquetsol}
where $U(Y,Z),V(Y,Z),W(Y,Z),P(Y,Z)$, each a periodic function of $Y$ with period $1$, obey the 2D problem
\begin{subequations}
\begin{align}
U_{YY}+U_{ZZ}-2\eps\im\beta U_Y &= \eps^2(\im\omega + \alpha^2+\beta^2) U -\eps\im\alpha P \\
V_{YY}+V_{ZZ}-2\eps\im\beta V_Y &= \eps^2(\im\omega + \alpha^2+\beta^2) V -\eps\im\beta P \\
W_{YY}+W_{ZZ}-2\eps\im\beta W_Y &= \eps^2(\im\omega + \alpha^2+\beta^2) W + P_Z \\
W_Z &= \eps\im\alpha U + \eps\im\beta V
\end{align}\label{Floquet3DStokes}
\end{subequations}
(as can be verified by just inserting \eqref{Floquetsol} into \eqref{3DStokes}), with boundary conditions $U(Y,F(Y))=V(Y,F(Y))=W(Y,F(Y))=0, U_Z(Y,Z_\text{test})=u_{z,\text{test}}, V_Z(Y,Z_\text{test})=v_{z,\text{test}}, P(Y,Z_\text{test})=p_{\text{test}}$, and of course $\eps Z_\text{test}=z_\text{test}$.

The velocity vector $(U,V,W)$ measured at the location $Z=Z_\text{test}$ will exponentially tend to become independent of $Y$ for sufficiently large $Z_\text{test}$, and can therefore be identified with a constant (its mean value over $Y$, for definiteness). Both the exact problem \eqref{Floquet3DStokes} and its homogenized version \eqref{h3DStokes} will thus produce outer velocities $u_\text{out,ex.}=\eps U_\text{out,ex.},v_\text{out,ex.}=\eps V_\text{out,ex.},w_\text{out,ex.}=\eps W_\text{out,ex.}$ and $u_\text{out,hom.}=\eps U_\text{out,hom.},v_\text{out,hom.}=\eps V_\text{out,hom.},w_\text{out,hom.}=\eps W_\text{out,hom.}$ at the location $Z=Z_\text{test}$ in response to a given outer forcing $u_{z,\text{test}},v_{z,\text{test}},p_{\text{test}}$. The homogenization error, defined as the difference between the outer velocities produced by the two problems for the same forcing, must tend to zero faster than the third power of $\eps$ (in fact proportionally to $\eps^4$ if the expansion is a series of integer powers as hereto assumed) when $\eps\rightarrow 0$. Verifying this convergence rate numerically is the definition of our test.

A few observations are in order. The homogenized problem \eqref{h3DStokes}, when considering $\alpha,\beta,\omega$ as variables, can be seen as the Fourier transform of the original homogenized problem; in this sense the equivalent boundary conditions \eqref{harmeffectivebc} that we are testing are nothing else than the Fourier transform, or the frequency-wavenumber response-function representation, of the equivalent boundary conditions \eqref{effectivebc}. In this frequency-wavenumber response \eqref{harmeffectivebc}, wavenumbers and frequency $\alpha,\beta,\omega$ only appear in the combinations $\eps\alpha,\eps\beta,\eps^2\omega$ everywhere; this is not surprising if one thinks that in the dimensional version of the problem $\eps$ becomes replaced by the riblet period $s^*$ (the outer length $L$ cancels out), and the dimensionless forms of $\alpha,\beta,\omega$ are $\alpha^* s^*,\beta^* s^*,\omega^* {s^*}^2/\nu^*$. All this implies that the inner-outer $\eps$-expansion is at the same time a three-ways Taylor series of the frequency-wavenumber response in powers of $\alpha,\beta,\omega$. Numerically evaluating this Taylor series would be an alternate, independent way to calculate the same expansion coefficients (in fact, an extension to non-zero $\alpha,\beta,\omega$ of the wavenumber-domain approach considered in \S\ref{LaplaceWN}--\ref{pressurewd}--\ref{StokesWN}), but with advantages and disadvantages that we shall not presently pursue.

\subsection{Numerical test}
Rather than trying to get the expansion \eqref{harmeffectivebc} in full again, we shall be content with verifying it for a selected sextuple $(\alpha,\beta,\omega,u_{z,\text{test}},v_{z,\text{test}},p_{\text{test}})$, for instance one obtained from a random-number generator, and a suitable sequence of values of $\eps$. The computer code to do so is available at \url{https://cplcode.net/Applications/article-CPLcodes/horbcs/}, and will be described hereafter.
As we said before, the likelihood of obtaining the correct rate of convergence by chance is sufficiently low to provide confidence in such a test.

To solve the test problem exactly, we can adopt a numerical discretization of \eqref{Floquet3DStokes} very similar to the one used in \S\ref{numerical}, and the same immersed-boundary description, since we are dealing with the same two-dimensional geometry, although in complex variables. A seemingly unsurmountable obstacle, however, is that we need to evaluate fine differences, close to machine precision, using a second-order numerical scheme that has its own truncation error (proportional to the second power of the step size $\delta z$ rather than to some power of $\eps$), and to make this truncation error smaller than the differences we are trying to evaluate requires a non-viable step size. 

The way out of this conundrum is to make the numerical discretizations of the coefficient-extraction problem and of the test problem (both equations and boundary conditions of each) not just similar but identical in every detail, in such a way that the whole asymptotic-expansion procedure becomes applicable to the difference equations, and not just to the differential equations, involved. Once this is achieved, the measured difference between the numerical outer velocity and its homogenized version will converge cubically with $\eps$ irrespective of $\delta z$. Although the value itself of this velocity will not be accurate to the same level, observing the expected convergence rate of its homogenisation error will certify the numerical procedure, and by induction the differential equations and more generally the theory from which the procedure was derived (if the rationale is accepted that with very high probability any random mistakes would cascade down the chain and disrupt such a high convergence rate).

If we start from the one-dimensional problem \eqref{h3DStokes} again, we may look at it as an initial-value problem like we did when introducing the Cauchy expansion. With the six quantities $u(0),v(0),w(0),u_z(0),v_z(0),p(0)$ assumed to be simultaneously known at the wall, the solution of \eqref{h3DStokes} can be written in power-series form as
\begin{subequations}
\begin{align}
\begin{split}
  u(z)={}&u(0)+u_z(0)z+[-\im\alpha p(0)+(\alpha^2+\beta^2+\im\omega)u(0)]z^2/2+{}\\
  &\{-\im\alpha[(\im\alpha u_z(0)+\im\beta v_z(0))-(\alpha^2+\beta^2+\im \omega)w(0)]+(\alpha^2+\beta^2+\im\omega)u_z(0)\}z^3/6+\cdots
  \end{split}\label{Cauchy1Du}\\
  \begin{split}
  v(z)={}&v(0)+v_z(0)z+[-\im\beta p(0)+(\alpha^2+\beta^2+\im\omega)v(0)]z^2/2+{}\\
  &\{-\im\beta[(\im\alpha u_z(0)+\im\beta v_z(0))-(\alpha^2+\beta^2+\im\omega)w(0)]+(\alpha^2+\beta^2+\im\omega)v_z(0)\}z^3/6+\cdots
  \end{split}\label{Cauchy1Dv}\\
  \begin{split}
  w(z)={}&w(0)+(\im\alpha u(0)+\im\beta v(0))z+(\im\alpha u_z(0)+\im\beta v_z(0))z^2/2{}+\\
  &\{\im\alpha[-\im\alpha p(0)+(\alpha^2+\beta^2+\im\omega)u(0)]+\im\beta[-\im\beta p(0)+(\alpha^2+\beta^2+\im\omega) v(0)]\}z^3/6+\cdots 
  \end{split}\label{Cauchy1Dw}\\
  \begin{split}
  p(z)={}&p(0)+[(\im \alpha u_z(0)+\im\beta v_z(0))-(\alpha^2+\beta^2+\im\omega)w(0)] z+(\alpha^2+\beta^2) p(0) z^2/2+{}\\
  &(\alpha^2+\beta^2)[(\im\alpha u_z(0)+\im\beta v_z(0))-(\alpha^2+\beta^2+\im\omega)w(0)]z^3/6+\cdots
  \end{split}\label{Cauchy1Dp}
\end{align}\label{Cauchy1D}
\end{subequations}
where all $z$-derivatives higher than those present in the initial conditions have been obtained from the equations \eqref{h3DStokes}, differentiated zero or more times as needed.

The second-order, staggered discretization of \eqref{h3DStokes} is most efficiently written after defining the narrow, centered finite-difference operator $D$ such that $Df(z)=[f(z+\delta z/2)-f(z-\delta z/2)]/\delta z$ (where $f$ is a mute symbol standing for any one of the variables involved). Then \eqref{h3DStokes} are transformed into their (staggered) discretization by simply substituting $D$ for $d/dz$ everywhere:
\begin{subequations}
\begin{align}
DDu &= (\im\omega + \alpha^2+\beta^2) u -\im\alpha p \\
DDv &= (\im\omega + \alpha^2+\beta^2) v -\im\beta p \\
Dw &= \im\alpha u + \im\beta v\\
Dp &= DDw - (\im\omega + \alpha^2+\beta^2) w 
\end{align}\label{discrheqs}
\end{subequations}
As is known \citep[\textit{e.g.}][]{GrahamKnuth}, an entire discrete calculus can be built around difference operators like $D$, including the equivalent of a Taylor series. However, the primitives of simple powers in this calculus are not powers themselves but polynomials, and it is through these polynomials that the Taylor series is composed. For the present purposes we shall only need such primitive polynomials up to third order; the following identities are easily verified:
\begin{subequations}
\begin{align}
D 1 &= 0,\\
D z &= 1,\\
D(z^2/2) &= z,\\
D(z^3/6 - z\delta z^2/24) &= z^2/2.\label{z3corr}
\end{align}\label{primpoly}
\end{subequations}

The first three of \eqref{primpoly} are formally identical to their differential counterparts. Third order \eqref{z3corr} is where they start to differ. It follows that an expansion of the solution of the discrete equations similar to \eqref{Cauchy1D} can be built by simply substituting $Du(0)$ for $u_z(0)$, $Dv(0)$ for $v_z(0)$, and $z^3/6 - z\delta z^2/24$ for $z^3/6$ in it.

This is not yet a useful expansion, however, because $Du,Dv$ and $w$ are staggered (displaced by $\delta z/2$) with respect to $u,v,p$, and thus the corresponding initial conditions are not needed at $z=0$ but at $z=\delta z/2$. Moreover, $z=0$ itself (the vertical position of the riblet tip) may not coincide with a gridpoint. If we want to write a power expansion which obeys the difference equations \eqref{discrheqs}, but at the same time satisfies initial conditions at the same (unstaggered) location as the differential problem, we have to follow a hybrid approach.

We can start by observing that, in \eqref{Cauchy1Du}--\eqref{Cauchy1Dv}, $[-\im\alpha p(0)+(\alpha^2+\beta^2+\im\omega)u(0)]$ now represents $DDu$, rather than $u_{zz}$, and so are the following second and third derivatives replaced by corresponding differences. However, the difference between a second difference and a second derivative is proportional to the fourth derivative, and therefore comparable to terms that have already been neglected under ``$+\cdots$''. The first two of \eqref{Cauchy1D} therefore apply unchanged to the solution of the difference equations. The same considerations apply to second and third derivatives in \eqref{Cauchy1Dw}--\eqref{Cauchy1Dp}. The coefficient of $z$ in \eqref{Cauchy1Dw} $(\im\alpha u(0)+\im\beta v(0))$, however, now represents $Dw$ rather than $w_z$, and we have
\be
w_z = Dw - (\delta z^2/24)DDDw  +\cdots
\ee{wzexpansion}
(an expansion in powers of $\delta z^2$ that can be obtained by reversing \eqref{z3corr}, or else by Taylor-expanding the Fourier transforms of these operators). Inserting \eqref{wzexpansion}, and its analog for $p_z$, into \eqref{Cauchy1D} finally produces

\begin{subequations}
\begin{align}
\begin{split}
  u(z)={}&u(0)+u_z(0)z+[-\im\alpha p(0)+(\alpha^2+\beta^2+\im\omega)u(0)]z^2/2+{}\\
  &\{-\im\alpha[(\im\alpha u_z(0)+\im\beta v_z(0))-(\alpha^2+\beta^2+\im \omega)w(0)]+(\alpha^2+\beta^2+\im\omega)u_z(0)\}z^3/6+\cdots
  \end{split}\label{discrCauchy1Du}\\
  \begin{split}
  v(z)={}&v(0)+v_z(0)z+[-\im\beta p(0)+(\alpha^2+\beta^2+\im\omega)v(0)]z^2/2+{}\\
  &\{-\im\beta[(\im\alpha u_z(0)+\im\beta v_z(0))-(\alpha^2+\beta^2+\im\omega)w(0)]+(\alpha^2+\beta^2+\im\omega)v_z(0)\}z^3/6+\cdots
  \end{split}\label{discrCauchy1Dv}\\
  \begin{split}
  w(z)={}&w(0)+(\im\alpha u(0)+\im\beta v(0))z+(\im\alpha u_z(0)+\im\beta v_z(0))z^2/2+{}\\
  &\{(\alpha^2+\beta^2) p(0)+(\alpha^2+\beta^2+\im\omega)[\im\alpha u(0)+\im\beta v(0)]\}(z^3/6-z\delta z^2/24)+\cdots 
  \end{split}\label{discrCauchy1Dw}\\
  \begin{split}
  p(z)={}&p(0)+[(\im\alpha u_z(0)+\im\beta v_z(0))-(\alpha^2+\beta^2+\im\omega)w(0)]z+(\alpha^2+\beta^2)p(0)z^2/2+{}\\
  &(\alpha^2+\beta^2)\{[\im\alpha u_z(0)+\im\beta v_z(0)]-(\alpha^2+\beta^2+\im\omega)w(0)\}(z^3/6-z\delta z^2/24)+\cdots
  \end{split}\label{discrCauchy1Dp}
\end{align}\label{discrCauchy1D}
\end{subequations}
as the Cauchy expansion of the solution of the difference equations \eqref{discrheqs}. As can be seen, $z^3/6$ has been replaced by \eqref{z3corr} in the $w$ and $p$ expansions, whose first-order terms are determined by the difference equations themselves through \eqref{wzexpansion}, but not in the $u$ and $v$ expansions, whose first-order terms are determined by the initial conditions.

Armed with the Cauchy expansion \eqref{discrCauchy1D} of the solution of the difference equations, or with its equivalent in inner variables which is straightforwardly obtained by substituting $\eps Z$ for $z$ and $\eps \delta Z$ for $\delta z$, we can now proceed to describe the test algorithm. Key to streamlining the procedure is to solve the homogenised problem first and the (discrete) exact problem last.

\subsubsection{Stage 1: expansion\label{stage1}}
Choose a riblet geometry, step sizes $\delta Y,\delta Z$, and an upper boundary $Z_\text{test}$ substitutive of infinity ($Z_\text{test}=3$ is a reasonable value, owing to exponential convergence with it), and perform the calculation of the coefficients in Table \ref{tab:coefficients} as before. Only take care to a) keep them in memory to machine precision, even if such precision is unwarranted; b) in the terms whose Cauchy expansion of $W$ includes $Z^3/6$ ($\overline{W_{32}}$, producing $d_3$, and $\overline{W_{36}}$, producing $g_3^{(p_{xx})}$, fit this description), replace $Z^3/6$ by $Z^3/6 - Z\delta Z^2/24$; c) in all places where a boundary condition on either $U_Z$ or $V_Z$ is required at $Z=Z_\text{test}$, and numerically a boundary condition on  $DU$ or $DV$ is imposed at $Z=Z_\text{test}-\delta Z/2$, use a finite difference instead of a derivative in its \textit{r.h.s} as well, despite the \textit{r.h.s} being analytical. None of these modifications affects the O$(\delta Z^2)$ accuracy with which the procedure approximates the true expansion coefficients of the differential problem, because each modification is itself O$(\delta Z^2)$, but their combined effect is to generate the exact $\eps$-expansion of the discrete problem, large or small that $\delta Z$ may be.

\subsubsection{Stage 2: homogenised problem\label{stage2}}
The solution of the homogenised problem involves randomly choosing three boundary conditions for $U_Z(Z_\text{test})$, $V_Z(Z_\text{test})$, $P(Z_\text{test})$, and then solving a boundary-value problem for the difference equations \eqref{discrheqs} with these three conditions at the upper end, and \eqref{harmeffectivebc} as boundary conditions at the lower end. It should be evident that this is equivalent to choosing three random conditions for $U_Z(0),V_Z(0),P(0)$, and then solving an initial-value problem with these three conditions and $U(0),V(0),W(0)$ obtained from \eqref{harmeffectivebc}. An additional difficulty, however, is that the difference problem requires staggered initial conditions, and moreover the position of the virtual wall (chosen as the origin $Z=0$ here, but in fact coinciding with the tip of each riblet in the actual geometry) may not coincide with a gridpoint. In order to perform this fractional initial step, the series expansion \eqref{discrCauchy1D} comes handy again. Putting it all together, we a) randomly choose and fix $\alpha,\beta,\omega,U_Z(0),V_Z(0),P(0)$, and then, for each of a sequence of values of $\eps$ and the corresponding $\eps\alpha,\eps\beta$ and $\eps^2\omega$, we b) obtain $U(0),V(0),W(0)$ from the equivalent boundary conditions \eqref{harmeffectivebc}, c) use \eqref{discrCauchy1D} to move from $U(0),V(0),W(0),U_Z(0),V_Z(0),P(0)$ to $U,V,P$ at the nearest gridpoint and $DU,DV,W$ at the nearest staggered gridpoint, d) march the difference equations \eqref{discrheqs} forward in $Z$, until we get to $U_\text{out,hom.}=U(Z_\text{test}),V_\text{out,hom.}=V(Z_\text{test}),W_\text{out,hom.}=W(Z_\text{test}-\delta Z/2),DU(Z_\text{test}-\delta Z/2),DV(Z_\text{test}-\delta Z/2),P(Z_\text{test})$. The solution will be the same as we had solved a boundary-value problem with these just-found values of $DU(Z_\text{test}-\delta Z/2),DV(Z_\text{test}-\delta Z/2),P(Z_\text{test})$.

\subsubsection{Stage 3: ``exact'' discrete problem\label{stage3}}
As the last step, for each previously chosen value of $\eps$, numerically solve the 2D discretisation of \eqref{Floquet3DStokes} on the true riblet geometry (a banded linear system of algebraic equations, which can be directly inverted to machine precision), having care to use the exact same staggered grid and immersed-boundary representation that was used in Stage 1. Do so with an upper boundary condition that specifies $DU(Y,Z_\text{test}-\delta Z/2),DV(Y,Z_\text{test}-\delta Z/2),P(Y,Z_\text{test})$ to equal their values as gotten from Stage 2 (exactly as written, with finite differences and not derivatives). From the solution of this 2D problem isolate $U(Y,Z_\text{test}),V(Y,Z_\text{test}),W(Y,Z_\text{test}-\delta Z/2)$ and take their averages over $Y$ to define $U_\text{out,ex.}$, $V_\text{out,ex.}$ and $W_\text{out,ex.}$.

\subsection{Test results}
The main output of this procedure was already displayed in Figure \ref{testerr}, a plot of the absolute value of the difference between the complex outer velocity vectors produced by the homogenized problem in \S\ref{stage2} and by the exact discrete problem in \S\ref{stage3},
\[
\sqrt{\left|U_\text{out,hom.}-U_\text{out,ex.}\right|^2+\left|V_\text{out,hom.}-V_\text{out,ex.}\right|^2+\left|W_\text{out,hom.}-W_\text{out,ex.}\right|^2}
\]
Please be aware that the error plotted in this figure is calculated in inner variables, thus an error O($\eps^4$) upon $u=\eps U$ appears as an error O($\eps^3$) upon $U$. Since $U$ is numerically of unit order of magnitude, this is also the relative error. 

It can be visually verified that the slope of the plot in Figure \ref{testerr}, to a very good accuracy, is $3$ for the third-order equivalent boundary condition \eqref{effectivebc}, $2$ for the second-order equivalent boundary condition obtained by truncating \eqref{effectivebc} to second order, and $1$ for the classical first-order equivalent boundary condition containing the two original protrusion heights only. For completeness, we specify here that this plot was obtained for the sawtooth riblets as geometrically defined in \S\ref{numerical}, and for the parameters $U_Z(0)=0.480125-0.754587\im, V_Z(0)=0.268092-1.27046\im, P(0)=0.592799-0.367651\im, \alpha=0.835223, \beta=0.777775, \omega=1.26823$, on a $40\times 120$ mesh. We also note that $\left|\eps\beta\right|$ is limited by the Floquet expansion \eqref{Floquetsol} to $\left|\eps\beta\right|\le\upi$, therefore the range of $\eps\le 1$ considered in Figure \ref{testerr} could never be much wider to the right than it presently is. 

\begin{figure}
\centerline{\input{testerrny.tex}}
\caption{Homogenisation error of the equivalent boundary conditions for different discretizations, with $\delta Z = \delta Y = 1/N_Y$.}\label{testerrny}
\end{figure}
The resolution used in obtaining the error plot is much coarser than the one used for Table \ref{tab:coefficients} (a $240\times 720$ mesh). Nonetheless this plot is a fair representation of the convergence properties of the $\eps$-expansion, just because the test was designed to work on the discrete equations independently of step size. In order to illustrate such independence, Figure \ref{testerrny} shows similar error plots, all calculated at third order in $\eps$, but with varying number $N_Y$ of discretization points per period (and always $\delta Z = \delta Y = 1/N_Y$). That the slope of such curves turns out to be $3$ irrespective of discretization, all the way down to $5$ grid points per period where geometrical accuracy is laughable, should be fairly evident. It goes without saying that this theoretical slope is only achieved in a limited range of $\eps$, bounded on the right by the finite convergence radius of the $\eps$-expansion, and on the left by errors so small they fall below the machine's floating-point truncation error. It should also not be a surprise that the effect of floating-point error increases noticeably with complexity of the calculation (number of discretization points), so that being able to perform the test on a coarse, though maybe physically inaccurate, discretization actually improves its sensitivity.

\begin{figure}
\centerline{\input{wrongc3.tex}}
\caption{Homogenisation error of the equivalent boundary conditions in the presence of programming or formulation mistakes.}\label{wrongc3}
\end{figure}
Finally, in order to show that our test is indeed capable to detect errors in either the formulation or the programming, Figure \ref{wrongc3} shows what happens if a single coefficient ($c_3$ in this case, one of those that are only nonzero for asymmetric riblets) takes on a wrong value: the slope visibly mutates from $3$ to $2$ with decreasing $\eps$. This is an actual example of a mistake which was present in the code before the test was developed, and that the test helped us to uncover.

One can also see from Figure \ref{wrongc3} why we need to be able to push the test all the way to machine precision: not because this region of small $\eps$ has any considerable interest in applications, but because if discretization error were present, only the rightmost part of the plot would be accessible, and the above programming mistake might have gone undetected. To further illustrate this point, the third curve in Figure \ref{wrongc3} shows what happens if the correction \eqref{z3corr} for the difference between a discrete and a continuous Taylor series is \emph{not} applied to the $d_3$ \eqref{def:d3} and $g_3^{(p_{xx})}$ \eqref{def:g3} coefficients: third-order convergence is again destroyed, and so is its sensitivity to possible incorrectness.

%% file: testerrny.tex
\setlength{\unitlength}{0.120450pt}
\ifx\plotpoint\undefined\newsavebox{\plotpoint}\fi
\ifx\transparent\undefined%
    \providecommand{\gpopaque}{}%
    \providecommand{\gptransparent}[2]{\color{.!#2}}%
\else%
    \providecommand{\gpopaque}{\transparent{1.0}}%
    \providecommand{\gptransparent}[2]{\transparent{#1}}%
\fi%
\begin{picture}(3000,1800)(0,0)
\miterjoin\buttcap
\color{black}
\sbox{\plotpoint}{\rule[-0.200pt]{0.400pt}{0.400pt}}%
\linethickness{0.4pt}%
\multiput(472,265)(24.907,0.000){95}{\usebox{\plotpoint}}
\put(2835,265){\usebox{\plotpoint}}
\Line(472,265)(513,265)
\Line(2835,265)(2794,265)
\put(431,265){\makebox(0,0)[r]{$1\times10^{-12}$}}
\multiput(472,265)(24.907,0.000){95}{\usebox{\plotpoint}}
\put(2835,265){\usebox{\plotpoint}}
\Line(472,265)(492,265)
\Line(2835,265)(2815,265)
\multiput(472,410)(24.907,0.000){74}{\usebox{\plotpoint}}
\put(2302,410){\usebox{\plotpoint}}
\multiput(2794,410)(24.907,0.000){2}{\usebox{\plotpoint}}
\put(2835,410){\usebox{\plotpoint}}
\Line(472,410)(492,410)
\Line(2835,410)(2815,410)
\multiput(472,555)(24.907,0.000){74}{\usebox{\plotpoint}}
\put(2302,555){\usebox{\plotpoint}}
\multiput(2794,555)(24.907,0.000){2}{\usebox{\plotpoint}}
\put(2835,555){\usebox{\plotpoint}}
\Line(472,555)(513,555)
\Line(2835,555)(2794,555)
\put(431,555){\makebox(0,0)[r]{$1\times10^{-10}$}}
\multiput(472,555)(24.907,0.000){74}{\usebox{\plotpoint}}
\put(2302,555){\usebox{\plotpoint}}
\multiput(2794,555)(24.907,0.000){2}{\usebox{\plotpoint}}
\put(2835,555){\usebox{\plotpoint}}
\Line(472,555)(492,555)
\Line(2835,555)(2815,555)
\multiput(472,700)(24.907,0.000){74}{\usebox{\plotpoint}}
\put(2302,700){\usebox{\plotpoint}}
\multiput(2794,700)(24.907,0.000){2}{\usebox{\plotpoint}}
\put(2835,700){\usebox{\plotpoint}}
\Line(472,700)(492,700)
\Line(2835,700)(2815,700)
\multiput(472,845)(24.907,0.000){95}{\usebox{\plotpoint}}
\put(2835,845){\usebox{\plotpoint}}
\Line(472,845)(513,845)
\Line(2835,845)(2794,845)
\put(431,845){\makebox(0,0)[r]{$1\times10^{-8}$}}
\multiput(472,845)(24.907,0.000){95}{\usebox{\plotpoint}}
\put(2835,845){\usebox{\plotpoint}}
\Line(472,845)(492,845)
\Line(2835,845)(2815,845)
\multiput(472,991)(24.907,0.000){95}{\usebox{\plotpoint}}
\put(2835,991){\usebox{\plotpoint}}
\Line(472,991)(492,991)
\Line(2835,991)(2815,991)
\multiput(472,1136)(24.907,0.000){95}{\usebox{\plotpoint}}
\put(2835,1136){\usebox{\plotpoint}}
\Line(472,1136)(513,1136)
\Line(2835,1136)(2794,1136)
\put(431,1136){\makebox(0,0)[r]{$1\times10^{-6}$}}
\multiput(472,1136)(24.907,0.000){95}{\usebox{\plotpoint}}
\put(2835,1136){\usebox{\plotpoint}}
\Line(472,1136)(492,1136)
\Line(2835,1136)(2815,1136)
\multiput(472,1281)(24.907,0.000){95}{\usebox{\plotpoint}}
\put(2835,1281){\usebox{\plotpoint}}
\Line(472,1281)(492,1281)
\Line(2835,1281)(2815,1281)
\multiput(472,1426)(24.907,0.000){95}{\usebox{\plotpoint}}
\put(2835,1426){\usebox{\plotpoint}}
\Line(472,1426)(513,1426)
\Line(2835,1426)(2794,1426)
\put(431,1426){\makebox(0,0)[r]{$0.0001$}}
\multiput(472,1426)(24.907,0.000){95}{\usebox{\plotpoint}}
\put(2835,1426){\usebox{\plotpoint}}
\Line(472,1426)(492,1426)
\Line(2835,1426)(2815,1426)
\multiput(472,1571)(24.907,0.000){95}{\usebox{\plotpoint}}
\put(2835,1571){\usebox{\plotpoint}}
\Line(472,1571)(492,1571)
\Line(2835,1571)(2815,1571)
\multiput(472,1716)(24.907,0.000){95}{\usebox{\plotpoint}}
\put(2835,1716){\usebox{\plotpoint}}
\Line(472,1716)(513,1716)
\Line(2835,1716)(2794,1716)
\put(431,1716){\makebox(0,0)[r]{$0.01$}}
\multiput(472,1716)(24.907,0.000){95}{\usebox{\plotpoint}}
\put(2835,1716){\usebox{\plotpoint}}
\Line(472,1716)(492,1716)
\Line(2835,1716)(2815,1716)
\multiput(472,265)(0.000,24.907){59}{\usebox{\plotpoint}}
\put(472,1716){\usebox{\plotpoint}}
\Line(472,265)(472,306)
\Line(472,1716)(472,1675)
\put(472,182){\makebox(0,0){$0.0001$}}
\Line(650,265)(650,285)
\Line(650,1716)(650,1696)
\Line(754,265)(754,285)
\Line(754,1716)(754,1696)
\Line(828,265)(828,285)
\Line(828,1716)(828,1696)
\Line(885,265)(885,285)
\Line(885,1716)(885,1696)
\Line(932,265)(932,285)
\Line(932,1716)(932,1696)
\Line(971,265)(971,285)
\Line(971,1716)(971,1696)
\Line(1006,265)(1006,285)
\Line(1006,1716)(1006,1696)
\Line(1036,265)(1036,285)
\Line(1036,1716)(1036,1696)
\multiput(1063,265)(0.000,24.907){59}{\usebox{\plotpoint}}
\put(1063,1716){\usebox{\plotpoint}}
\Line(1063,265)(1063,306)
\Line(1063,1716)(1063,1675)
\put(1063,182){\makebox(0,0){$0.001$}}
\Line(1241,265)(1241,285)
\Line(1241,1716)(1241,1696)
\Line(1345,265)(1345,285)
\Line(1345,1716)(1345,1696)
\Line(1418,265)(1418,285)
\Line(1418,1716)(1418,1696)
\Line(1476,265)(1476,285)
\Line(1476,1716)(1476,1696)
\Line(1522,265)(1522,285)
\Line(1522,1716)(1522,1696)
\Line(1562,265)(1562,285)
\Line(1562,1716)(1562,1696)
\Line(1596,265)(1596,285)
\Line(1596,1716)(1596,1696)
\Line(1626,265)(1626,285)
\Line(1626,1716)(1626,1696)
\multiput(1654,265)(0.000,24.907){59}{\usebox{\plotpoint}}
\put(1654,1716){\usebox{\plotpoint}}
\Line(1654,265)(1654,306)
\Line(1654,1716)(1654,1675)
\put(1654,182){\makebox(0,0){$0.01$}}
\Line(1831,265)(1831,285)
\Line(1831,1716)(1831,1696)
\Line(1935,265)(1935,285)
\Line(1935,1716)(1935,1696)
\Line(2009,265)(2009,285)
\Line(2009,1716)(2009,1696)
\Line(2066,265)(2066,285)
\Line(2066,1716)(2066,1696)
\Line(2113,265)(2113,285)
\Line(2113,1716)(2113,1696)
\Line(2153,265)(2153,285)
\Line(2153,1716)(2153,1696)
\Line(2187,265)(2187,285)
\Line(2187,1716)(2187,1696)
\Line(2217,265)(2217,285)
\Line(2217,1716)(2217,1696)
\multiput(2244,265)(0.000,24.907){59}{\usebox{\plotpoint}}
\put(2244,1716){\usebox{\plotpoint}}
\Line(2244,265)(2244,306)
\Line(2244,1716)(2244,1675)
\put(2244,182){\makebox(0,0){$0.1$}}
\Line(2422,265)(2422,285)
\Line(2422,1716)(2422,1696)
\Line(2526,265)(2526,285)
\Line(2526,1716)(2526,1696)
\Line(2600,265)(2600,285)
\Line(2600,1716)(2600,1696)
\Line(2657,265)(2657,285)
\Line(2657,1716)(2657,1696)
\Line(2704,265)(2704,285)
\Line(2704,1716)(2704,1696)
\Line(2743,265)(2743,285)
\Line(2743,1716)(2743,1696)
\Line(2778,265)(2778,285)
\Line(2778,1716)(2778,1696)
\Line(2808,265)(2808,285)
\Line(2808,1716)(2808,1696)
\multiput(2835,265)(0.000,24.907){59}{\usebox{\plotpoint}}
\put(2835,1716){\usebox{\plotpoint}}
\Line(2835,265)(2835,306)
\Line(2835,1716)(2835,1675)
\put(2835,182){\makebox(0,0){$1$}}
\polygon(472,1716)(472,265)(2835,265)(2835,1716)
\put(2507,679){\makebox(0,0)[r]{$N_Y=80$}}
\color[rgb]{0.58,0.00,0.83}
\Line(2548,679)(2753,679)
\polyline(472,604)(472,604)(590,596)(708,548)(826,523)(945,492)(1063,490)(1181,456)(1299,423)(1417,486)(1535,571)(1654,661)(1772,748)(1890,834)(2008,921)(2126,1007)(2244,1093)(2362,1177)(2481,1260)(2599,1336)(2717,1394)(2835,1427)
\put(472,604){\makebox(0,0){$+$}}
\put(590,596){\makebox(0,0){$+$}}
\put(708,548){\makebox(0,0){$+$}}
\put(826,523){\makebox(0,0){$+$}}
\put(945,492){\makebox(0,0){$+$}}
\put(1063,490){\makebox(0,0){$+$}}
\put(1181,456){\makebox(0,0){$+$}}
\put(1299,423){\makebox(0,0){$+$}}
\put(1417,486){\makebox(0,0){$+$}}
\put(1535,571){\makebox(0,0){$+$}}
\put(1654,661){\makebox(0,0){$+$}}
\put(1772,748){\makebox(0,0){$+$}}
\put(1890,834){\makebox(0,0){$+$}}
\put(2008,921){\makebox(0,0){$+$}}
\put(2126,1007){\makebox(0,0){$+$}}
\put(2244,1093){\makebox(0,0){$+$}}
\put(2362,1177){\makebox(0,0){$+$}}
\put(2481,1260){\makebox(0,0){$+$}}
\put(2599,1336){\makebox(0,0){$+$}}
\put(2717,1394){\makebox(0,0){$+$}}
\put(2835,1427){\makebox(0,0){$+$}}
\put(2650,679){\makebox(0,0){$+$}}
\color{black}
\put(2507,596){\makebox(0,0)[r]{$N_Y=40$}}
\color[rgb]{0.00,0.62,0.45}
\Line(2548,596)(2753,596)
\polyline(472,379)(472,379)(590,384)(708,312)
\Line(708,312)(786,265)
\polyline(844,265)(945,394)(1063,352)(1181,330)(1299,419)(1417,511)(1535,597)(1654,685)(1772,772)(1890,859)(2008,945)(2126,1032)(2244,1117)(2362,1202)(2481,1284)(2599,1360)(2717,1420)(2835,1458)
\put(472,379){\makebox(0,0){$\times$}}
\put(590,384){\makebox(0,0){$\times$}}
\put(708,312){\makebox(0,0){$\times$}}
\put(945,394){\makebox(0,0){$\times$}}
\put(1063,352){\makebox(0,0){$\times$}}
\put(1181,330){\makebox(0,0){$\times$}}
\put(1299,419){\makebox(0,0){$\times$}}
\put(1417,511){\makebox(0,0){$\times$}}
\put(1535,597){\makebox(0,0){$\times$}}
\put(1654,685){\makebox(0,0){$\times$}}
\put(1772,772){\makebox(0,0){$\times$}}
\put(1890,859){\makebox(0,0){$\times$}}
\put(2008,945){\makebox(0,0){$\times$}}
\put(2126,1032){\makebox(0,0){$\times$}}
\put(2244,1117){\makebox(0,0){$\times$}}
\put(2362,1202){\makebox(0,0){$\times$}}
\put(2481,1284){\makebox(0,0){$\times$}}
\put(2599,1360){\makebox(0,0){$\times$}}
\put(2717,1420){\makebox(0,0){$\times$}}
\put(2835,1458){\makebox(0,0){$\times$}}
\put(2650,596){\makebox(0,0){$\times$}}
\color{black}
\put(2507,513){\makebox(0,0)[r]{$N_Y=20$}}
\color[rgb]{0.34,0.71,0.91}
\Line(2548,513)(2753,513)
\Line(562,265)(590,278)
\Line(590,278)(685,265)
\polyline(894,265)(945,308)(1063,308)(1181,398)(1299,482)(1417,570)(1535,657)(1654,744)(1772,830)(1890,917)(2008,1004)(2126,1090)(2244,1176)(2362,1260)(2481,1341)(2599,1417)(2717,1480)(2835,1524)
\put(590,278){\makebox(0,0){$\ast$}}
\put(945,308){\makebox(0,0){$\ast$}}
\put(1063,308){\makebox(0,0){$\ast$}}
\put(1181,398){\makebox(0,0){$\ast$}}
\put(1299,482){\makebox(0,0){$\ast$}}
\put(1417,570){\makebox(0,0){$\ast$}}
\put(1535,657){\makebox(0,0){$\ast$}}
\put(1654,744){\makebox(0,0){$\ast$}}
\put(1772,830){\makebox(0,0){$\ast$}}
\put(1890,917){\makebox(0,0){$\ast$}}
\put(2008,1004){\makebox(0,0){$\ast$}}
\put(2126,1090){\makebox(0,0){$\ast$}}
\put(2244,1176){\makebox(0,0){$\ast$}}
\put(2362,1260){\makebox(0,0){$\ast$}}
\put(2481,1341){\makebox(0,0){$\ast$}}
\put(2599,1417){\makebox(0,0){$\ast$}}
\put(2717,1480){\makebox(0,0){$\ast$}}
\put(2835,1524){\makebox(0,0){$\ast$}}
\put(2650,513){\makebox(0,0){$\ast$}}
\color{black}
\put(2507,430){\makebox(0,0)[r]{$N_Y=10$}}
\color[rgb]{0.90,0.62,0.00}
\Line(2548,430)(2753,430)
\polyline(892,265)(945,304)(1063,390)(1181,477)(1299,564)(1417,651)(1535,738)(1654,825)(1772,912)(1890,999)(2008,1085)(2126,1172)(2244,1257)(2362,1342)(2481,1423)(2599,1500)(2717,1565)(2835,1610)
\put(945,304){\raisebox{-.8pt}{\makebox(0,0){$\Box$}}}
\put(1063,390){\raisebox{-.8pt}{\makebox(0,0){$\Box$}}}
\put(1181,477){\raisebox{-.8pt}{\makebox(0,0){$\Box$}}}
\put(1299,564){\raisebox{-.8pt}{\makebox(0,0){$\Box$}}}
\put(1417,651){\raisebox{-.8pt}{\makebox(0,0){$\Box$}}}
\put(1535,738){\raisebox{-.8pt}{\makebox(0,0){$\Box$}}}
\put(1654,825){\raisebox{-.8pt}{\makebox(0,0){$\Box$}}}
\put(1772,912){\raisebox{-.8pt}{\makebox(0,0){$\Box$}}}
\put(1890,999){\raisebox{-.8pt}{\makebox(0,0){$\Box$}}}
\put(2008,1085){\raisebox{-.8pt}{\makebox(0,0){$\Box$}}}
\put(2126,1172){\raisebox{-.8pt}{\makebox(0,0){$\Box$}}}
\put(2244,1257){\raisebox{-.8pt}{\makebox(0,0){$\Box$}}}
\put(2362,1342){\raisebox{-.8pt}{\makebox(0,0){$\Box$}}}
\put(2481,1423){\raisebox{-.8pt}{\makebox(0,0){$\Box$}}}
\put(2599,1500){\raisebox{-.8pt}{\makebox(0,0){$\Box$}}}
\put(2717,1565){\raisebox{-.8pt}{\makebox(0,0){$\Box$}}}
\put(2835,1610){\raisebox{-.8pt}{\makebox(0,0){$\Box$}}}
\put(2650,430){\raisebox{-.8pt}{\makebox(0,0){$\Box$}}}
\color{black}
\put(2507,347){\makebox(0,0)[r]{$N_Y=5$}}
\color[rgb]{0.94,0.89,0.26}
\Line(2548,347)(2753,347)
\polyline(769,265)(826,306)(945,392)(1063,480)(1181,567)(1299,654)(1417,741)(1535,828)(1654,915)(1772,1002)(1890,1088)(2008,1175)(2126,1261)(2244,1347)(2362,1432)(2481,1514)(2599,1592)(2717,1659)(2835,1707)
\put(826,306){\makebox(0,0){$\blacksquare$}}
\put(945,392){\makebox(0,0){$\blacksquare$}}
\put(1063,480){\makebox(0,0){$\blacksquare$}}
\put(1181,567){\makebox(0,0){$\blacksquare$}}
\put(1299,654){\makebox(0,0){$\blacksquare$}}
\put(1417,741){\makebox(0,0){$\blacksquare$}}
\put(1535,828){\makebox(0,0){$\blacksquare$}}
\put(1654,915){\makebox(0,0){$\blacksquare$}}
\put(1772,1002){\makebox(0,0){$\blacksquare$}}
\put(1890,1088){\makebox(0,0){$\blacksquare$}}
\put(2008,1175){\makebox(0,0){$\blacksquare$}}
\put(2126,1261){\makebox(0,0){$\blacksquare$}}
\put(2244,1347){\makebox(0,0){$\blacksquare$}}
\put(2362,1432){\makebox(0,0){$\blacksquare$}}
\put(2481,1514){\makebox(0,0){$\blacksquare$}}
\put(2599,1592){\makebox(0,0){$\blacksquare$}}
\put(2717,1659){\makebox(0,0){$\blacksquare$}}
\put(2835,1707){\makebox(0,0){$\blacksquare$}}
\put(2650,347){\makebox(0,0){$\blacksquare$}}
\color{black}
\polygon(472,1716)(472,265)(2835,265)(2835,1716)
\put(72,990){\rotatebox{-270}{\makebox(0,0){homogenisation error}}}
\put(1653,58){\makebox(0,0){$\eps$}}
\end{picture}

%% file: wrongc3.tex
\setlength{\unitlength}{0.120450pt}
\ifx\plotpoint\undefined\newsavebox{\plotpoint}\fi
\ifx\transparent\undefined%
    \providecommand{\gpopaque}{}%
    \providecommand{\gptransparent}[2]{\color{.!#2}}%
\else%
    \providecommand{\gpopaque}{\transparent{1.0}}%
    \providecommand{\gptransparent}[2]{\transparent{#1}}%
\fi%
\begin{picture}(3000,1800)(0,0)
\miterjoin\buttcap
\color{black}
\sbox{\plotpoint}{\rule[-0.200pt]{0.400pt}{0.400pt}}%
\linethickness{0.4pt}%
\multiput(472,265)(24.907,0.000){95}{\usebox{\plotpoint}}
\put(2835,265){\usebox{\plotpoint}}
\Line(472,265)(513,265)
\Line(2835,265)(2794,265)
\put(431,265){\makebox(0,0)[r]{$1\times10^{-12}$}}
\multiput(472,265)(24.907,0.000){95}{\usebox{\plotpoint}}
\put(2835,265){\usebox{\plotpoint}}
\Line(472,265)(492,265)
\Line(2835,265)(2815,265)
\multiput(472,410)(24.907,0.000){53}{\usebox{\plotpoint}}
\put(1769,410){\usebox{\plotpoint}}
\multiput(2794,410)(24.907,0.000){2}{\usebox{\plotpoint}}
\put(2835,410){\usebox{\plotpoint}}
\Line(472,410)(492,410)
\Line(2835,410)(2815,410)
\multiput(472,555)(24.907,0.000){95}{\usebox{\plotpoint}}
\put(2835,555){\usebox{\plotpoint}}
\Line(472,555)(513,555)
\Line(2835,555)(2794,555)
\put(431,555){\makebox(0,0)[r]{$1\times10^{-10}$}}
\multiput(472,555)(24.907,0.000){95}{\usebox{\plotpoint}}
\put(2835,555){\usebox{\plotpoint}}
\Line(472,555)(492,555)
\Line(2835,555)(2815,555)
\multiput(472,700)(24.907,0.000){95}{\usebox{\plotpoint}}
\put(2835,700){\usebox{\plotpoint}}
\Line(472,700)(492,700)
\Line(2835,700)(2815,700)
\multiput(472,845)(24.907,0.000){95}{\usebox{\plotpoint}}
\put(2835,845){\usebox{\plotpoint}}
\Line(472,845)(513,845)
\Line(2835,845)(2794,845)
\put(431,845){\makebox(0,0)[r]{$1\times10^{-8}$}}
\multiput(472,845)(24.907,0.000){95}{\usebox{\plotpoint}}
\put(2835,845){\usebox{\plotpoint}}
\Line(472,845)(492,845)
\Line(2835,845)(2815,845)
\multiput(472,991)(24.907,0.000){95}{\usebox{\plotpoint}}
\put(2835,991){\usebox{\plotpoint}}
\Line(472,991)(492,991)
\Line(2835,991)(2815,991)
\multiput(472,1136)(24.907,0.000){95}{\usebox{\plotpoint}}
\put(2835,1136){\usebox{\plotpoint}}
\Line(472,1136)(513,1136)
\Line(2835,1136)(2794,1136)
\put(431,1136){\makebox(0,0)[r]{$1\times10^{-6}$}}
\multiput(472,1136)(24.907,0.000){95}{\usebox{\plotpoint}}
\put(2835,1136){\usebox{\plotpoint}}
\Line(472,1136)(492,1136)
\Line(2835,1136)(2815,1136)
\multiput(472,1281)(24.907,0.000){95}{\usebox{\plotpoint}}
\put(2835,1281){\usebox{\plotpoint}}
\Line(472,1281)(492,1281)
\Line(2835,1281)(2815,1281)
\multiput(472,1426)(24.907,0.000){95}{\usebox{\plotpoint}}
\put(2835,1426){\usebox{\plotpoint}}
\Line(472,1426)(513,1426)
\Line(2835,1426)(2794,1426)
\put(431,1426){\makebox(0,0)[r]{$0.0001$}}
\multiput(472,1426)(24.907,0.000){95}{\usebox{\plotpoint}}
\put(2835,1426){\usebox{\plotpoint}}
\Line(472,1426)(492,1426)
\Line(2835,1426)(2815,1426)
\multiput(472,1571)(24.907,0.000){95}{\usebox{\plotpoint}}
\put(2835,1571){\usebox{\plotpoint}}
\Line(472,1571)(492,1571)
\Line(2835,1571)(2815,1571)
\multiput(472,1716)(24.907,0.000){95}{\usebox{\plotpoint}}
\put(2835,1716){\usebox{\plotpoint}}
\Line(472,1716)(513,1716)
\Line(2835,1716)(2794,1716)
\put(431,1716){\makebox(0,0)[r]{$0.01$}}
\multiput(472,1716)(24.907,0.000){95}{\usebox{\plotpoint}}
\put(2835,1716){\usebox{\plotpoint}}
\Line(472,1716)(492,1716)
\Line(2835,1716)(2815,1716)
\multiput(472,265)(0.000,24.907){59}{\usebox{\plotpoint}}
\put(472,1716){\usebox{\plotpoint}}
\Line(472,265)(472,306)
\Line(472,1716)(472,1675)
\put(472,182){\makebox(0,0){$0.0001$}}
\Line(650,265)(650,285)
\Line(650,1716)(650,1696)
\Line(754,265)(754,285)
\Line(754,1716)(754,1696)
\Line(828,265)(828,285)
\Line(828,1716)(828,1696)
\Line(885,265)(885,285)
\Line(885,1716)(885,1696)
\Line(932,265)(932,285)
\Line(932,1716)(932,1696)
\Line(971,265)(971,285)
\Line(971,1716)(971,1696)
\Line(1006,265)(1006,285)
\Line(1006,1716)(1006,1696)
\Line(1036,265)(1036,285)
\Line(1036,1716)(1036,1696)
\multiput(1063,265)(0.000,24.907){59}{\usebox{\plotpoint}}
\put(1063,1716){\usebox{\plotpoint}}
\Line(1063,265)(1063,306)
\Line(1063,1716)(1063,1675)
\put(1063,182){\makebox(0,0){$0.001$}}
\Line(1241,265)(1241,285)
\Line(1241,1716)(1241,1696)
\Line(1345,265)(1345,285)
\Line(1345,1716)(1345,1696)
\Line(1418,265)(1418,285)
\Line(1418,1716)(1418,1696)
\Line(1476,265)(1476,285)
\Line(1476,1716)(1476,1696)
\Line(1522,265)(1522,285)
\Line(1522,1716)(1522,1696)
\Line(1562,265)(1562,285)
\Line(1562,1716)(1562,1696)
\Line(1596,265)(1596,285)
\Line(1596,1716)(1596,1696)
\Line(1626,265)(1626,285)
\Line(1626,1716)(1626,1696)
\multiput(1654,265)(0.000,24.907){59}{\usebox{\plotpoint}}
\put(1654,1716){\usebox{\plotpoint}}
\Line(1654,265)(1654,306)
\Line(1654,1716)(1654,1675)
\put(1654,182){\makebox(0,0){$0.01$}}
\Line(1831,265)(1831,285)
\Line(1831,1716)(1831,1696)
\Line(1935,265)(1935,285)
\Line(1935,1716)(1935,1696)
\Line(2009,265)(2009,285)
\Line(2009,1716)(2009,1696)
\Line(2066,265)(2066,285)
\Line(2066,1716)(2066,1696)
\Line(2113,265)(2113,285)
\Line(2113,1716)(2113,1696)
\Line(2153,265)(2153,285)
\Line(2153,1716)(2153,1696)
\Line(2187,265)(2187,285)
\Line(2187,1716)(2187,1696)
\Line(2217,265)(2217,285)
\Line(2217,1716)(2217,1696)
\multiput(2244,265)(0.000,24.907){2}{\usebox{\plotpoint}}
\put(2244,306){\usebox{\plotpoint}}
\multiput(2244,555)(0.000,24.907){47}{\usebox{\plotpoint}}
\put(2244,1716){\usebox{\plotpoint}}
\Line(2244,265)(2244,306)
\Line(2244,1716)(2244,1675)
\put(2244,182){\makebox(0,0){$0.1$}}
\Line(2422,265)(2422,285)
\Line(2422,1716)(2422,1696)
\Line(2526,265)(2526,285)
\Line(2526,1716)(2526,1696)
\Line(2600,265)(2600,285)
\Line(2600,1716)(2600,1696)
\Line(2657,265)(2657,285)
\Line(2657,1716)(2657,1696)
\Line(2704,265)(2704,285)
\Line(2704,1716)(2704,1696)
\Line(2743,265)(2743,285)
\Line(2743,1716)(2743,1696)
\Line(2778,265)(2778,285)
\Line(2778,1716)(2778,1696)
\Line(2808,265)(2808,285)
\Line(2808,1716)(2808,1696)
\multiput(2835,265)(0.000,24.907){59}{\usebox{\plotpoint}}
\put(2835,1716){\usebox{\plotpoint}}
\Line(2835,265)(2835,306)
\Line(2835,1716)(2835,1675)
\put(2835,182){\makebox(0,0){$1$}}
\polygon(472,1716)(472,265)(2835,265)(2835,1716)
\put(2507,513){\makebox(0,0)[r]{correct $3^\text{rd}$ order}}
\color[rgb]{0.58,0.00,0.83}
\Line(2548,513)(2753,513)
\polyline(472,379)(472,379)(590,384)(708,312)
\Line(708,312)(786,265)
\polyline(844,265)(945,394)(1063,352)(1181,330)(1299,419)(1417,511)(1535,597)(1654,685)(1772,772)(1890,859)(2008,945)(2126,1032)(2244,1117)(2362,1202)(2481,1284)(2599,1360)(2717,1420)(2835,1458)
\put(472,379){\makebox(0,0){$+$}}
\put(590,384){\makebox(0,0){$+$}}
\put(708,312){\makebox(0,0){$+$}}
\put(945,394){\makebox(0,0){$+$}}
\put(1063,352){\makebox(0,0){$+$}}
\put(1181,330){\makebox(0,0){$+$}}
\put(1299,419){\makebox(0,0){$+$}}
\put(1417,511){\makebox(0,0){$+$}}
\put(1535,597){\makebox(0,0){$+$}}
\put(1654,685){\makebox(0,0){$+$}}
\put(1772,772){\makebox(0,0){$+$}}
\put(1890,859){\makebox(0,0){$+$}}
\put(2008,945){\makebox(0,0){$+$}}
\put(2126,1032){\makebox(0,0){$+$}}
\put(2244,1117){\makebox(0,0){$+$}}
\put(2362,1202){\makebox(0,0){$+$}}
\put(2481,1284){\makebox(0,0){$+$}}
\put(2599,1360){\makebox(0,0){$+$}}
\put(2717,1420){\makebox(0,0){$+$}}
\put(2835,1458){\makebox(0,0){$+$}}
\put(2650,513){\makebox(0,0){$+$}}
\color{black}
\put(2507,430){\makebox(0,0)[r]{with wrong $c_3$}}
\color[rgb]{0.00,0.62,0.45}
\Line(2548,430)(2753,430)
\polyline(472,383)(472,383)(590,399)(708,427)(826,484)(945,543)(1063,601)(1181,659)(1299,717)(1417,775)(1535,834)(1654,892)(1772,951)(1890,1011)(2008,1071)(2126,1133)(2244,1196)(2362,1261)(2481,1327)(2599,1390)(2717,1441)(2835,1471)
\put(472,383){\makebox(0,0){$\times$}}
\put(590,399){\makebox(0,0){$\times$}}
\put(708,427){\makebox(0,0){$\times$}}
\put(826,484){\makebox(0,0){$\times$}}
\put(945,543){\makebox(0,0){$\times$}}
\put(1063,601){\makebox(0,0){$\times$}}
\put(1181,659){\makebox(0,0){$\times$}}
\put(1299,717){\makebox(0,0){$\times$}}
\put(1417,775){\makebox(0,0){$\times$}}
\put(1535,834){\makebox(0,0){$\times$}}
\put(1654,892){\makebox(0,0){$\times$}}
\put(1772,951){\makebox(0,0){$\times$}}
\put(1890,1011){\makebox(0,0){$\times$}}
\put(2008,1071){\makebox(0,0){$\times$}}
\put(2126,1133){\makebox(0,0){$\times$}}
\put(2244,1196){\makebox(0,0){$\times$}}
\put(2362,1261){\makebox(0,0){$\times$}}
\put(2481,1327){\makebox(0,0){$\times$}}
\put(2599,1390){\makebox(0,0){$\times$}}
\put(2717,1441){\makebox(0,0){$\times$}}
\put(2835,1471){\makebox(0,0){$\times$}}
\put(2650,430){\makebox(0,0){$\times$}}
\color{black}
\put(2507,347){\makebox(0,0)[r]{without \eqref{z3corr}}}
\color[rgb]{0.34,0.71,0.91}
\Line(2548,347)(2753,347)
\polyline(472,380)(472,380)(590,386)(708,357)(826,406)(945,467)(1063,522)(1181,580)(1299,638)(1417,696)(1535,755)(1654,813)(1772,873)(1890,934)(2008,997)(2126,1065)(2244,1138)(2362,1215)(2481,1294)(2599,1367)(2717,1426)(2835,1461)
\put(472,380){\makebox(0,0){$\ast$}}
\put(590,386){\makebox(0,0){$\ast$}}
\put(708,357){\makebox(0,0){$\ast$}}
\put(826,406){\makebox(0,0){$\ast$}}
\put(945,467){\makebox(0,0){$\ast$}}
\put(1063,522){\makebox(0,0){$\ast$}}
\put(1181,580){\makebox(0,0){$\ast$}}
\put(1299,638){\makebox(0,0){$\ast$}}
\put(1417,696){\makebox(0,0){$\ast$}}
\put(1535,755){\makebox(0,0){$\ast$}}
\put(1654,813){\makebox(0,0){$\ast$}}
\put(1772,873){\makebox(0,0){$\ast$}}
\put(1890,934){\makebox(0,0){$\ast$}}
\put(2008,997){\makebox(0,0){$\ast$}}
\put(2126,1065){\makebox(0,0){$\ast$}}
\put(2244,1138){\makebox(0,0){$\ast$}}
\put(2362,1215){\makebox(0,0){$\ast$}}
\put(2481,1294){\makebox(0,0){$\ast$}}
\put(2599,1367){\makebox(0,0){$\ast$}}
\put(2717,1426){\makebox(0,0){$\ast$}}
\put(2835,1461){\makebox(0,0){$\ast$}}
\put(2650,347){\makebox(0,0){$\ast$}}
\color{black}
\polygon(472,1716)(472,265)(2835,265)(2835,1716)
\put(72,990){\rotatebox{-270}{\makebox(0,0){homogenisation error}}}
\put(1653,58){\makebox(0,0){$\eps$}}
\end{picture}